\documentclass[prx,twocolumn,longbibliography,superscriptaddress,preprintnumbers]{revtex4-1}
\usepackage{graphicx}
\usepackage{latexsym}
\usepackage{amssymb}
\usepackage{amsmath}
\usepackage{amsfonts}
\usepackage{upgreek}
\usepackage{bm}
\usepackage{multirow}
\usepackage{enumitem}
\usepackage{xcolor}
\usepackage{hyperref}
\usepackage{physics}
\usepackage{array}
\usepackage{comment}
\usepackage{lipsum}
\usepackage[normalem]{ulem}

\hypersetup{
colorlinks = true,
linkcolor = [rgb]{0.70,0.13,0.13},
citecolor = [rgb]{0.13,0.55,0.13},
urlcolor = [rgb]{0.25, 0.41, 0.88}}
\newcommand{\U}{\mathrm{U}}
\newcommand{\SU}{\mathrm{SU}}

\newcommand{\sgn}{\operatorname{sgn}}

\newcommand{\appref}[1]{Appendix.\,\ref{#1}}
\newcommand{\eqnref}[1]{Eq.\,\eqref{#1}}
\newcommand{\figref}[1]{Fig.\,\ref{#1}}
\newcommand{\tabref}[1]{Tab.\,\ref{#1}}

\newcommand{\rd}{\partial}

\newcommand{\cc}{c^{\vphantom{\dagger}}}

\newcommand{\hc}{\,\textrm{H.c.}}

\newcommand{\Ovbs}{{O}_\textrm{VBS}}
\newcommand{\Ozafm}{{O}_\textrm{zAFM}}
\newcommand{\Heff}{H_\textrm{eff}}

\begin{document}

\title{Landau-Forbidden Quantum Criticality in Rydberg Quantum Simulators}

\author{Jong Yeon Lee}\thanks{jongyeon@kitp.ucsb.edu}
\affiliation{Kavli Institute for Theoretical Physics, University of California, Santa Barbara, CA 93106, USA}
\author{Joshua Ramette}
\affiliation{Department of Physics, Massachusetts Institute of Technology, Cambridge, MA 02139, USA}
\affiliation{Research Laboratory of Electronics, Massachusetts Institute of Technology, Cambridge, MA 02139, USA}
\author{Max A. Metlitski}
\affiliation{Department of Physics, Massachusetts Institute of Technology, Cambridge, MA 02139, USA}
\author{Vladan Vuletic}
\affiliation{Department of Physics, Massachusetts Institute of Technology, Cambridge, MA 02139, USA}
\affiliation{Research Laboratory of Electronics, Massachusetts Institute of Technology, Cambridge, MA 02139, USA}
\author{Wen Wei Ho}
\affiliation{Department of Physics, Stanford University, Stanford, CA 94305, USA}
\author{Soonwon Choi}
\affiliation{Department of Physics, Massachusetts Institute of Technology, Cambridge, MA 02139, USA}
\affiliation{Laboratory for Nuclear Science, Massachusetts Institute of Technology, Cambridge, MA 02139, USA}

\date{\today}

\preprint{MIT-CTP/5452}

\begin{abstract} 
The Landau-Ginzburg-Wilson theory of phase transitions precludes a continuous transition between two phases that spontaneously break distinct symmetries. However, quantum mechanical effects can intertwine the symmetries, giving rise to an exotic phenomenon called deconfined quantum criticality (DQC). In this work, we study the ground state phase diagram of a one-dimensional array of individually trapped neutral atoms interacting strongly via Rydberg states, and demonstrate through extensive numerical simulations that it hosts a variety of symmetry-breaking phases and their transitions including DQC. We show how an enlarged, emergent continuous symmetry arises at the DQCs, which can be experimentally observed in the joint distribution of two distinct order parameters,  obtained within measurement snapshots in the standard computational basis. Our findings highlight quantum simulators of Rydberg atoms not only as promising platforms to experimentally realize such exotic phenomena, but also as unique ones  allowing access to physical properties not obtainable in traditional experiments.
\end{abstract}
\maketitle

The modern theory of continuous phase transitions is rooted in the Landau-Ginzburg-Wilson~(LGW) framework.
The central idea is to describe   phases and their transitions using  order parameters: local observables measuring spontaneous symmetry-breaking~(SSB).
In recent years, however, new kinds of critical behavior beyond this paradigm have been shown to exist.
For example, quantum phase transitions~(QPT) between phases with and without topological order are characterized not by   symmetry-breaking but rather by singular changes in   patterns of long-range quantum entanglement.
Another example is the continuous QPT 
between distinct SSB phases of certain two-dimensional magnets\,\cite{Senthil2004, Senthil2004_Science}. Such a scenario is generally forbidden within the LGW framework since there is no {\it a priori} reason why the order parameter of one phase vanishes {\it concomitantly} as the order parameter of another develops.

\begin{figure}[!t]
    \centering
    \includegraphics[width = 0.49 \textwidth]{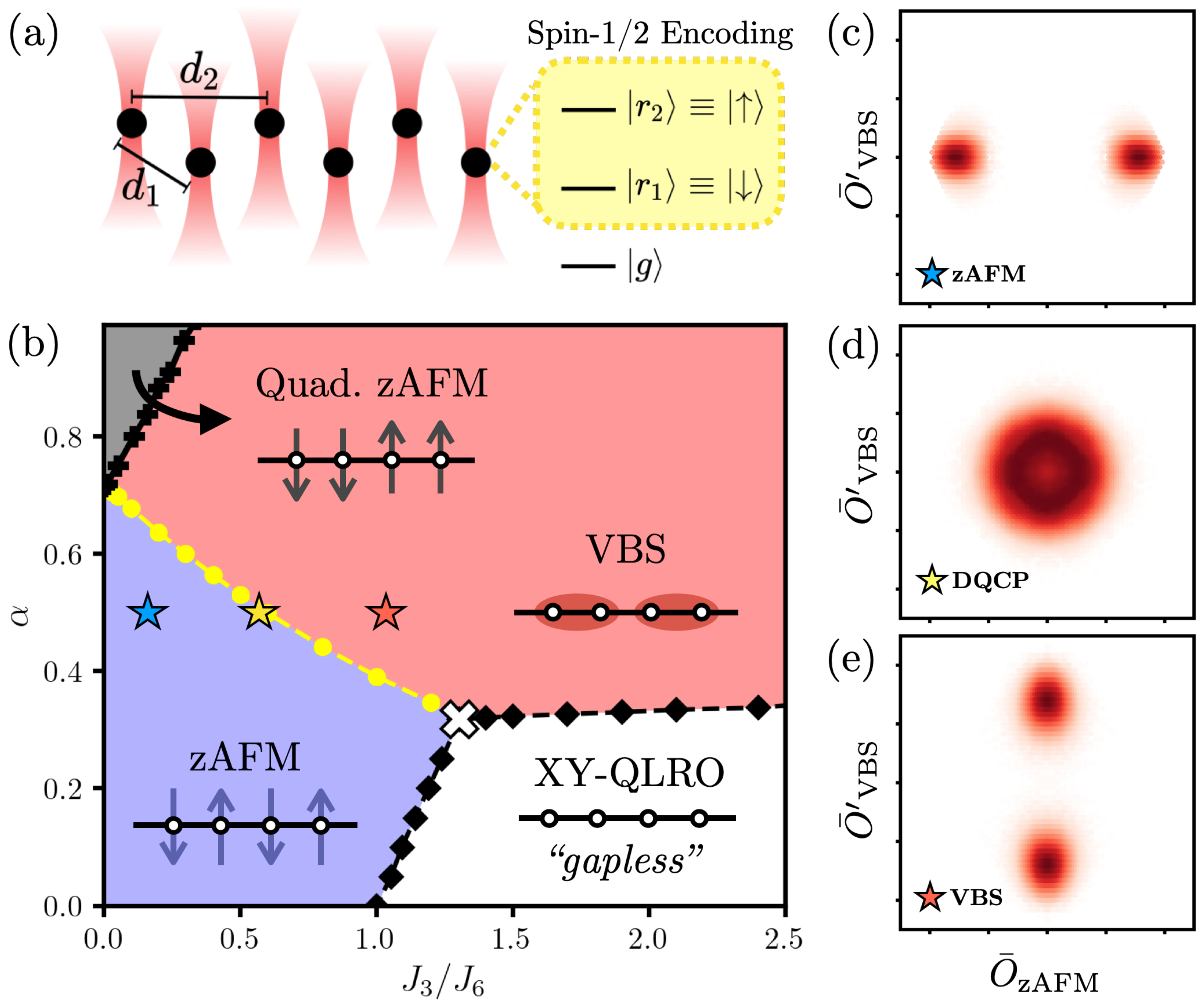}
    \caption{\label{fig:1} (a)~Zig-zag arrangement of atoms and  spin-$\frac{1}{2}$ encoding utilizing Rydberg states.
    (b)~Ground state phase diagram of  $\Heff$.
    Shaded regions depict different quantum orders  as described in the main text. Solid points are extracted numerically by finite-size-scaling method.
    Dashed~(solid) lines depict (dis)continuous QPTs. Yellow~(black) dashed lines are deconfined quantum~(BKT) critical points.
    (c,d,e) Joint distribution of zAFM and VBS order parameters over $2\times10^4$ $z$-basis snapshots, computed   at the three different markers in the phase diagram~($L$\,$=$\,$192$).
    The ring-shaped distribution in (d) is the hallmark of an emergent $\U(1)$ symmetry arising at the DQCP. 
    }
    \vspace{-20pt}
\end{figure}

Deconfined quantum criticality~(DQC) is a unifying framework proposed to explain such unconventional behavior: instead of order parameters, these critical points are described by emergent fractionalized degrees of freedom interacting via deconfined gauge fields.
This can lead to interesting measurable consequences in macroscopic phenomena, such as emergent symmetries and accompanying  conserved currents\,\cite{Nahum2015SO5, Thorngren2017, Wang2017, Ma2017DQCP, MaDQCP2019}.
However, despite numerous experimental proposals and attempts\,\cite{ Kuklov2008DQCP, Chen2009DQCP, Lou2009DQCP, Charrier2010DQCP, LoopModel2011DQCP, Harada2013DQCP, Block2013DQCP, DQCP_correction2013, Meng2017DQCP, Sato2017DQCP, Ma2017DQCP, Shao2017DQCP, Ippoliti2018DQCP,Lee2018FQH, Sandvik2018O4,  Nahum2018O4, Lee2019DQCP, Huang2019, Jiang2019DQCP, Mudry2019DQCP, KZ_DQCP, Robert2021DQCP, Zou2020, aliceaIsing}, DQC is to date still a largely theoretical concept, and an unambiguous experimental observation remains to be made.

In this Letter, we propose programmable quantum simulators based on arrays of Rydberg atoms as promising platforms to  realize and verify DQC. These are systems of  atoms individually trapped by optical tweezers, and pumped by lasers to highly excited Rydberg states  through which they interact.
Owing to their wide programmability, a host of interesting quantum many-body phenomena can be simulated\,\cite{scar2017, SPT2019, KZ2019, SL2021, Saffman}.
Here, we similarly leverage their programmability to present a realistic model of interacting spin-$\frac{1}{2}$ particles in 1D, and show 
that a host of SSB phases and QPTs, including DQC, arise~[\figref{fig:1}(a,b)]. 
Furthermore, we demonstrate the emergence of an enlarged, continuous symmetry --- a smoking gun signature of DQC --- is readily observable in experiments through the joint distribution of two order parameters over global measurement snapshots~[\figref{fig:1}(c-e)]. 

\begin{figure*}[!t]
    \centering
    \includegraphics[width = 1 \textwidth]{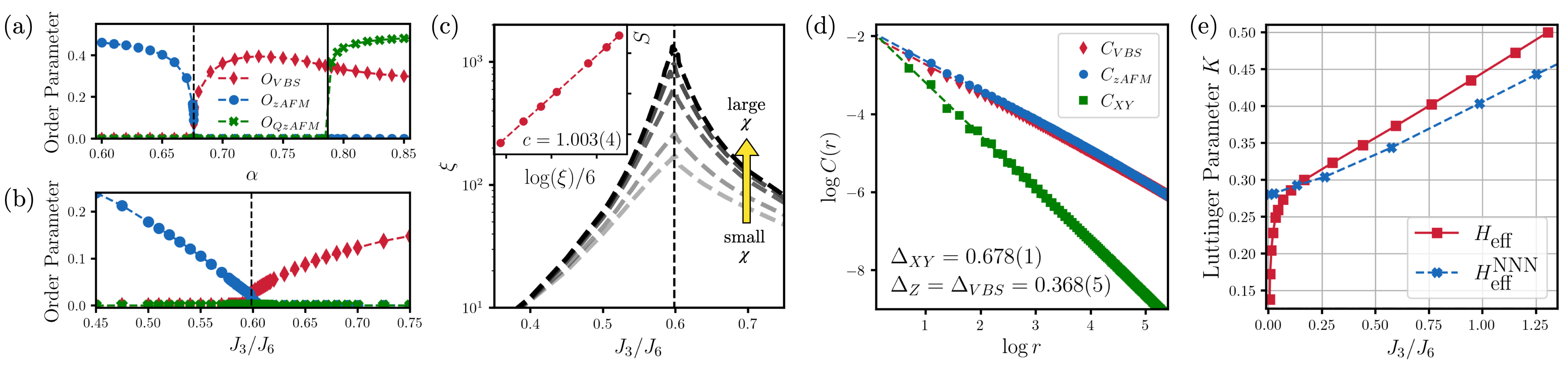}
    \caption{\label{fig:2}
    (a,b) Magnitudes of spatially averaged VBS, zAFM, and QzAFM order parameters along different cuts in the phase diagram. (a) $J_3/J_6$\,$=$\,$0.1$. 
    The system is zAFM(VBS) ordered to the left(right) of the critical point $\alpha_{c1}$\,$\sim$\,$0.677$. Past   $\alpha_{c2}$\,$\sim$\,$0.787$ the system transitions to QzAFM.
    (b) $\alpha$\,$=$\,$0.5$. The system is zAFM(VBS) ordered to the left(right) of the critical point $J_3/J_6$\,$\sim$\,$0.5968$, whereupon both order parameters vanish with increasing bond dimension in DMRG numerics (and therefore correlation length $\xi$)~\cite{SM}. 
    (c) Divergence of correlation lengths at the critical point of (b) with increasing bond dimension $\chi$\,$=$\,$70$\,$,$\,$100$\,$,$\,$200$\,$,$\,$300$\,$,$\,$400$.
    [Inset]  Scaling of   entanglement entropy versus (log) correlation length yields a slope $c$\,$=$\,$1$. 
    (d) Correlation functions behavior  at the critical point of (b). (e) Luttinger parameter along the 
    DQCP line in the phase diagram of $\Heff$, and its NNN truncation $\Heff^\textrm{NNN}$. 
    }
\end{figure*}

{\it Model.}---We study an array of neutral atoms trapped in optical tweezers and arranged in a 1D zig-zag structure~(\figref{fig:1}a), with periodic boundary conditions imposed by closing the chain into a ring.
An effective spin-$\frac{1}{2}$ degree of freedom $\{\ket{\uparrow}$,$\ket{\downarrow}\}$ is taken to be encoded by distinct highly excited Rydberg states of each atom.
Then, the effective many-body Hamiltonian for each total $z$-magnetization of spins becomes:
\begin{align} \label{eq:H}
    & \Heff = J_3 \sum_r \Big[ (X_r X_{r+1} + Y_r Y_{r+1}) + \alpha (X_r X_{r+2}  \\
    &+ Y_r Y_{r+2})  \Big] + J_6 \sum_r \Big[ Z_r Z_{r+1} + \alpha^2 Z_r Z_{r+2}  \Big] + H_\textrm{LR}. \nonumber
\end{align}
Above, $X,Y,Z$ are standard Pauli-matrices; $J_3$\,$,$\,$J_6$ quantify   strengths of spin-exchange and Ising interactions for nearest-neighbor pairs of atoms respectively; $\alpha$\,$\equiv$\,$(d_1/d_2)^3$ and $\alpha^2$ govern the relative strengths of the different nearest-neighbor~(NN) to next nearest-neighbor~(NNN) couplings, where $d_1~(d_2)$ is the NN~(NNN) atomic distance. $H_\textrm{LR}$ contains long-range terms beyond NNN arising from both dipolar and van der Waals~(vdW) interactions, which decay with distance as $1/r^3$ and $1/r^6$ respectively.
$\Heff$ is distinct from conventional Hamiltonians previously realized in Rydberg simulators~\cite{scar2017, SPT2019, KZ2019, SL2021}: it contains \textit{both} Ising and exchange couplings. 
Importantly, we assume the ability to independently tune  parameters ($\alpha$\,$,$\,$J_3/J_6$)  over a wide range of values; 
we will demonstrate how to achieve this experimentally later.

{\it Quantum phases.}---We aim to ascertain the ground state phase diagram of $\Heff$ over ($\alpha$\,$,$\,$J_3/J_6$) at zero magnetization. 
By inspecting its symmetries:~translation symmetry with a spin-$\frac{1}{2}$ per unit cell,  $\U(1)_z$\,$\times$\,$\mathbb{Z}^x_2$ symmetry (spin rotation[spin-flip] about the $z[x]$-axis respectively), and site-centered inversion symmetry ${\cal I}$, one can already declare that all phases must either be SSB or gapless.
This stems from the  Lieb-Schultz-Mattis theorem~\cite{LSM_original, LSM, Hastings2005, Oshikawa2000}, which forbids a gapped disordered phase under such symmetry  considerations.

Salient features of the phase diagram can be understood upon truncating $\Heff$ to at most NNN terms, i.e.,~ignoring $H_{LR}$~\cite{Parreira1997, Laflorencie2005, absenceCSB2017}.  
When $J_3$\,$=$\,$0$, the system is purely classical~\cite{Zarubin2020}. There is however a competition (tuned by $\alpha$) between NN Ising interactions, which  induce antiferromagnetic~(zAFM) order in the $z$-direction that spontaneously breaks the $\mathbb{Z}_2^x$ spin-flip symmetry, and NNN Ising interactions, which induce instead so-called quadrupled antiferromagnetic order~(QzAFM),  further breaking  ${\cal I}$.
These phases are separated by a first-order transition at   $\alpha$\,$=$\,$1/\sqrt{2}$ (modified with $H_{LR}$).
When $\alpha$\,$=$\,$0$, the model reduces to the familiar XXZ model, which hosts zAFM order at $J_3$\,$<$\,$J_6$, and a symmetric but gapless  XY phase with quasi-long range order~(XY-QLRO) at $J_3$\,$>$\,$J_6$.
A final limiting case is when $\alpha$\,$=$\,$1/2$ and $J_3/J_6$\,$\rightarrow$\,$\infty$, called the Majumdar-Ghosh point~\cite{MG1970}. There, the valence bond solid~(VBS) states describing dimerized patterns of spin singlets   are the ground states, which spontaneously break ${\cal I}$. Caricatures of the different orders are shown in Fig.~\ref{fig:1}b.

We numerically verify the presence of all these phases for the full model $\Heff$ with long-range interactions.
Concretely, we consider the order parameters: 
\begin{align} \label{eqn:order_pams}
{O}_\textrm{zAFM}(r)& \equiv  e^{i\pi r} Z_r, \quad\,\,  {O}_\textrm{QzAFM}(r) \equiv  e^{i\pi r/2} Z_r,\nonumber \\
{O}_\textrm{VBS}(r) &\equiv   e^{i\pi r} \qty[\vec{S}_{r+1} \cdot \vec{S}_r - \vec{S}_{r} \cdot  \vec{S}_{r-1} ].
\end{align}
which measure  violations of symmetries: $\mathbb{Z}_2^x$ with wavevector $\pi$ and $\pi/2$, and ${\cal I}$ respectively. We also consider their correlations $C_a(r)$\,$\equiv$\,$\expval{O_a(0) O_a(r)}$, and     $C_{XY}(r)$\,$\equiv$\,$\expval{X(0) X(r)}$\,$=$\,$\expval{Y(0) Y(r)}$ detecting ordering in the easy-plane.
Employing a density-matrix renormalization group~(DMRG) algorithm for infinite systems\,\cite{DMRG1, DMRG2, DMRG3}, we compute Eq.~\eqref{eqn:order_pams} along various cuts of the phase diagram.

Focusing first along a vertical cut $J_3/J_6$\,$=$\,$0.1$~(Fig.~\ref{fig:2}a), we see that for $\alpha$\,$<$\,$\alpha_{c1}$\,$=$\,$0.678(1)$, the system is zAFM ordered, evinced by a non-zero  ${O}_\textrm{zAFM}$ and vanishing ${O}_\textrm{VBS}$\,\cite{iDMRG_misc}. When $\alpha_{c1}$\,$<$\,$\alpha$\,$<$\,$\alpha_{c2}$\,$=$\,$0.787(1)$, the converse happens, indicating the system is  VBS ordered. For $\alpha$\,$>$\,$\alpha_{c2}$, another phase appears wherein ${O}_\textrm{VBS}$ remains non-zero, while ${O}_\textrm{QzAFM}$ appears in  discontinuous fashion; this is the QzAFM phase. The derivative of the ground state energy  across this transition is seen  not to be smooth, indicating that it is a first-order transition.
For the horizontal cut $\alpha$\,$=$\,$0.5$, we see     the system has zAFM(VBS) order to the left(right) of $J_3/J_6$\,$=$\,$0.5968$~(Fig.~\ref{fig:2}b). 
Interestingly, the two order parameters, ${O}_\textrm{zAFM}$ and ${O}_\textrm{VBS}$, appear to   vanish/appear continuously precisely at this same point---indication   this QPT is unconventional. Further evidence of its continuous nature is provided by a divergent correlation length seen in DMRG simulations with increasing accuracy, enabled by increasing bond dimension~(\figref{fig:2}c); 
scaling of the von Neumann entanglement entropy with correlation length also yields a central charge $c$\,$=$\,$1$ \cite{Calabrese_2009}, indicative of an underlying CFT. 
Lastly, on the horizontal cut $\alpha$\,$=$\,$0.2$~(see~\cite{SM}),  at small $J_3/J_6$ we observe that $O_\textrm{zAFM}$ is non-zero as expected, while it goes very smoothly to zero for larger $J_3/J_6$, with no obvious discontinuity in any of its derivatives.
This suggests that the QPT crossed is of Berezinskii-Kosterlitz-Thouless~(BKT) type~\cite{BKT_1,BKT_2,BKT_misc}.
Plots of the zAFM, VBS and XY correlation functions in the large $J_3/J_6$ regime yield that they all decay with power laws, with   $C_\textrm{XY}(r)$   decaying slowest~\cite{SM}; we thus identify this to be the gapless XY-QLRO phase.

Using these methods, the full topology of the phase diagram can be ascertained,   depicted  in \figref{fig:1}b; we more carefully determined the precise phase boundaries via the method of level spectroscopy, see~\cite{Nomura_1994, OshikawaKT2021, SM} for details.

{\it Deconfined quantum criticality (DQC)}---We hone in on the continuous QPT between the zAFM and XY phases, which  above investigations already strongly suggest  is an  example of DQC\,\cite{Haldane1982DQCP, Huang2019, Jiang2019DQCP, Mudry2019DQCP}.
More insight is given by a field theory analysis: 
using the Jordan-Wigner transformation followed by bosonization~\cite{Giamarchi2004}, we obtain the continuum  Hamiltonian\,\cite{SM} 
\begin{equation} \label{eq:eff_H}
    H \propto \int_0^L \dd x \qty[ \frac{1}{K} (\rd_x \phi)^2 + K (\rd_x \theta )^2 ] + g_4 \cos4\phi + \cdots.
\end{equation}
Above, $K$ is the so-called Luttinger parameter; $\phi$\,$,$\,$\theta$ are   bosonic fields obeying $[\phi(x)$\,$,$\,$\rd \theta(x')]$\,$=$\,$i\pi \delta(x-x')$, so that original~(spin) order parameters are expressed as:
\begin{align}\label{eq:mapping}
    \Ozafm \sim \cos 2\phi, \qquad \Ovbs \sim \sin 2\phi, \nonumber \\
    \,\,{O}_\textrm{xAFM} \sim \cos \theta, \qquad \,\, {O}_\textrm{yAFM} \sim \sin \theta.
\end{align}
The  microscopic $\U(1)_z$ spin-rotation symmetry manifests   as the transformation $\theta$\,$\mapsto$\,$\theta$\,$+$\,$\alpha$ for arbitrary $\alpha$, translation symmetry 
as $\phi$\,$\mapsto$\,$\phi$\,$+$\,$\pi/2$ and $\theta$\,$\mapsto$\,$\theta$\,$+$\,$\pi$, and site-centered inversion as $\phi$\,$\mapsto$\,$-\phi$.
Therefore, symmetry-allowed terms beyond the   parenthesis  in \eqref{eq:eff_H} have the structure $\cos 4n\phi$.
Now, for $K$\,$>$\,$1/2$, it can be shown that all such terms are irrelevant under renormalization group~(RG) flow so that the system is gapless (specifically, a Luttinger liquid), corresponding to the XY-QLRO phase~\cite{LR_misc}.
However, for $1/8$\,$<$\,$K$\,$<$\,$1/2$, the $n$\,$=$\,$1$ term is relevant, so that non-zero $g_4$ leads to   condensation of $\phi$\,$=$\,$0$ or $\pi/4$ depending on sign,  corresponding to the (gapped) zAFM and VBS phases. Crucially, at the  critical point $g_4$\,$=$\,$0$, an enlarged $\U(1)$ symmetry, associated with $\phi$\,$\mapsto$\,$\phi$\,$+$\,$\beta$ for arbitrary $\beta$,
is seen to emerge (recall higher order terms can be ignored~\cite{SM}).
This emergent symmetry, characteristic of a DQCP, implies that the   ground state is invariant under   a continuous transformation that rotates  $\Ozafm$ into $\Ovbs$ and back.
Consequently,  $C_\textrm{zAFM}(r)$ and $C_\textrm{VBS}(r)$  are expected to exhibit power-law decays with identical exponents, as verified in \figref{fig:2}d. 

The boundary between the zAFM and VBS phases is in fact a line of DQCPs~(yellow line of \figref{fig:1}b). 
Along this line, we numerically find the Luttinger parameter $K$ varies  from $1/2$ at the tricritical point~(white cross of \figref{fig:1}b) to  $\approx$\,$0.137$ at the smallest value of $J_3/J_6$ we could reliably simulate, see~\figref{fig:2}e. 
We expect that $K$ still decreases for even smaller $J_3/J_6$ down to $1/8$, whereupon the DQC becomes destabilized 
as the next-order term in Eq.~\eqref{eq:eff_H} becomes relevant, which is expected to drive  a discontinuous transition or phase coexistence~\cite{SM}. 
Interestingly, interactions further than NNN appear  crucial to the small values of $K$ observed~(see \figref{fig:2}e and~\cite{SM}).

\begin{figure*}[!t]
    \centering
    \includegraphics[width = 1 \textwidth]{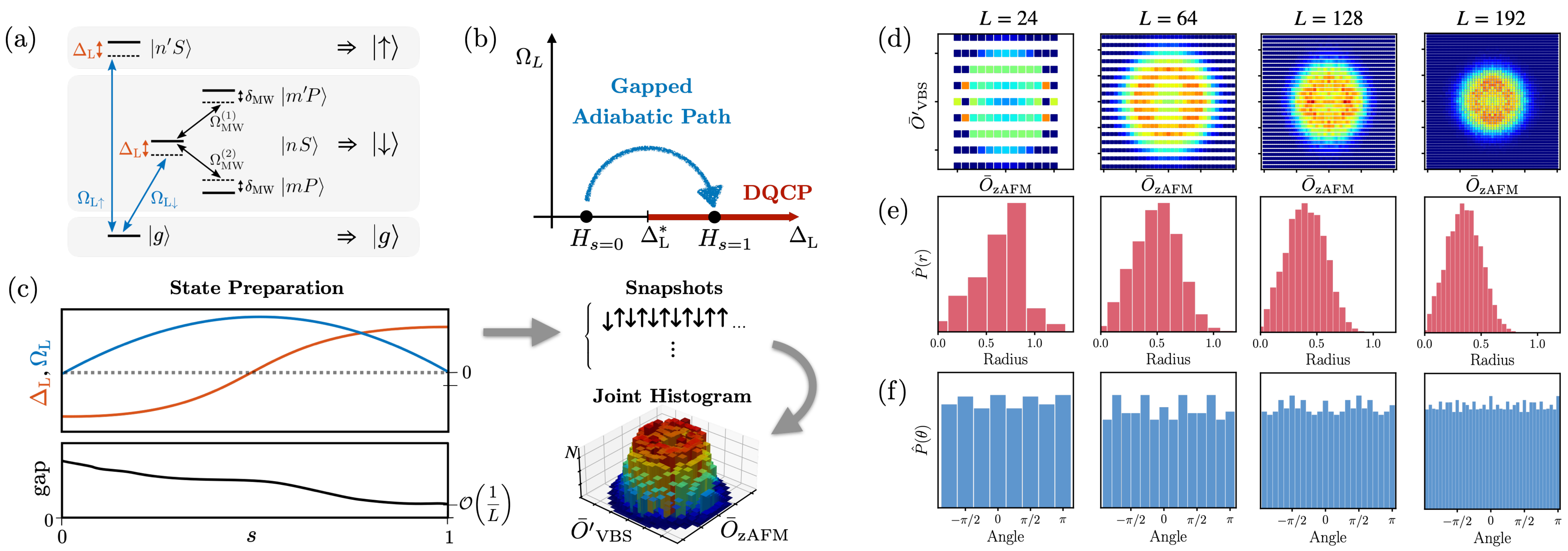}
    \vspace{-15pt}
    \caption{\label{fig:3} (a) Proposed spin-state encoding  enabling tunable $J_3/J_6$ couplings. 
    We admix $\ket{mP}$ and $\ket{m'P}$ into $\ket{nS}$ via   microwave drives of strengths $\Omega_\textrm{MW}^{(1,2)}$ to form $\ket{\downarrow}$.
    Lasers with Rabi frequencies $\Omega_{\textrm{L} \uparrow}$/$\Omega_{\textrm{L} \downarrow}$ couple  the   ground state $\ket{g}$ to   respective Rydberg spin states  with detuning $\Delta_\textrm{L}$, used in the state preparation protocol.
    (b) Schematic phase diagram of $H(s)$\,$\equiv$\,$\Heff$\,$+$\,$H_\ell(s)$ and adiabatic path taken.
    (c) Ramp profiles of $\Omega_L(s)$\,$,$\,$\Delta(s)$ considered for the state preparation, and accompanying many-body gap. 
    (d,e,f) JPDs of order parameters $\bar{O}_\textrm{zAFM}$\,$,$\,$\bar{O}'_\textrm{VBS}$, and corresponding radial and angular distributions derived from $2$\,$\times$\,$10^4$ simulated measurement outcomes at the DQCP ($\alpha$\,$,$\,$J_3/J_6)$\,$\simeq$\,$(0.5$\,$,$\,$0.597)$ for various system sizes.
    }
\end{figure*}

{\it Experimental protocol.}---In order to realize the above physics in the laboratory, we have to address three challenges: 
(i)~engineering $\Heff$ with tunable parameters, 
(ii)~devising an efficient protocol to   prepare a critical ground state,
and (iii)~providing a measurement and data processing procedure to identify signatures of DQC.

%

Tunable $\alpha$ is easily achieved by geometrically rearranging atoms using optical tweezers. For tunable $J_3/J_6$, we propose encoding each spin state 
as an \emph{admixture} of Rydberg states with different parities.
As a concrete example, we choose $\ket{\uparrow}$ and $\ket{\downarrow}$ both as states of the same parity, denoted as $S$, and we further dress $\ket{\downarrow}$ with two nearby states of opposite parity, denoted as $P$, using independent off-resonant microwave drives~(\figref{fig:3}a).
Without admixing, vdW interactions between $S$ states give rise to $1/r^6$-decaying Ising-like couplings as already demonstrated in multiple experiments~\cite{cat_state, scar2017, KZ2019, SL2021}.
Admixing $P$ states generally introduces $1/r^3$-decaying dipolar interactions that contain both spin exchange and Ising couplings.
Here, by judiciously choosing two different $P$ states, it is possible to engineer a  negligible diagonal dipole moment of the dressed state,  while keeping substantial off-diagonal (transition) dipole moment such that only exchange couplings are realized~\cite{adjust_dipolar2021}.
In this way, one can tune $J_3/J_6$ over a wide range, from nearly zero to greater than unity, with even a modest amount of admixture~\cite{SM}.
This realizes  $\Heff$ up to a uniform global Zeeman field, i.e., $\propto \sum_j Z_j $, which is inconsequential as long as our state preparation protocol lands us in the desired magnetization sector.
%
We note that utilizing other microwave dressing schemes is possible~\cite{SM} and also that exact engineering of $\Heff$ is not needed as the existence of DQC is robust against perturbations.


To prepare the DQCP ground state, we propose an adiabatic protocol. 
Three remarks are in order:
First, in experiments, atoms are 
typically initialized in their respective electronic ground states $\ket{g}^{\otimes L}$; 
thus, a state preparation protocol   necessarily involves  an extended Hilbert space of three internal states $\{\ket{g}, \ket{\uparrow}, \ket{\downarrow}\}$ per atom.
Second, we desire to prepare the ground state of $\Heff$ in the zero magnetization sector, which may not be the global ground state considered over all magnetization sectors.
Finally, given   finite coherence times in experiments, the many-body gap should ideally remain large throughout the adiabatic passage so that   state preparation can be completed as quickly as possible while minimizing diabatic losses. 

We present a many-body trajectory that satisfies all three criteria: $H(s) = \Heff + H_\ell (s)$ with $s\in [0, 1]$, where $\Heff$ is assumed tuned to a desired DQCP, and  
\begin{align} \label{eq:H_prep_laser}
H_\ell(s) &= \sum_{i}\Omega_{\textrm{L}}(s)  \Big(\ket{\sigma(i)}_i \bra{g} + h.c.\Big) + \Delta_{\textrm{L}}(s) \sum_{i}\ket{g}_i\bra{g} \nonumber
\end{align}
%
%
%
represents lasers coupling 
$\ket{g}$ to spin states $\ket{\sigma(i)}$ with $\sigma(i) = \uparrow(\downarrow)$ for even (odd) sites $i$, 
characterized by time-dependent Rabi frequencies $\Omega_{\textrm{L}}(s)$\,and detunings\,$\Delta_L(s)$~(\figref{fig:3}a-c). 
Now, under a sufficiently slow, smooth ramp  up of $\Delta_\textrm{L}$ from a large negative to positive value while $\Omega_\textrm{L}$ is switched on and off, all population from $\ket{g}$ will be transferred to the spin states (tantamount to an adiabatic rapid passage~\cite{beterov2020}).
%
Furthermore, as we prove in~\cite{SM}, $H(s)$ harbors two independent conserved quantities throughout the entire evolution, which ensures that the 
final 
state has   zero magnetization, provided all population in $\ket{g}$ is transferred.
Such a protocol thus ensures that the instantaneous ground state of $H(s$\,$=$\,$0)$ is $\ket{g}^{\otimes L}$ while that of $H(s$\,$=$\,$1)$ is the target DQCP.
Finally, the choice of staggered couplings explicitly breaks translation symmetry 
except at the start and end of the trajectory, opening  the many-body gap away from the  DQCP, which we numerically observe~(\figref{fig:3}b,c).
 
To demonstrate the protocol's feasibility, we consider $\Omega_L(s)$\,$=$\,$J_6 \sin(\pi s)$ and $\Delta_L(s)$\,$=$\,$-2J_6 \cos(\pi s)$, fixing $(\alpha$\,$,$\,$J_3/J_6)$\,$=$\,$(0.5,0.597)$~(i.e.,~a DQCP).
Up to $L$\,$=$\,$12$, we can perform exact simulations with realistic values $J_6$\,$\sim$\,$2\pi$\,$\times$\,$25$MHz~(used in~\cite{cat_state}) assuming a linear ramp  $s(t)$\,$=$\,$t/T$, which reveals that a state with  many-body overlap$\sim$\,$0.99$ with the exact groundstate can be prepared with the state-preparation time $T$\,$=$\,$60/J_6$\,$\sim$\,$0.4\upmu$s, well within typical Rydberg lifetimes\,$\sim$\,$150 \upmu$s~\cite{cat_state}.
Furthermore, based on the Kibble-Zurek scaling ansatz~\cite{Kibble_1976, Zurek1985, KZ2019}, we find that the condition for the adiabaticity is $T\gtrsim L^{3-4K}$. 
Combined with exact numerical results, we estimate that a system of $L$\,$=$\,$24$ can be prepared with a state-preparation time $T$\,$\sim$\,$1\mu$s, and $L$\,$=$\,$64$ with $T$\,$\sim$\,$5\mu$s.

The smoking-gun signature of DQC is the emergent symmetry unifying different order parameters. 
We now argue this can be directly observed in Rydberg simulators. 
Na\"ively, an explicit way to verify the emergent symmetry is to   measure arbitrary linear combinations of order parameters $O_\eta$\,$=$\,$O_\textrm{zAFM} \cos\eta$\,$+$\,$O_\textrm{VBS} \sin \eta$ and to show that the distribution of $O_\eta$ behaves identically for any $\eta$ upon potential rescaling of $O_\textrm{zAFM}$ and $O_\textrm{VBS}$.
This approach, however, is infeasible with existing experimental technologies as measuring $O_{\eta\neq0}$ requires applying highly-complicated unitary rotations before performing measurements in the standard $z$-basis.
%
%
Instead, we can consider  ${O'}_\textrm{VBS}(r)$\,$\equiv$\,$(-1)^r(Z_{r+1} Z_r$\,$-$\,$Z_rZ_{r-1})$, 
which behaves identically to $O_\textrm{VBS}(r)$ under  symmetry transformations relevant to $H_\textrm{eff}$,  and hence serves as an alternative, but bona~fide VBS order parameter~\cite{Singlet_misc}.
%
%
%
%
%
%
Now  $\bar{O}_\textrm{zAFM}$ and ${\bar{O}'}_\textrm{VBS}$~(bar denotes spatial averaging) are \emph{simultaneously} evaluable within  global measurement snapshots in the standard $z$-basis~(\figref{fig:3}c). Such measurements in fact give access to the entire statistical properties of $\bar{O}_\textrm{zAFM}$ and ${\bar{O}'}_\textrm{VBS}$, captured by their joint probability distribution~(JPD).
Figure~\ref{fig:3}(d-f) illustrates the JPD and corresponding radial/angular distributions,  derived from    simulated snapshots at a DQCP   for various system sizes~\cite{SM}. Already at  $L$\,$=$\,$24$,  the rotational invariance between the order parameters can   be gleaned, which becomes increasingly prominent with larger sizes.
Note that this ring distribution would not arise if the transition were instead  characterized only by a simple co-existence of   zAFM and VBS orders: the JPD would have four distinct peaks, amounting to overlaying distributions of \figref{fig:1}c,e.


{\it Conclusion and outlook.}---In this work, we have studied a realistic 1D model of interacting neutral atoms, and showed that it hosts interesting quantum phases and   transitions, including deconfined quantum criticality. We also proposed an experimental protocol to image the emergent symmetry associated with   DQC, paving the way for a novel, categorical verification of this long sought-after, unconventional quantum criticality in Rydberg quantum simulators.

Interestingly, our finding of DQC described by a 1D Luttinger-liquid  with a small Luttinger parameter $K$ can possibly be leveraged to realize   another exotic physics: higher-dimensional non-Fermi liquids (NFL)~\cite{NFL_review1,NFL_review2}.
We can imagine an array of critical 1D chains in parallel with non-zero interchain tunnelings but parametrically negligible interchain Ising interactions. For small enough $K<1/4$, the system should remain gapless with no proper quasiparticles, a characteristic feature of NFLs~\cite{crosssliding_NFL, SlidingLL2001, Leviatan2020}. This thus represents a concrete blueprint to experimentally construct a family of 2D NFLs, opening doors for a systematic study of such physics.

\begin{acknowledgments}
{\it Acknowledgments.}~JYL is supported by GBMF 8690 and NSF PHY-1748958. JR and VV is supported by NSF QLCI-CI-2016244, NSF PHY-1734011, and DOE 032054-0000. MM is supported by the NSF DMR-1847861. WWH is supported in part by the Stanford Institute of Theoretical Physics. Computing resources were administered by the Center for Scientific Computing (CSC) and funded by the National Science Foundation (CNS-1725797). This work was performed in part at the Aspen Center for Physics, which is supported by NSF PHY-1607611.
\end{acknowledgments}

\appendix

\setcounter{equation}{0}
\setcounter{figure}{0}
\setcounter{table}{0}

\makeatletter
\renewcommand{\thefigure}{S\arabic{figure}}
\renewcommand*{\bibnumfmt}[1]{[S#1]}
\setcounter{subsection}{0}

\begin{widetext}
 
\begin{center}
\textbf{\large Supplementary Material for \\ \vspace{7pt}
``Landau-Forbidden Quantum Criticality in Rydberg Atom Arrays''}

\vskip6mm

{\noindent \normalsize  Jong Yeon Lee, Joshua Ramette, Max A. Metlitski, Vladan Vuletic, Wen Wei Ho, Soonwon Choi}

\vskip3mm

\end{center}

In this supplementary material, we present 
details on the various methods used to ascertain the phase diagram, as well as details on the experimental protocol to realize deconfined quantum criticality (DQC). 
In \appref{app:ED}, we elaborate on the exact diagonalization study of the phase diagram, in particular explaining the  method of level spectroscopy, used to determine precisely the location of the quantum phase transitions (QPT). This includes both the Berezinskii-Kosterlitz-Thouless~(BKT) transitions as well as deconfined quantum critical points (DQCP). We also explain how this technique can be used to extract the Luttinger parameter along the DQCP line of our phase diagram. 
In \appref{app:bosonization}, we review and apply the technique of bosonization, which allows us to derive the field theoretic description underlying the DQCP of our model.
In \appref{app:degeneracy}, we discuss the relation between the  possible values of the Luttinger parameter $K$ attainable which is compatible with DQC. 
In \appref{app:model}, we explain in  detail how to experimentally engineer our model Rydberg Hamiltonian,  using the spin-1/2 encoding scheme introduced in the main text. 
In \appref{app:preparation}, we describe a general state preparation protocol for realizing ground states of interacting Rydberg-encoded spin models that can be used to target a ground state in a desired magnetization sector in the presence of the magnetization conservation symmetry.
In \appref{app:observable}, we explain in more detail how the signature of   DQC can be gleaned from the joint probability distribution (JPD) of the VBS and zAFM order parameters within finite-size systems.
Finally, in \appref{app:long_range}, we discuss how we simulate long-range Hamiltonian in the DMRG simulations.


\tableofcontents








\section{Details on Exact Diagonalization Study of the Phase Diagram}\label{app:ED}

For a thorough investigation of the model Hamiltonian, we desire (i) to identify the possible quantum phases at zero temperature, (ii) to locate the phase boundaries (i.e., critical points) of the model accurately in the thermodynamic limit, and (iii) to extract critical exponents at continuous phase transitions.

However, in numerics we are limited to simulating only finite size systems. Thus in order to locate phase boundaries precisely, finite-size scaling of order parameters~\footnote{More precisely, one investigates the finite-size scaling behaviors of an \emph{squared} order parameter $O$ detecting symmetry-breaking, since $\expval{O}=0$ in the ground state of a finite size system, which is symmetric.} or derivatives of energy are often employed. In our case, we can alternatively leverage the fact that the critical points we are analyzing are between quantum phases with different symmetry properties,
like zAFM-VBS, which spontaneously break spin-flip and site inversion respectively, and zAFM-XY, the latter of which does not break any symmetries but exhibits  {characteristic power-law correlations}. 
In such situations, the so-called method of level spectroscopy~\cite{Nomura_1994, OshikawaKT2021} can be employed.

The key idea is to determine the phase transition point in the thermodynamic limit, by extrapolating from the level crossing of first excited states in finite-size systems, see \figref{fig:schematic}.
The crucial point is that unlike the ground state obtained in finite size systems which is generally symmetric across parameter space, the numerically obtained first excited states from each side of symmetry broken (or quasi-long-range ordered) phases belong to different symmetry sectors, i.e., they transform differently under different symmetry generators of the system.
Therefore, the first excited states from two sides of the phase do not exhibit an avoided level-crossing, 
and instead directly cross.
We can identity such a crossing point as the `critical point' of a finite size system, see \figref{fig:schematic}. 
Then, by performing a finite size scaling of the locations of these crossing points, we can extract the  transition point in the thermodynamic limit with very good accuracy, which often tends to converge very fast with system size.
In the following sections, we elaborate on the method  as well as its theoretical foundations~\cite{DiFrancesco:1997nk}. Finally, we  argue why this method is particularly useful for analyzing the BKT phase transition~\cite{BKT_1, BKT_2}, and also for DQC.

\begin{figure}[!t]
    \centering
\includegraphics[width = 0.6 \textwidth]{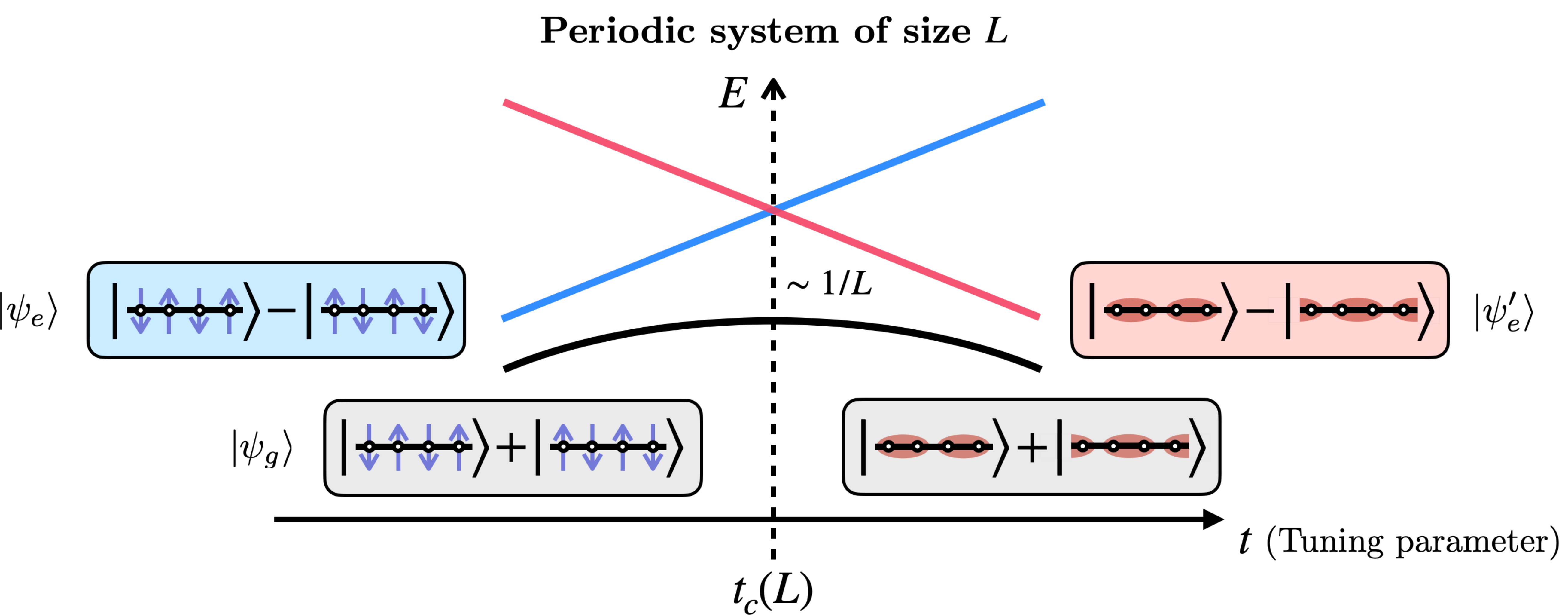}
    \caption{\label{fig:schematic} Schematic diagram of exact energy eigenvalues for the ground state $\ket{\psi_g}$ as well as the first and second excited states,  $\{ \ket{\psi_e}, \ket{\psi_e'}\}$ across the deconfined quantum critical point $t_c(L)$ in  finite size  systems of size $L$, assuming periodic boundary conditions. Note that at finite size, there is a gap scaling as $1/L$ at the critical point. Although the phases on either side of the phase transition (zAFM and VBS) 
    are expected to spontaneously-break (different) symmetries, note that the ground state $\ket{\psi_g}$ is always unique and hence symmetric.
    For example, deep in the zAFM phase, the ground state is well-captured by the symmetric superposition of two antiferromagnetic classical configurations (i.e., a ``cat state''). Deep in the VBS, the ground state is well-captured by the symmetric superposition of two dimerized singlet patterns (also a ``cat state'').
    One sees that in both instances, $\ket{\psi_g}$ transforms trivially under both spin-flip transformation $\mathbb{Z}_2^x$ and site-centered inversion transformation $\cal I$, and in fact, this property holds true for all values of the tuning parameter $\alpha$.
    Instead, the signature of spontaneous symmetry-breaking shows up in the  excited states: 
    $\ket{\psi_e} \mapsto - \ket{\psi_e}$ under the spin-flip transformation while it remains invariant under site-centered inversion; and conversely $\ket{\psi_e'} \mapsto - \ket{\psi_e'}$ under site-centered inversion but remains invariant under the spin-flip transformation. Therefore, all three states belong to different symmetry sectors. Importantly, there  is therefore necessarily a direct (i.e., non-avoided) crossing between the excited states $\ket{\psi_e}$ and $\ket{\psi_e'}$, whose energy difference to $\ket{\psi_g}$ should go as $\sim 1/L$ as expected at a continuous phase transition; an emergent $U(1)$ symmetry of the DQC implies that this crossing point occurs at the true critical point in the thermodynamic limit. }
\end{figure}

\subsection{Level Spectroscopy Method from State-Operator Correspondence}

The level spectroscopy method can be rigorously understood from conformal field theory (CFT), a long-wavelength description 
appropriate for many systems at criticality.
Consider a one-dimensional periodic system with size $L$ and the CFT Hamiltonian $H_\textrm{CFT}$, a lattice Hamiltonian \emph{exactly} realizing a certain conformal field theory. The celebrated \emph{state-operator correspondence}~\cite{DiFrancesco:1997nk} posits that the energy spectrum of $H_\textrm{CFT}$ is related to the scaling dimensions of the operators appearing in the CFT  via the following relation:
\begin{equation} \label{eq:CFT_correspondence}
    H_\textrm{CFT} \ket{\psi_\alpha} = E^\textrm{CFT}_\alpha \ket{\psi_\alpha} \quad \rightarrow \quad 
    E^\textrm{CFT}_\alpha  = \frac{2\pi v}{L} \qty(\Delta_\alpha-\frac{c}{12}),
\end{equation}
where $\ket{\psi_\alpha}$ is an eigenstate with label $\alpha$, $v$ is the speed of gapless degrees of freedom, $c$ is the central charge, and $\Delta_\alpha$ is the scaling dimension of the corresponding operator. From this one sees that the hierarchy of the excited energy spectrum of critical Hamiltonians with periodic boundary conditions directly corresponds to the hierarchy of the scaling dimensions of the operators of the corresponding conformal field theory.

However, this relation is only for the \emph{exact} CFT Hamiltonian. In practice, a generic lattice Hamiltonian $H_\textrm{lattice}$ at criticality does not realize the exact CFT Hamiltonian, but instead is the sum of $H_\textrm{CFT}$ and extra irrelevant perturbations $\lambda \int \dd{x} O_\textrm{CFT}(x)$, where $O_\textrm{CFT}(x) = \sum_\beta a_\beta \psi_\beta(x)$ and $\psi_\beta(x)$ is an operator with scaling dimension $\Delta_\beta$. Such perturbations give a correction to \eqnref{eq:CFT_correspondence} for finite-size systems, which can be calculated as such:
\begin{equation}
    E^\textrm{lattice}_\alpha  = E_\alpha^\textrm{CFT} + \lambda \int \dd{x} \bra{\psi_\alpha} O_\textrm{CFT}(x) \ket{\psi_\alpha}.
\end{equation}
The expectation value of operators within states can be obtained through the so-called operator product expansion (OPE), which yields\footnote{This follows from a radial quantization in the CFT through a logarithmic mapping which transforms the plane to the cylinder.} 
\begin{equation}
    \bra{\psi_\alpha} \psi_\beta \ket{\psi_\alpha} = C_{\alpha \beta \alpha}  \qty(\frac{2\pi}{L})^{\Delta_\beta},
\end{equation}
where $C_{\alpha\beta\alpha}$ is the so-called OPE coefficient, which is of order one. We  see that the leading correction to $E_\alpha^\textrm{lattice}$ at large $L$ is dominated by the smallest $\Delta_\beta$  among the symmetry-allowed irrelevant perturbations. After performing the integration, we obtain
\begin{equation} \label{eq:cft_lattice}
    E_\alpha^\textrm{lattice} = \frac{2\pi v}{L} \qty(\Delta_\alpha-\frac{c}{12}) + \frac{C_{\alpha \beta}}{L^{\Delta_\beta - 1}}.
\end{equation}
where the proportionality constant $C_{\alpha \beta}$ depends on the OPE coefficient $C_{\alpha \beta \alpha}$ and the magnitude of perturbation $\lambda$.
Now, assume that we perturb the system away from the critical point by tuning operator $O_t$ carrying scaling dimension $\Delta_t$, with a coefficient strength $\bar{t} = (t-t_c)$ where $t_c$ is the true transition point at thermodynamic limit (which we desire to determine). Let the energy correction from  this tuning operator be $K_{\alpha t}(L) \bar{t}/L$. Then we get
\begin{equation}
    E_\alpha^\textrm{lattice}(t,L) = \frac{2\pi v}{L} \qty(\Delta_\alpha-\frac{c}{12}) + \frac{C_{\alpha \beta}}{L^{\Delta_\beta - 1}} + \frac{K_{\alpha t}(L)}{L} \bar{t} \equiv \frac{2\pi v}{L} \qty(\Delta^\textrm{lattice}_\alpha(t,L)-\frac{c}{12}).
\end{equation}

Now that we understand the energy spectrum of the lattice Hamiltonian at criticality (assumed described by CFT), we can use this structure to identify the critical point $t_c$. The method is applicable for  critical points with two operators $O_a$ and $O_b$, such that (i) they have the same scaling dimensions $\Delta_a = \Delta_b$, and (ii) their corresponding states in the state-operator correspondence separate out in energy  as we tune away from the criticality. If these two conditions are met, their energy crossing point $t$ in a finite size system $L$ can be determined as such: 
\begin{equation}
    \Delta^\textrm{lattice}_a(t,L) =  \Delta^\textrm{lattice}_b (t,L) \quad \Rightarrow \quad t =t_c - \frac{C_{a \beta} - C_{b \beta}}{K_{a t} - K_{b t}} \frac{1}{L^{\Delta_\beta - 2}}.
\end{equation}
In general, $K_{ab}(L) \equiv K_{a t} - K_{b t}$ and $C_{ab}(L) \equiv C_{a \beta} - C_{b \beta}$ are functions of $L$, which would provide a further correction. Perturbative CFT arguments imply that the $L$ dependence of $K$ is determined by the scaling dimension of tuning operator $\Delta_t$ ($\Delta_t < 2$) such that $K_{ab}(L) \sim L^{2-\Delta_t}$. Therefore, the energy crossing point $t$ at system size $L$ is given by 
\begin{equation} \label{eq:finite_scaling}
    t(L) = t_c - \frac{T(L)}{L^{\Delta_\beta - \Delta_t}}, \qquad T(L) \equiv C_{ab}(L) \frac{L^{2-\Delta_t}}{K_{ab}(L)}.
\end{equation}
In most cases, $C_{ab}(L) \sim \textrm{const}$ and $K_{ab}(L) \sim L^{2-\Delta_t}$ and therefore $T(L)$ can be taken as a fixed constant. However, if the CFT has a marginally irrelevant perturbation, $C_{ab}(L)$ and $T(L)$ can acquire a logarithmic dependence on $L$. In practice, for small systems we study here the logarithmic dependence is so small that its presence barely affects the precise determination of $t_c$. 
Generally, there is more than one irrelevant operator in the lattice critical Hamiltonian, and the finite-size correction will contain contributions from all of them.
To summarize, assuming a phase transition to be identified has an underlying CFT description and two operators with the same scaling dimensions but which move oppositely in energy upon tuning away from criticality, one can precisely locate the critical points using \eqnref{eq:finite_scaling}.

Given the relation in Eq.~\ref{eq:finite_scaling}, we can perform the finite size scaling in the following way.
For a given system size, we extract the level-crossing point $t(L)$ of the first excited states. We choose the operators $O_\alpha$ and $O_\beta$ corresponding to the operators with different quantum numbers with the same (lowest) scaling dimension. Then, using the first expression in  Eq.~\ref{eq:finite_scaling}, we fit the extracted $t(L)$ using the fitting function $t(L) = t_c - a L^{-b}$ with three fitting parameters $t_c$, $a$ and $b$.

\begin{figure}[!t]
    \centering
    \includegraphics[width = 1 \textwidth]{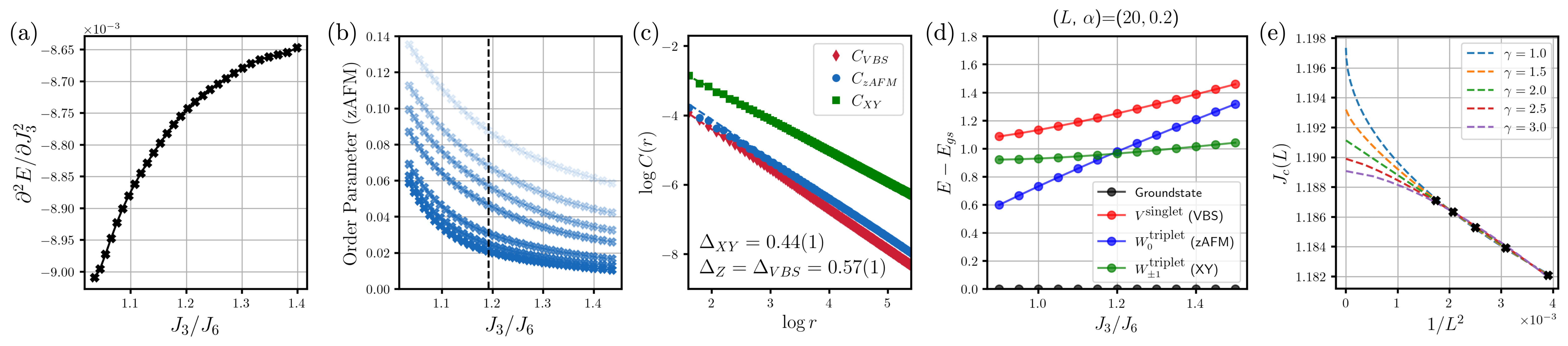}
    \vspace{-10pt}
    \caption{\label{fig:KT_data} Behavior of the system for $\alpha=0.2$ and $J_3/J_6 \in [1.0,1.4]$. (a) Second derivative of energy. (b) zAFM order parameters under iDMRG simulations with   bond dimension $\chi=70,100,200,300,400$ (bond dimension increasing with darkness). Note that unlike at the DQCP, neither the order parameter nor the second derivative of energy changes sharply at the transition.
    (c) Correlation functions $C_{XY}$, $C_\textrm{zAFM}$ and $C_\textrm{VBS}$ at $(J_{3}/J_6,\alpha) = (1.435,0.2)$ corresponding to the XY-QLRO phase. All correlation functions exhibit power-law behavior, with the XY correlation function decaying the slowest.
    (d) Low-lying spectrum (i.e., eigenvalues) of $\Heff$ for system size $L=20$. We use the crossing between two energies $W_0^\textrm{triplet}$ and $W_{\pm 1}^\textrm{triplet}$ to extract the location of the BKT transition.
    (e) Finite size scaling of the crossing point $J_c(L) \equiv J_3/J_6$ at $\alpha=0.2$ between the energies corresponding to zAFM and XY operators with $L=16,18,20,22,24$. 
    In order to verify our expectation on leading irrelevant operators, we fitted with a range of $\gamma\equiv \Delta_\beta - \Delta_t$ in \eqnref{eq:finite_scaling}.
    The curve with $\gamma=2$ fits the best, which is the value expected for the BKT transition, giving $J_c(L\rightarrow \infty) = 1.19114(2)$. Logarithmic corrections can modify this value only slightly.
    }
\end{figure}

\subsection{BKT transitions}

To identify the possible quantum phases of the model Hamiltonian as well as their phase transitions, we performed iDMRG simulations, tracking order parameters across different parameter cuts. However, the phase transitions out of the gapless XY phase (either into the zAFM or VBS phase) pose some challenges. In \figref{fig:KT_data}a,b, we plot the second derivative of the ground state energy and behavior of the order parameter across the phase transition between the XY-QLRO phase and zAFM phase. We note that both the second derivative of energy and the order parameter behave smoothly across the phase transition.

In fact, such smooth behavior of the energy and order parameter is consistent with the Berezinskii-Kosterlitz-Thouless~(BKT) transitions~\cite{BKT_1,BKT_2}, a topological phase transition from a gapped to a gapless phase. BKT transitions are famous for their  logarithmic corrections as well as being a transition with \emph{infinite smoothness}, which makes it difficult to identify their precise location based on conventional methods developed for second-order continuous phase transitions. 
However, it turns out that the level spectroscopy method combined with scaling of the finite size crossing points provides a great solution to identify the phase boundaries accurately, due to the cancellation of  logarithimic corrections, which we will elaborate on below.




Let us first understand the nature of  logarithmic corrections at a BKT transition. At such a transition, the presence of a marginally irrelevant operator generates an anomalous dimension for the operators whose corresponding states appear in the energy spectrum. 
Consider the following perturbation $H'$ on top of the Gaussian Hamiltonian, appearing in the continuum description of our model~\eqnref{eq:critical_ham}:
\begin{equation}
H' = \int \frac{\dd x}{2 \pi} \qty[-y_K \qty( (\partial_t \phi)^2 + (\partial_x \phi)^2 ) + y_V \cos 4\phi ].
\end{equation}
The coefficients $y_K$ and $y_V$ appear in front of  marginal operators and flow under RG according to 
\begin{equation}
    d y_K/d l = - y_V^2, \quad d y_V/dl = - y_K y_V, \quad l = \log L, 
\end{equation}
where $L$ is the length scale of the RG flow. In order to understand the BKT transition, we introduce new variables $s = y_K - \eta y_V$ and $t = y_K + \eta y_V$ such that the BKT transition is characterized by $s=0$ (Note that there are two such boundaries corresponding to $\eta = \pm 1$. For $y_V >0$ the transition is at $\eta = 1$ and for $y_V < 0$ the transition is at $\eta = -1$). 
At the BKT transition, $t$ can still be non-zero; in fact, it obeys the following RG flow: 
\begin{equation}
    d t/dl = - t^2/2 \quad \Rightarrow \quad t \sim 1/\log L. 
\end{equation}
This implies that even if $s=0$, $\abs{y_{K,V}} \sim 1/\log L$, which vanishes slowly as $L \rightarrow \infty$ but nonetheless gives a significant correction to the physical quantities at finite $L$. (Such a correction can be considered as a shift in the Luttinger parameter from the value $1/2$ in our model; note that for a classical BKT transition, $y_V$ is for the perturbation $\cos 2 \phi$ and the Luttinger parameter at the transition is 2.)

In terms of $y_K$ and $y_V$, the scaling dimensions of the operators corresponding to the VBS, zAFM, and XY order parameters are~\cite{OshikawaKT2021}:
\begin{align} \label{eq:log_correction}
    \Delta_{XY} &\approx \frac{1}{2} - \frac{1}{4} y_K \nonumber \\
    \Delta_\textrm{zAFM} &\approx \frac{1}{2} + \frac{1}{4} y_K + \frac{1}{2} y_V \nonumber \\
    \Delta_\textrm{VBS} &\approx \frac{1}{2} + \frac{1}{4} y_K - \frac{1}{2} y_V,    
\end{align}
where $\Delta_{XY} \equiv \Delta_{e^{\pm i \theta}}$, $\Delta_{zAFM} \equiv \Delta_{\cos 2\phi}$, and $\Delta_{VBS} \equiv \Delta_{\sin 2\phi}$ as defined in Section \ref{app:bosonization}. 
As $y_{K,V} \sim 1/\log L$, for any finite size system scaling dimensions acquire significant corrections, and so do corresponding excited state energies in \eqnref{eq:CFT_correspondence}. 
However, a careful examination shows that for $\eta=1$ ($y_K = y_V \sim 1/\log L$), anomalous dimensions for $\Delta_{XY}$ and $\Delta_\textrm{VBS}$  match, and for $\eta = -1$ ($y_K = - y_V$), anomalous dimensions for $\Delta_{XY}$ and $\Delta_\textrm{zAFM}$ match:
\begin{align}
                &y_K = y_V \quad \Rightarrow \quad \Delta_{VBS} = \frac{1}{2}  - \frac{1}{4} y_V = \Delta_\textrm{XY} \quad \textrm{(VBS to XY-QLRO transition)} \nonumber\\
        &y_K = -y_V \quad \Rightarrow \quad \Delta_{zAFM} = \frac{1}{2}  + \frac{1}{4} y_V = \Delta_\textrm{XY} \quad \textrm{(zAFM to XY-QLRO transition)} 
\end{align}
This matching is not a coincidence, but is rather due to the enhancement of the $U(1)$ symmetry to $\SU(2)$ at the BKT transition. For the zAFM to XY-QLRO transition, the zAFM and X, Y components of the XY order parameter transform as a vector under this symmetry. Similarly, for the VBS to XY-QLRO transition, the VBS and XY order parameters form a vector. At the transition point all excited states can be labeled by an $\SU(2)$ spin quantum number $S$; we denote by $V$ the singlet $S=0$, and by $W$ the triplet $S = 1$.  Therefore, even in the presence of logarithmic corrections,
the crossing for the pair of (zAFM, XY) energy levels can be identified as the BKT transition point for the zAFM-XY transition, and the crossing for the pair of (VBS, XY) energy levels can be identified as the BKT transition point for the VBS-XY transition.

In \figref{fig:KT_data}d, we numerically plot the energy levels through the zAFM-XY transition (for a particular sweep), where we observe a crossing between the first and second excited states, which correspond to the zAFM/XY CFT operators through the operator-state correspondence. Although we expect all excited states for VBS, zAFM, and XY order parameters to have the same energy in thermodynamic limit, logarithmic corrections in \eqnref{eq:log_correction} separate the VBS excited state from the others.

To  proceed, we obtain the energy spectrum of a given Hamiltonian by exact diagonalization technique (for larger system sizes, one can perform finite DMRG for excited states). Then, by examining their quantum numbers, we can identify the corresponding CFT operators in the given excited states energy spectrum. 
In \eqnref{eq:symmetry}, we elaborate on how bosonic fields $(\theta, \phi)$ transform under the microscopic symmetries of the system. Since CFT operators corresponding to the order parameters are represented in terms of $(\theta, \phi)$, we can immediately identify their quantum numbers under the spin rotation in $z$-direction $S^z_\textrm{tot}$, translation ($k$), bond-centered inversion ($\cal I_{\rm bond}$), and spin-flip ($g \equiv \prod_i X_i$) symmetries as in \tabref{tab:order_qm}.
\begin{table}[!h]
    \centering
\renewcommand{\arraystretch}{1.25}
\begin{tabular}{|>{\centering}m{8em}|*{4}{>{\centering\arraybackslash}m{4em}|}}
\hline 
Operators  &  $S^z_\textrm{tot}$ & $k$ & $\cal I_{\rm bond}$ & $g$  \\
\hline 
VBS  & $0$ & $\pi$ & 1 & 1 \\
\hline
zAFM  & $0$ &$\pi$ & -1 & -1 \\
\hline
XY  & $\pm 1/2$ &$\pi$ & -1 & N/A \\
\hline
\end{tabular}
\caption{\label{tab:order_qm} Quantum numbers carried by CFT operators corresponding to each order parameters. Since we block-diagonalize the Hamiltonian in total spin-$z$ and momentum quantum numbers first, spin-flip quantum number is well-defined only in the zero total spin-$z$ sector.}
\end{table}

However, it is not enough to identify CFT operators through the excited states energy spectrum. Note that the states corresponding to the CFT primary operators are generated by inserting operators with respect to the groundstate. Therefore, an eigenstate corresponding to a certain order parameter will carry the sum of quantum numbers of the groundstate and the order parameter.

In order to identify the quantum numbers of the groundstate across the phase transition, we can perform a perturbative analysis. Consider the groundstate manifold in the e.g. zAFM symmetry broken phase. In the fixed-point limit, the groundstate manifold simply consists of two symmetry broken states, $\ket{\uparrow \downarrow \uparrow ...}$ and $\ket{\downarrow \uparrow \downarrow ...}$. However, when we perturb the system, two symmetry broken states become hybridized, forming symmetric and anti-symmetric combinations as eigenstates of the system.  
From degenerate perturbation theory, we can show that two ground states are connected at order  $(L/2)$. If $L=4n$, since couplings are all positive, the matrix element generated is always negative. Therefore, the symmetric superposition has lower energy. However, for $L=4n+2$, the $2n+1$-th order perturbative contribution connecting two ground states is positive, and the anti-symmetric superposition has a lower energy. Since two ground states are exchanged under translation, bond-centered inversion, and spin flips, the symmetric superposition carries trivial quantum numbers while the anti-symmetric superposition carries $-1$ quantum numbers for those symmetry operators. The quantum number carried by the groundstate is summarized in \tabref{tab:gs_qm}. 
\begin{table}[!h]
    \centering
\renewcommand{\arraystretch}{1.25}
\begin{tabular}{|>{\centering}m{8em}|*{4}{>{\centering\arraybackslash}m{4em}|}}
\hline 
System size  &  $S^z_\textrm{tot}$ & $k$ & $\cal I_{\rm bond}$ & $g$  \\
\hline 
$L=4n$  & $0$ & $0$ & 1 & 1 \\
\hline
$L=4n+2$  & $0$ & $\pi$ & $-1$ & $-1$ \\
\hline
\end{tabular}
\caption{\label{tab:gs_qm} Quantum numbers carried by the groundstate for different system sizes.}
\end{table}
Therefore, using \tabref{tab:order_qm}  and \tabref{tab:gs_qm}, one can obtain the quantum numbers of the state corresponding to a desired CFT operator in any finite-size system. For example, when $L=4n+2$, the VBS state will carry the quantum number $(S^z,k,{\cal I_{\rm bond}}, g) = (0,0,-1,-1)$. Based on this understanding, we can perform exact diagonalization of the Hamiltonian, which is block-diagonalized by symmetry quantum numbers, to identify energies corresponding to specific CFT operators.

Finally, we discuss the result of finite-size scaling analysis for BKT transitions using \eqnref{eq:finite_scaling}. 
At the BKT transition, $\Delta_\beta=4$ is the lowest scaling dimension of the  irrelevant symmetry-allowed perturbations, which break the emergent $\SU(2)$ symmetry and can thus lift the the degeneracy of zAFM and XY order parameters. We find two such operators with Lorentz spin $\ell = \pm2$ and $S=2$: $O^{ab}_{3,1} =(L_{-2} J^a_{-1} \bar{J}^b_{-1} + a \leftrightarrow b- {\rm trace})|0\rangle$ and  $O^{ab}_{1,3} = (\bar{L}_{-2} J^a_{-1} \bar{J}^b_{-1}  + a \leftrightarrow b - {\rm trace})|0\rangle$, and two operators with Lorentz spin $\ell= \pm 4$ and $S =2$: $O^{ab}_{4,0} = (J^a_{-2} J^b_{-2} +  a \leftrightarrow b - {\rm trace})|0\rangle$ and $O^{ab}_{0,4} = (\bar{J}^a_{-2} \bar{J}^b_{-2}+  a \leftrightarrow b - {\rm trace})|0\rangle$. Here we are using the state operator correspondence and $\SU(2)_1$ current algebra notation, where $J^a_{n}$ ($\bar{J}^a_n$) is the generators of $\SU(2)$ left (right) current algebra, and $L_n$ ($\bar{L}_n$) generators of the left (right) Virasoro algebra.
Inversion symmetry fixes a linear combination of the operators with $\ell = \pm 2$ and a linear combination  of operators with $\ell = \pm 4$. $U(1)$ symmetry picks out the operator with $m = 0$ out of the five $S = 2$ operators. Thus, we find two operators with $\Delta = 4$ that break $\SU(2)$, but preserve the microscopic symmetries, and are not total derivatives (i.e., global descendants). Note in particular that there is no such operator with $\ell = 0$: the only non-descendant with this quantum number is $T \bar{T}$, which is an $\SU(2)$ singlet and does not lift the degeneracy between zAFM and XY operators.

Both $O_{1,3}$, $O_{3,1}$ and $O_{4,0}$, $O_{0,4}$ are expected to acquire anomalous dimensions due to the presence of marginally irrelevant operator at the KT transition. Likewise, the marginal operator driving the KT transition (with coefficient $s$) acquires an anomalous dimension. Thus, we expect logarithmic corrections to $C(L)$ and $K(L)$ in \eqnref{eq:finite_scaling}. However, the effect of logarithmic correction is negligible at small system sizes we examine.
Using ED up to system size $L=24$, which can be done with a standard laptop in a few minutes, we determine the BKT transition points with high accuracy as shown in \figref{fig:KT_data}e.


\begin{figure}[!t]
    \centering
    \includegraphics[width = 0.9 \textwidth]{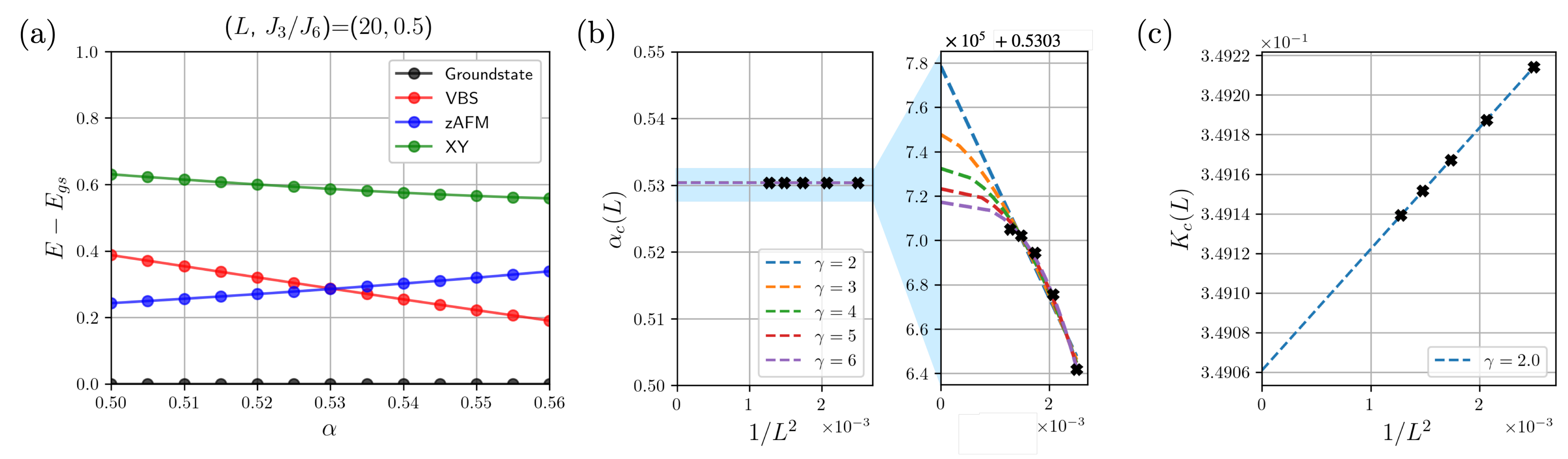}
    \vspace{-10pt}
    \caption{\label{fig:ED_DQCP} (a) Exact diagonalization spectrum of the model for $J_3/J_6=0.5$ and $\alpha \in [0.50,0.56]$ at $L=20$. Here we plot first four lowest energies. Each excited state can be labeled by relevant quantum numbers -- how it transforms under the $\U(1)_{xy}$ spin rotation, $g$ spin flip, and ${\cal I}$ site-centered inversion, and accordingly, gives the scaling dimension of the associated CFT operator. (b) Finite size scaling of the crossing point $\alpha_c(L)$ between the VBS and zAFM operators with $L=20,22,24,26,28$. Here, we demonstrate the linear regression between $\alpha_c(L)$ and $1/L^\gamma$ for $\gamma = 2,2.5,3,3.5$. In the ideal scenario, $\gamma =2$, which can change in practice. However, the results are quite insensitive to the choice of $\gamma$; 
    we obtain $\alpha_c(L\rightarrow \infty) = 0.530374(3)$. (c) Finite size scaling of the Luttinger parameter. Here, we demonstrate the linear regression between $K_c(L)$ and $1/L^2$, where we obtain $K_c(L\rightarrow \infty) = 0.349061(1)$.} 
\end{figure}

\subsection{Deconfined Quantum Critical Points}

In the main manuscript, we have shown iDMRG simulation results that indicate the presence of a continuous phase transition between VBS and zAFM phases. Since VBS and zAFM break different symmetry groups, such a phase transition corresponds to DQC. From the bosonization analysis in Section \ref{app:bosonization}, we can show that its conformal field theory description is again captured by \eqnref{eq:critical_ham} as in the case of the BKT transitions. However, unlike the BKT transition, the DQCP has no marginal operator that contributes logarithmic corrections to RG flows of operators. Therefore, the general finite-size scaling relation of  \eqnref{eq:finite_scaling} should suffice for the DQCP.

At the DQCP, CFT operators with the lowest scaling dimensions are given by $\cos 2\phi$ and $\sin 2\phi$, which correspond to the zAFM and VBS order parameters respectively. Therefore, the crossing between eigenstates corresponding to the VBS and zAFM operators is expected, as shown in \figref{fig:ED_DQCP}a. However, note that the line of deconfined quantum criticality is characterized by the Luttinger parameter $K$, which is an unknown quantity to be determined. For DQCP, we have $\Delta_\beta = 2 + 4K$ from $\delta L = [(\rd_x \phi)^2+(\rd_t \phi)^2] \cos 4\phi$,
which can lift the degeneracy of $\Ozafm$ and $\Ovbs$.
The scaling dimension of the tuning parameter $\Delta_t = 4K < 2$. 
Therefore, we expect the scaling behavior (\eqnref{eq:finite_scaling}) with $\gamma \equiv \Delta_\beta - \Delta_t = 2$. However, note that this is the expectation from the ideal scenario. In practice, the presence of other irrelevant operators, potentially with larger magnitudes, can alter the scaling behavior at small system sizes in the way dependent on critical exponents including Luttinger parameter $K$. Therefore, we performed finite size scaling with a range of $\gamma$.
It turns out that using different values of $\gamma$ for the fitting does not affect much the resulting thermodynamic limit of the crossing point, as we demonstrate in \figref{fig:ED_DQCP}(b) at $J_3/J_6=0.5$ for the system size upto $L=28$. Therefore, using the level spectroscopy method, we can determine the phase boundary accurately, as depicted in  Fig.~1b of the main text. Further, one can employ the relation \eqnref{eq:CFT_correspondence} at the DQCP to determine the value of the Luttinger parameter. Since $\dim[e^{i2\phi}] = K$ while $\dim[e^{i\theta}] = 1/4K$, it follows from \eqnref{eq:cft_lattice} that at the DQCP, 
\begin{align}\label{eq:fs_lutt} 
    \qty(\frac{E_\textrm{zAFM}-E_\textrm{gs}}{E_{XY}-E_\textrm{gs}})^{1/2} &= \sqrt{ \qty(K + \frac{C_1}{L^{\Delta_\beta-2}} + \dots) \cdot \qty(\frac{1}{4K} + \frac{C_2}{L^{\Delta_\beta-2}} + \dots)^{-1} } = 2K + {\cal O}(1/L^{\Delta_\beta-2}).
\end{align}
By performing a finite-size scaling of the above quantity, we can obtain the Luttinger parameter $K$ as in \figref{fig:ED_DQCP}(c). Note that one can perform the following finite-size scaling for the quasiparticle velocity as well:
\begin{align}\label{eq:fs_vel} 
    \qty(\frac{L}{2\pi})  \sqrt{ (E_\textrm{zAFM}-E_\textrm{gs}) \cdot (E_{XY}-E_\textrm{gs}) } &= v \sqrt{ \qty(K + \frac{C_1}{L^{\Delta_\beta-2}} + \dots) \cdot \qty(\frac{1}{4K} + \frac{C_2}{L^{\Delta_\beta-2}} + \dots) } = \frac{v}{2} + {\cal O}(1/L^{\Delta_\beta-2}).
\end{align}

\section{Bosonization of Rydberg Hamiltonian}  
\label{app:bosonization}

\subsection{Mapping into Continuum Description}

In this section, we review and apply the technique of bosonization in the context of the Rydberg   Hamitonian $\Heff$ considered in the main text, to derive our field-theoretic description of the system. We first utilize a Jordan-Wigner transformation to map spin-1/2s to spinless fermions; the map is given by $S^z_i = c^\dagger_i \cc_i - 1/2$ and $S^+_i = \cc_i \prod_{i<j}(1-2 c^\dagger_j \cc_j)$. The  effective Hamiltonian, truncated to terms with range of at most next nearest-neighbor interactions, can thus be expressed as follows:
\begin{align}
    \Heff &= -\frac{J_3 }{2} \sum_i \qty( c_i^\dagger \cc_{i+1} + \hc) - \alpha J_3  \sum_i \qty[ c_i^\dagger \qty(n_{i+1} - \frac{1}{2}) \cc_{i+2} + \hc ] \nonumber \\
    & \quad + J_6 \sum_i \qty[ \qty(n_i - \frac{1}{2}) \qty(n_{i+1} - \frac{1}{2}) +\alpha^2 \qty(n_i - \frac{1}{2}) \qty(n_{i+2} - \frac{1}{2}) ], 
\end{align}
where $n_i = c^\dagger_i \cc_i$.

In order to understand the low energy, long-wavelength physics of the ground state of this model at half-filling, we can first start from the non-interacting limit $J_6=0, \alpha = 0$. 
In such a scenario, 
the ground state is one where fermions form a Fermi sea, up to the Fermi wavevector $k_F = \pm \pi/2$. Labeling fermions near $k_F = \pi/2$ as $\psi_R$ (right-propagating) and $k_F=-\pi/2$ as $\psi_L$ (left-propagating), we can express the above Hamiltonian in the continuum limit as $\hat{H} = \int   \hat{h}(x) \dd x$, where the Hamiltonian density $\hat{h}(x)$  is given as
\begin{align}
    \hat{h}(x)/\epsilon  & =
    \sum_{\eta=L,R}\eta \psi_\eta^\dagger(r) (-i \rd_x)  \psi_\eta(r)
    + \xi \, \rho(r) \rho(r+\epsilon) + \alpha^2 \xi \, \rho(r) \rho(r+2\epsilon)
     \nonumber \\
     & + \alpha  \, \sum_{\eta,\eta'=L,R} \qty[ \psi^\dagger_{\eta}(r) \cdot   \rho(r+\epsilon) \cdot \psi_{\eta'}(r+2\epsilon) + \hc ].
\end{align}
Above, $\rho(r) = \sum_{\eta=L,R} \psi^\dagger_\eta(r) \psi^{\vphantom{\dagger}}_\eta(r)$ is the total density of the fermions, and $\epsilon$ is the length scale associated with lattice spacing, $\xi = J_6/J_3$, and we often identify $\eta = R = 1$ and $\eta = L = -1$ following the convention.

Next, we invoke \emph{bosonization} \cite{Giamarchi2004}, a well-established technique that allows us to    express fermionic variables $\psi_{\eta}$ in 1D in terms of bosonic variables $\Phi_{\eta}$: 
\begin{equation}
    \psi_\eta (x)= \hat{F}_\eta  e^{i \eta k_F x} e^{i\Phi_\eta}, \qquad [\Phi_\eta(x),\Phi_{\eta'}(x')] = i \pi  \delta_{\eta,\eta'} \sgn(x-x'),
\end{equation}
where $\hat{F}_\eta$ are so-called Klein factors. One can further introduce two variables $\theta = (\Phi_L + \Phi_R)/2$ and $\phi=(\Phi_L - \Phi_R)/2$, to equivalently express
\begin{equation}
    \psi_\eta (x)= \hat{F}_\eta  e^{i \eta k_F x} e^{i(\theta - \eta \phi)}, \qquad [\phi(x),\theta(x')] = i \frac{\pi}{2} \sgn(x-x').
\end{equation}
Under   bosonization, we therefore obtain the following Hamiltonian, upon keeping the leading order contributions:
\begin{equation} \label{eq:ham_boson}
    H = \frac{u}{2\pi} \int_0^L \dd x \qty[ \frac{1}{K} (\rd_x \phi)^2 + K (\rd_x \theta )^2 ] + g_4 \cos 4\phi + g_8 \cos 8\phi + ...
\end{equation}
where 
\begin{align} \label{eq:LL_parametes}
    u &= \epsilon  \sqrt{ \qty( 1 - \frac{4}{\pi}  \alpha) \cdot \qty( 1 + \frac{4}{\pi}\qty(\xi + \alpha))},  \nonumber \\
    K &= \qty(\frac{1 - \frac{4 \alpha}{\pi}}{1 + \frac{4}{\pi}(\xi +\alpha)})^{1/2}, \nonumber \\
    g_4 &= \frac{J_3}{4\pi^2 \epsilon} \qty( -\xi+ \alpha^2 \xi + 2 \alpha),
\end{align}
where the above parametrization of $(u,K,g_4)$ is correct at most perturbatively in $\xi$ and $\alpha$. 
This is the field theoretic model as presented in the main text. Note that in deriving this field theoretic analysis, we have ignored the long-range interaction terms present in the microscopic Hamiltonian $H_\textrm{eff}$; this is because even if we were to include any long-range interactions between physical operators (e.g.~$\rd \phi$ or $e^{ia \phi}$) which decay as $1/r^n$ with $n \geq 3$, such interaction terms are irrelevant in 1D under the renormalization group flow. Such an expectation is also supported by many other analytical and numerical studies~\cite{Parreira1997, Laflorencie2005, absenceCSB2017}.

\subsection{Symmetry Actions}

We now discuss symmetries  of the system and how they manifest themselves in different representations.
Now, the original microscopic spin Hamiltonian has   translation symmetry $T_x$, spin rotation symmetry $\U(1)_z \times \mathbb{Z}^x_2$ (spin rotation  about the $z$-axis and  spin-flip about the $x$-axis respectively), site-centered inversion ${\cal I}$, and time-reversal symmetry ${\cal T}$. These symmetries act on  microscopic spin variables and bosonic field theory variables in the following way: 
\begin{eqnarray}\label{eq:symmetry}
    T_x:&\quad S^a_i \mapsto S^a_{i+1} \quad & \quad \phi \mapsto \phi+\pi/2, \quad \theta \mapsto \theta+\pi \nonumber \\
    U(1)_z:& \quad \vec{S}_i \mapsto R(\alpha \hat{z}) \vec{S}_{i+1} \quad & \quad \theta \mapsto \theta+\alpha \nonumber \\
    \mathbb{Z}_2^x:& \quad S^{y,z}_i \mapsto -S^{y,z}_i \quad & \quad \phi \mapsto - \phi + \pi/2, \quad \theta \mapsto -\theta \nonumber \\
    {\cal I}:& \quad S^{a}_i \mapsto S^{a}_{-i} \quad & \quad \phi \mapsto - \phi  \nonumber \\
    {\cal T}:& \quad S^{a}_i \mapsto -S^{a}_i \quad & \quad \phi \mapsto - \phi + \pi/2, \quad \theta \mapsto \theta+\pi, \quad i \mapsto -i
\end{eqnarray}
Due to these symmetries, other than higher-order kinetic terms (e.g. $(\rd \phi)^4$, $(\rd \theta)^4$, ...) the only symmetry allowed terms in \eqnref{eq:ham_boson} have the structure of $\cos 4n \phi$, where $n \in \mathbb{N}$. In terms of bosonic fields $(\theta, \phi)$, original spin operators as well as the dimerization operator in continuum can be expressed as (at leading order)~\cite{Giamarchi2004}:
\begin{align} \label{eq:operator_mapping}
    S^x &\sim  \frac{1}{\sqrt{8\pi \epsilon}} \qty( (-1)^x \cos \theta - i \sin \theta \cos 2\phi ), \nonumber \\
    S^y &\sim  \frac{-1}{\sqrt{8\pi \epsilon}} \qty(  (-1)^x \sin \theta + i \cos \theta \cos 2\phi ), \nonumber \\
    S^z &\sim -\frac{1}{\pi} \rd \phi  + \frac{(-1)^x}{\pi \epsilon}\cos 2\phi, \nonumber \\
    D &\sim (-1)^x \vec{S}(x) \cdot \vec{S}(x+1) = \frac{1}{2\pi \epsilon}  \sin 2\phi, 
\end{align}
which is consistent with the symmetry transformation rules in \eqnref{eq:symmetry}.

\subsection{Renormalization Group Analysis}
\label{sec:rg}

Next, let us analyze the possible phases realized by the continuum Hamiltonian Eq.~\eqref{eq:ham_boson} within the renormalization group (RG) framework. If we first ignore the interaction terms $\cos 4n \phi$, then we obtain the following quadratic Hamiltonian
\begin{equation} \label{eq:critical_ham}
    H_0 = \frac{u}{2\pi} \int_0^L \dd x \qty[ \frac{1}{K} (\rd_x \phi)^2 + K (\rd_x \theta )^2 ]. 
\end{equation}
As the corresponding action is Gaussian, $H_0$ describes an exactly solvable model, and the scaling dimension of a generic operator can be derived:
\begin{equation}
    \dim[e^{i(a \phi + b \theta)}] = \frac{a^2 K}{4} + \frac{b^2}{4K} \quad \implies \quad \dim[\cos 4n \phi] = 4n^2 K.
    \label{eq:scaling_dim}
\end{equation}
Additionally, correlation functions are given as
\begin{align}
\label{eq:correlations}
    \expval{S_+(0) S_-(r)} &\sim (-1)^r \frac{C_1}{r^{\frac{1}{2K}}} +  \frac{C_2}{r^{2K + \frac{1}{2K}}} \nonumber \\
    \expval{S^z(0) S^z(r)} \sim \expval{D(0) D(r)} &\sim (-1)^r \frac{C_1}{r^{{2K}}} +  \frac{C_2}{r^{2}}. 
\end{align}
We see that all correlation functions decay as power laws; this is thus a gapless phase called a Luttinger liquid. 

Next, we can analyze the effect of the interactions $\cos 4n \phi$,  treating them as     perturbations away from the quadratic limit. In a one-dimensional system, an operator flows to zero under RG flow if its scaling dimension is larger than 2. When $K > 1/2$, \eqnref{eq:scaling_dim} implies that $\cos 4n\phi$ is irrelevant for all $n\geq 1$  and so the theory remains Gaussian, with   correlation functions decaying  as power laws. From Eq.~\eqref{eq:correlations}, we see that correlation decays slower in the $xy$ channel than the $z$ channel, and hence we identify this as the XY quasi long-range ordered (QLRO) phase.

When $1/8<K < 1/2$, the $\cos 4 \phi$ term becomes relevant while higher $n$ terms remain irrelevant. Therefore, the $\phi$-field can condense, and the value it takes ($0$ or $\pi/4$), depends on the sign of $g_4$, which is in turn determined by microscopic parameters $(\xi,\alpha)$. Specifically, if $g_4 < 0$, the $\phi$-field condenses at $\phi=0$ and $(-1)^x \expval{S^z} \sim  \cos 2\phi \neq 0$ and the system develops $z$AFM order, breaking $\mathbb{Z}_2^x$ spin-flip symmetry. If $g_4 > 0$, the $\phi$-field condenses at $\phi=\pi/4$ and $D \sim \sin 2\phi \neq 0$ and the system develops VBS order, breaking the site-centered inversion symmetry ${\cal I}$. Note that translation symmetry is broken in both cases.

Empirically, we find that the Luttinger parameter $K$ along the DQCP line of our model decreases as we move towards the classical limit $(J_3 \to 0)$.
An interesting question is whether $K$ of the system can decrease to an indefinitely small value while still exhibiting DQC.
Our field theory analysis suggests that the deconfined transition can become destabilized if additional symmetry-allowed perturbations become relevant, such as the $\cos 8\phi$ term (since we would then generally need two independent tuning knobs to ensure both $\cos 4\phi$ and $\cos 8\phi$ vanish). The presence of such a term, if relevant, would give rise to a non-trivial effect. Let $g_8$ be a coefficient for the term $\cos 8\phi$. For $g_8 < 0$, this term is minimized at $\phi= n\pi /4$. Combined with the effect of $g_4 \cos 4\phi$, one would expect a first-order phase transition between zAFM and VBS phases since it acts like a cubic anisotropy at Ising criticality. On the other hand, for $g_8 > 0$, this term is minimized at $\phi = (2n+1)\pi/8$. At such values of $\phi$, both $S^z$ and $D$ take non-zero values. Therefore, together with $g_4 \cos 4\phi$, this term would drive the system away from the DQCP physics into a new phase that breaks both  spin-flip $\mathbb{Z}_2^x$ and site-centered inversion ${\cal I}$ symmetries~\footnote{Note that the above field-theoretic analysis cannot capture the QzAFM phase described in the main text; while the QzAFM phase breaks the translation symmetry in such a way that it is invariant under multiple of $T_x^4$, the spin variables written in terms of $(\phi,\theta)$ in \eqnref{eq:operator_mapping} are always symmetric under $T_x^2$.}.  
The above analysis shows that for either sign for $g_8$, if $\cos 8\phi$ term is relevant, the DQCP physics would disappear. 
Therefore, as $\cos 8\phi$'s scaling dimension is $\dim[\cos 8\phi] = 16 K$, we expect that $K$ along the DQCP line should have a lower bound: $K \geq 1/8$.

\begin{figure}[!t]
    \centering
    \includegraphics[width = 0.97 \textwidth]{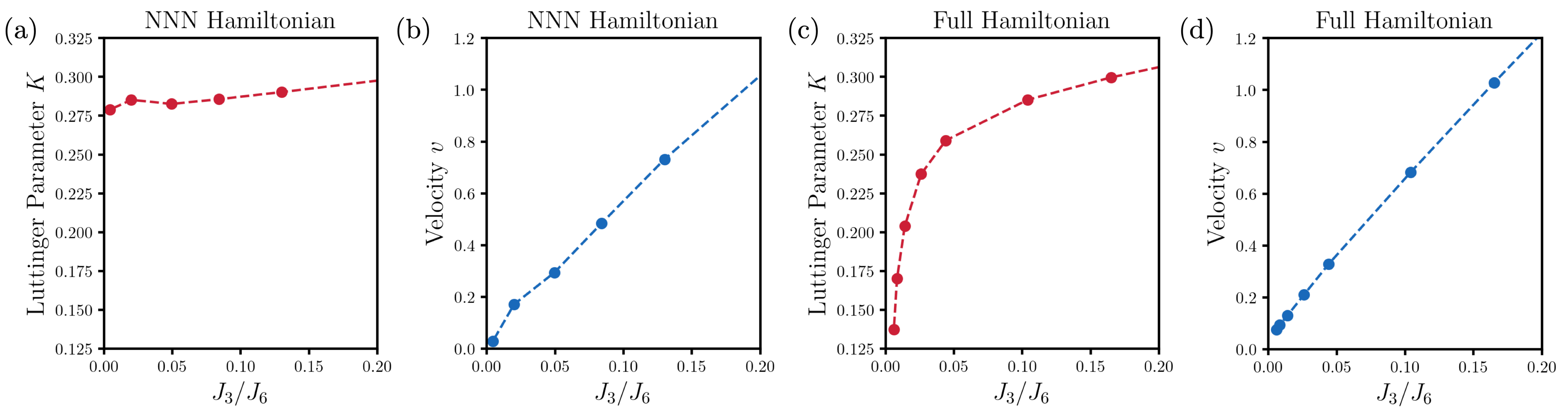}
    \vspace{-10pt}
    \caption{\label{fig:NNNvsLR} Luttinger parameter $K$ and quasiparticle velocity $v$ extracted by finite size scaling in \eqnref{eq:fs_lutt} and \eqnref{eq:fs_vel} for (a,b) NNN Hamiltonian $\Heff^\textrm{NNN}$ and (c,d) full Rydberg Hamiltonian $\Heff$ along the DQCP line. While $v \rightarrow 0$ for $\Heff^\textrm{NNN}$ as $J_3 \rightarrow 0$, $v$ approaches a finite value for $\Heff$ as $J_3 \rightarrow 0$. } 
\end{figure}

\section{Termination of DQCP along the critical line} \label{app:degeneracy}

In the main text, we have argued that   DQC exists along a line of critical points with varying Luttinger parameters up to a small value of $J_3/J_6$ ($\sim 0.005$). On the other hand, we know that at exactly $J_3/J_6=0$, there is a first-order transition from zAFM to QzAFM phases at a finite value of $\alpha$. 
This means that the DQCP line necessarily  terminates, 
either at some small but non-zero value of $J_3/J_6$, or   exactly only at $J_3/J_6=0$.
In this section, we would like to discuss the possible physical mechanisms behind this termination of the DQCP line, in particular focusing on the role of the range of interactions.

First of all, let us understand the first-order nature of the zAFM-QzAFM transition. Consider the model Hamiltonian truncated up to next-nearest-neighbor (NNN) interactions at $J_3/J_6=0$, which has only Ising-type interactions:
\begin{align}
    H &= \sum_i Z_i Z_{i+1} + \alpha^2 \sum_i Z_i Z_{i+2} \nonumber \\
    &=  \frac{1}{2} \sum_i \qty( Z_i I_{i+1} Z_{i+2} + Z_i Z_{i+1} I_{i+2} + I_i Z_{i+1} Z_{i+2} + I_i I_{i+1} I_{i+2} )    + \qty( \alpha^2-1/2) \sum_i Z_i Z_{i+2} \nonumber \\
    &= 2 \sum_i \qty( P_i^{\uparrow\uparrow\uparrow} + P_i^{\downarrow\downarrow\downarrow}) + (\alpha^2-1/2) \sum_i Z_i Z_{i+2}
\end{align}
where $P^{abc}_i$ is the projection onto the the spin-configuration $(a,b,c)_{i,i+1,i+2}$. With this rewriting of the Hamiltonian, we observe that at $\alpha^2=1/2$, the Hamiltonian is nothing but the sum of local projectors onto up-up-up and down-down-down configurations. Since there are extensively many configurations projected out by $(P_i^{\uparrow\uparrow\uparrow} + P_i^{\downarrow\downarrow\downarrow})$, at $\alpha^2=1/2$, there is an extensive degeneracy scaling like $\text{dim} \sim (\frac{1+\sqrt{5}}{2})^L$ (the degeneracy for given $L$ follows the Fibonacci sequence).
When $\alpha^2< 1/2$, the second term prefers the configuration where every other spins are aligned;  within the degenerate manifold, this corresponds to the up-down-up or down-up-down configuration, which corresponds to zAFM order. When $\alpha^2>1/2$, the second term prefers the configuration where spin direction changes every two sites; within the degenerate manifold, this corresponds to the up-up-down, up-down-down, down-down-up, or down-up-up configuration, which corresponds to quadrupled zAFM (QzAFM) order. 
Since these two classical configurations exhibit level crossing at $\alpha^2=1/2$, 
the phase transition upon tuning $\alpha^2$ must be first-order in nature.


Now, let us return to the field theoretic description of DQCP. In \appref{app:bosonization}, we pointed out that the DQCP develops an instability toward a first order transition or coexistence when the Luttinger parameter $K$ goes below $1/8$ so that the $\cos 8\phi$ term becomes relevant. However, in the main Fig.2e, we demonstrated that the DQCP line of the truncated model $\Heff^\textrm{NNN}$ has a Luttinger parameter which saturates around $\sim0.28$ in the limit $J_3 \rightarrow 0$ as in \figref{fig:NNNvsLR}(a). In fact, we observed  similar behavior along the DQCP line of the XXZ model (with interactions up to NNN)
\begin{align} \label{eq:XXZ}
    & \Heff = \sum_i \Big[ (X_i X_{i+1} + Y_i Y_{i+1}) + \alpha (X_i X_{i+2}  + Y_i Y_{i+2})  \Big] + \Delta \sum_i \Big[ Z_i Z_{i+1} + \alpha Z_i Z_{i+2}  \Big], \nonumber
\end{align}
which has a qualitatively similar phase diagram. In this case, we numerically observed that as $1/\Delta \rightarrow 0$, along the DQC line between VBS and zAFM phases, the Luttinger parameter does not go below $0.3$. In other words, it appears the DQC line terminates at the point $J_3 =0, \alpha^2 = 1/2$  where the zAFM, QzAFM and VBS phases meet, without the Luttinger parameter $K$ going below $1/8$. This is inconsistent with the termination mechanism described in \appref{sec:rg}. 
However, note that $H^{\rm NNN}_{\rm eff}$ has an extensive degeneracy at $J_3 = 0$ and is, thus, highly fine-tuned. Indeed, we find that as $J_3 \to 0$, the quasiparticle velocity along the DQC line goes to zero, see \figref{fig:NNNvsLR}(b), so at $J_3 =0$ the entire gapless spectrum collapses into the ground state and one obtains an extensive degeneracy.


When we introduce further neighbor interactions, we find that the extensive degeneracy at $J_3 = 0$ gets lifted, and the transition between zAFM and QzAFM phases has a degeneracy of 6 (2 from zAFM and 4 from QzAFM). For example, if we introduce a perturbation $g_3 \sum_i Z_i Z_{i+3}$, then it only lowers the energy of the zAFM state by $g_3$ while it does not affect the energy of the QzAFM state. To compensate this effect, the transition happens at $\alpha_c^2 > 1/2$, and the extensive degeneracy gets reduced to 6 at such $\alpha_c^2$ (still first-order). The first order zAFM to QzAFM transition must then persist for a finite range of $J_3$ and the DQC line must terminate at finite $J_3$. Our current numerics do not allow us to resolve the precise details of  the DQC line termination in this case. It is possible that the zAFM to VBS transition turns first order or broadens into a co-existence region before terminating at the QzAFM phase boundary according to the field-theoretic scenario in section \appref{app:bosonization}. Alternatively, it may remain second order all the way up to a first order transition to the QzAFM phase. Numerically, we find that in the full model $H_{\rm eff}$ the value of the Luttinger parameter along the DQC line becomes as small as $K \approx 0.14$ for the smallest values of $J_3$ where it can be reliably extracted, see \figref{fig:NNNvsLR}(c). This is much smaller than the smallest value $K \approx 0.3$ obtained in the NNN model, leaving the possibility that the field theoretic scenario for the termination of the DQC line with $K$ approaching $1/8$ might be realized here.



\section{Model Rydberg Hamiltonian with exchange and blockade interactions}
\label{app:model}

In this section we describe how to realize our model Hamiltonian, $\Heff = H_{XY} + H_{ZZ}$, containing both  resonant dipole-dipole spin exchange $H_{XY}$ and van der Waals blockade Ising-like interaction $H_{ZZ}$. First, we explain how the van der Waals blockade gives rise to the Ising couplings relevant to our model. Then, we explain how to adjust the relative strength of $H_{XY}$ and $H_{ZZ}$, i.e. $J_3/J_6$, by controlling the amount of admixture in microwave-dressed Rydberg states. For simplicity, our model Hamiltonian $H_\textrm{eff}$ is designed to contain only $1/R^3$ terms in $H_{XY}$ and $1/R^6$ terms in $H_{ZZ}$, although as we show below, depending on the Rydberg states and admixing schemes chosen, both $1/R^3$ and $1/R^6$ contributions may appear in both $H_{XY}$ and $H_{ZZ}$. While in that case the exact location of the phase boundaries may change, we expect the qualitative features to remain unchanged. As direct exchange (a first order interaction) can generally be made stronger than blockade (a second order interaction), realizing the model as we describe means the tuning $J_3$ can allow access to a broad range of $J_3/J_6$ values in the phase diagram (See \figref{fig:MW}b). If first order and/or second order contributions are present in both $H_{XY}$ and $H_{ZZ}$, the tuning range may be more limited, and we may not access entire phase diagram presented in Fig.1b.

\subsection{Physical Origin of XY and ZZ terms}

Each internal atomic state $\ket{\alpha}, \ket{\beta},....$ of a given atom is specified by quantum numbers $n, l, j, m_j$: principal, angular momentum, total angular momentum, and $z$-projected angular momentum. In general, Rydberg states with the same $n$ but different $l$ have different energy since the amount of screening can change depending on the angular momentum $l$, so-called quantum defect \cite{Browaeys_2016}. Even among the states with the same angular momentum $l$, the degeneracy can be further split by additional perturbation such as external electric/magnetic fields or spin-orbit coupling.

In order to understand the interaction between a pair of atoms in respective Rydberg states, we should first understand their non-interacting energy. Note that the eigenstates of two non-interacting atoms can be written as pair states $\ket{ab}\equiv\ket{a}\otimes \ket{b}$ with energies $E_{ab} = E_{a} + E_{b}$, which form the basis of their Hilbert space.
Then, the two-body interaction terms arise from the dipole-dipole interaction Hamiltonian, which can couple these pair states, with $\hat{n}$ the unit vector along the axis between the atoms, $\vec{d_1}, \vec{d_2}$ the dipole operators for the two atoms \cite{Browaeys_2016}:
\begin{equation}
    V_\textrm{dd} = \frac{1}{4 \pi \epsilon_0} \frac{\vec{d_1} \cdot \vec{d_2} - 3 (\vec{d_1} \cdot \hat{n})(\vec{d_2} \cdot \hat{n})}{R^3}.
\end{equation}
Consider a spin encoded $\ket{\uparrow} = \ket{\alpha}$, $\ket{\downarrow} = \ket{\beta}$ in the bare-atom basis, where throughout, a strong magnetic field creates a Zeeman splitting so that the degeneracy of the Rydberg states of our interest characterized by $(n,l,j,m_j)$ is completely lifted and these energy levels are all isolated.

XY-type exchange interactions originate from the dipole-dipole coupling of pair states $\ket{\alpha \beta}, \ket{\beta \alpha }$, which can be nonzero for example when $\ket{\alpha}$ is an $nS$ state and $\ket{\beta}$ is an $nP$ state. 
As the dipole operator takes non-zero value between the states with different parity, i.e., $\bra{nS}\vec{d}_i\ket{nP} \neq 0$, there can be a dipole-dipole interaction between these states. Therefore, as $E_{\alpha \beta} = E_{\beta \alpha}$, these states couple resonantly and we obtain a direct flip-flop exchange interaction as the following
\begin{equation}
    \bra{\alpha_1 \beta_2} V_{dd} \ket{\beta_1 \alpha_2} \neq 0 \quad \Rightarrow \quad \sim(XX + YY)
\end{equation}

ZZ-type Ising interactions may arise if $V_{dd}$ has nonzero diagonal matrix elements, i.e., $\bra{\alpha} V_{dd} \ket{\alpha} \neq 0$.
However, in order to achieve a wide tunability of $J_3/J^6$, we  try to suppress this type of mechanism and instead rely on a different mechanism to realize the same type of interactions.
Namely, ZZ-type Ising interactions
originate when a pair state such as $\ket{\alpha \alpha}$ has a nonzero dipole-dipole coupling to another pair state $\ket{\gamma \lambda}$ outside of the spin subspace. As the energy difference between $\ket{\alpha \alpha}$ and $\ket{\gamma \lambda}$ is generally large (typically many GHz and much larger than the dipole-dipole coupling strength), this coupling results in a second order energy shift of $\ket{\alpha \alpha}$ of strength $\frac{|\bra{\alpha \alpha}V_\textrm{dd}\ket{\gamma \lambda}|^2}{E_{\alpha \alpha} - E_{\gamma \lambda}} \propto 1/R^6$. This pair state level shift is equivalent to a ZZ term plus single body Z terms. 
The strengh of interactions arising from this mechanism could be substantial, leading to so-called Rydberg blockade.
Note that the dipole-dipole coupling of $\ket{\alpha \alpha}$ with $\ket{\beta \beta}$ gives rise to a negligible Zeeman field.

\subsection{Controlling $J_3/J_6$ with Microwave-mixed Rydberg states}

As the atomic dipole operators in $V_{\textrm{dd}}$ give zero matrix element between atomic states of the same parity, encoding the spins as  microwave-dressed Rydberg states, which gives the controllable admixture of Rydberg states with different parities, can enable tunability of $J_3/J_6$.  The second order processes described above result in a background $1/R^6$ interaction, on top of which we can introduce exchange couplings by using a microwave drive to control the amount of admixture of states of distinct parity. By tuning these first order exchange couplings from zero to much larger than the  second order Ising-type interactions, we can explore a wide range of relative coupling strengths $J_3/J_6$.

We first explain the presence of the background second order terms $H_{ZZ}$. Each Rydberg pair state experiences an energy shift with a corresponding $C_6$ coefficient:
\begin{align} 
    H_{ZZ} & = \sum_{i<j} \frac{C_6^{\uparrow \uparrow}}{d_{ij}^6} n_i^\uparrow n_j^\uparrow + \frac{C_6^{\downarrow \downarrow}}{d_{ij}^6} n_i^\downarrow n_j^\downarrow + \frac{C_6^{\uparrow \downarrow}}{d_{ij}^6} (n_i^\uparrow n_j^\downarrow + n_i^\downarrow n_j^\uparrow ),
\end{align}
where $n^\uparrow, n^\downarrow$ are the projectors $\ket{\uparrow}\bra{\uparrow}$, $\ket{\downarrow}\bra{\downarrow}$, and $d_{ij}$ are the physical distances between the atoms. Substituting $n^\uparrow = \frac{1}{2}(P^R + Z)$, $n^\downarrow = \frac{1}{2}(P^R - Z)$, where $P^R$ is the projector onto the Rydberg spin subspace, gives:
\begin{align}   
    H_{ZZ} & = \frac{1}{4} \sum_{i<j} \frac{(C_6^{\uparrow \uparrow} + C_6^{\downarrow \downarrow} - 2 C_6^{\uparrow \downarrow})}{d_{ij}^6} Z_i Z_j + \frac{1}{4} \sum_{i < j} \frac{(C_6^{\uparrow \uparrow} - C_6^{\downarrow \downarrow})}{d_{ij}^6} (Z_i  P^R_j + P^R_i Z_j) + \frac{1}{4} \sum_{i < j} \frac{(C_6^{\uparrow \uparrow} + C_6^{\downarrow \downarrow} + 2 C_6^{\uparrow \downarrow})}{d_{ij}^6} P^R_i P^R_j.
\end{align}

Considering the 1D geometry we introduced and separating out the terms up to next-to-nearest neighbor from the long range terms $H_{LR}$:
\begin{align}  \label{eq:Rydberg}
    H_{ZZ} = & \frac{1}{4} \frac{(C_6^{\uparrow \uparrow} + C_6^{\downarrow \downarrow} - 2 C_6^{\uparrow \downarrow})}{d_{1}^6} \sum_{i}  (Z_i Z_{i+1} + \alpha^2 Z_i Z_{i+2}) \nonumber \\
    & + \frac{1}{4} \frac{(C_6^{\uparrow \uparrow} - C_6^{\downarrow \downarrow})}{d_{1}^6} \sum_{i} \Big[ (Z_i  P^R_{i+1} + P^R_i  Z_{i+1})  + \alpha^2 (Z_i  P^R_{i+2} + P^R_i Z_{i+2}) \Big] \nonumber  \\
    & + \frac{1}{4} \frac{(C_6^{\uparrow \uparrow} + C_6^{\downarrow \downarrow} + 2 C_6^{\uparrow \downarrow})}{d_{1}^6} \sum_{i}  \Big[P^R_i P^R_{i+1} + \alpha^2 P^R_i P^R_{i+2}\Big] + H_{LR}.
\end{align}

While the full form for $H_{ZZ}$ shown above is necessary for a quantitative treatment when modelling atoms with other internal levels in addition to $\ket{\uparrow}$, $\ket{\downarrow}$, it takes a simpler form when restricting the atomic Hilbert space to the Rydberg spin subspace, in which case each projector $P^R$ becomes an identity operator. The final sum in the expression above is then an overall energy offset and the sums of $\sum_i Z_i$ become total spin operators which just give another energy offset for a fixed magnetization sector. Defining an effective van der Waals coefficient $C_6 \equiv \frac{1}{4} (C_6^{\uparrow \uparrow} + C_6^{\downarrow \downarrow} - 2 C_6^{\uparrow \downarrow})$ so that $J_6 \equiv C_6/d_1^6$, we obtain the model for $H_{ZZ}$ relevant for exploring the DQC ground state phase diagram for fixed magnetization sector:
\begin{align}
    H_{ZZ} = J_6 \sum_{i} \Big[ Z_i Z_{i+1} + \alpha^2 Z_i Z_{i+2} \Big] +H_{LR}^{(ZZ)}
\end{align}


To tune $J_3$ on top of this background, we first choose the two Rydberg spin states such that the dipole moment between two states vanish, e.g. $\ket{nS}$ and $\ket{n'S}$, to encode ``bare'' spin-1/2 $\{\ket{\downarrow}, \ket{\uparrow}\}$. Direct dipolar exchange interaction is then forbidden by selection rules, and for simplicity we also choose states with negligible second order exchange couplings \footnote{appreciable second order exchange couplings \cite{Whitlock_2017} can arise via dipole-dipole coupling of, for example, both $\ket{nSn'S}$ and $\ket{n'SnS}$ to states like $\ket{nPnP}$, if $\Delta E = E_{nSn'S} + E_{n'SnS} - 2 E_{nPnP} \approx 0 $. As this could conspire to give nonzero exchange couplings even when the microwave drives are zero, making it difficult to access the regime of small $J_3/J_6$, for simplicity we assume this is negligible. We can ensure access to the small $J_3/J_6$ regime in a few ways, including either by choosing states such that $\Delta E$ is sufficiently large (for example in alkalis by choosing $\ket{\uparrow}$ to be a $D$ state with very different quantum defect), or such that the dipole coupling on the microwave driven transition is small}. Now, we want to dress the state $\ket{\downarrow}$ in a controlled manner so that the the dressed $\ket{\downarrow}$ has a non-zero dipole moment with $\ket{\uparrow}$. We apply a microwave drive $\Omega_{1}$ with detuning $\delta_{\textrm{MW}}$ to admix another Rydberg state with distinct parity, such as $\ket{m'P}$, into $\ket{nS}$. 
Whereas $\ket{\uparrow} = \ket{n'S}$, $\ket{\downarrow}$ becomes a dressed state
\begin{equation}
    \ket{\downarrow} = c_0 \ket{nS} + c_1 \ket{m'P}, \qquad \abs{c_1}^2 + \abs{c_2}^2 = 1
\end{equation}

As there is a dipolar exchange coupling $\bra{n'Sm'P} V_{dd} \ket{m'Pn'S} = {C_3'}/{d_{ij}^3}$, with the admixture of $\ket{m'P}$ into $\ket{nS}$, there is now also an inherited exchange interaction between $\ket{\downarrow \uparrow}$ and $\ket{\uparrow \downarrow}$ such that $\bra{\uparrow \downarrow} V_{dd} \ket{\downarrow \uparrow} = |c_1|^2 {C_3'}/{d_{ij}^3}$. This leads to a controllable exchange interaction Hamiltonian acting on the Rydberg spins:
\begin{align}
    H_{XY} &= |c_1|^2 \sum_{i<j} \frac{C_3'}{d_{ij}^3} \left( \sigma^+_i \sigma^-_j + \sigma^-_i \sigma^+_j \right),
\end{align}
where $\sigma^+, \sigma^-$ operate on $\ket{\uparrow}, \ket{\downarrow}$. For the ring geometry, we separate the terms up to the next-to-nearest neighbor interactions from the long range terms $H_{LR}$, substitute $\sigma^{\pm} = (X \pm i Y)/2$, and define $J_3(c_1) \equiv \frac{C_3'}{4 d_{1}^3} |c_1|^2$, obtaining:
\begin{align}
    H_{XY} &= J_3 \sum_{i} \Big[  \left( X_i X_{i+1} + Y_i Y_{i+1} \right) + \alpha \left( X_i X_{i+2} + Y_i Y_{i+2} \right) \Big]  + H_{LR}^{(XY)}
\end{align}

Finally, since the spin down $\ket{\downarrow}$ state is an admixture of S and P orbitals, 
the microwave admixing will cause an additional level shift of the pair state $\ket{\downarrow \downarrow}$. Let  $\bra{nSm'P} V_{dd} \ket{m'PnS} = {C_3}/{d_{ij}^3}$. Then, we observe that $\bra{\downarrow \downarrow} V_{dd} \ket{\downarrow \downarrow} = 4 |c_1|^2 |c_0|^2 {C_3}/{d_{ij}^3}$, which corresponds to a ZZ operator of $|c_1|^2 |c_0|^2 {C_3}/{d_{ij}^3}$.
This term gives a $1/R^3$ ZZ interaction denoted $H'_{ZZ}$ as the following:
\begin{equation}
    H'_{ZZ} = 4J_3 \frac{C_3}{C_3'} \abs{c_0}^2 \sum_i \qty[ Z_i Z_{i+1} + \alpha Z_i Z_{i+2} ] + H'^{(ZZ)}_\textrm{LR}, \qquad \abs{c_0}^2 = 1 - \abs{c_1}^2
\end{equation}
whose dependence on the atomic separation $d_{ij}$ is similar to that of $H_{XY}$. Combining the effects from $H_{ZZ}$, $H_{XY}$, and $H_{ZZ}'$ gives us the following tunable total Hamiltonian:
\begin{equation} \label{eq:H_one}
    H_\textrm{tot}(\alpha, \abs{c_1}) = H_{XY} + H_{ZZ} + H_{ZZ}',
\end{equation}
which is a function of the geometric factor $\alpha$ and the amount of admixture $\abs{c_1}$.  
We note that $H_{XY}$ and $H'_{ZZ}$ have tunable, $1/d^3$ decaying strengths, whereas $H_{ZZ}$ has $1/d^6$ decaying strengths.
Compared to the Hamiltonian proposed in the main text, this Hamiltonian is somewhat inconvenient to parametrize; in particular, when working in the regime of small small $J_3/J_6$, the Ising part of $H_{\textrm{tot}}$ is dominated by the $1/R^6$ terms from $H_{ZZ}$, whereas at large $J_3/J_6$, the Ising part is dominated by the $1/R^3$ terms from $H_{ZZ}'$, so that two parameters, the ratio between exchange and Ising couplings and the distance dependence of the interactions, are no longer independent.




\begin{figure}[!t]
    \centering
    \includegraphics[width = 0.8 \textwidth]{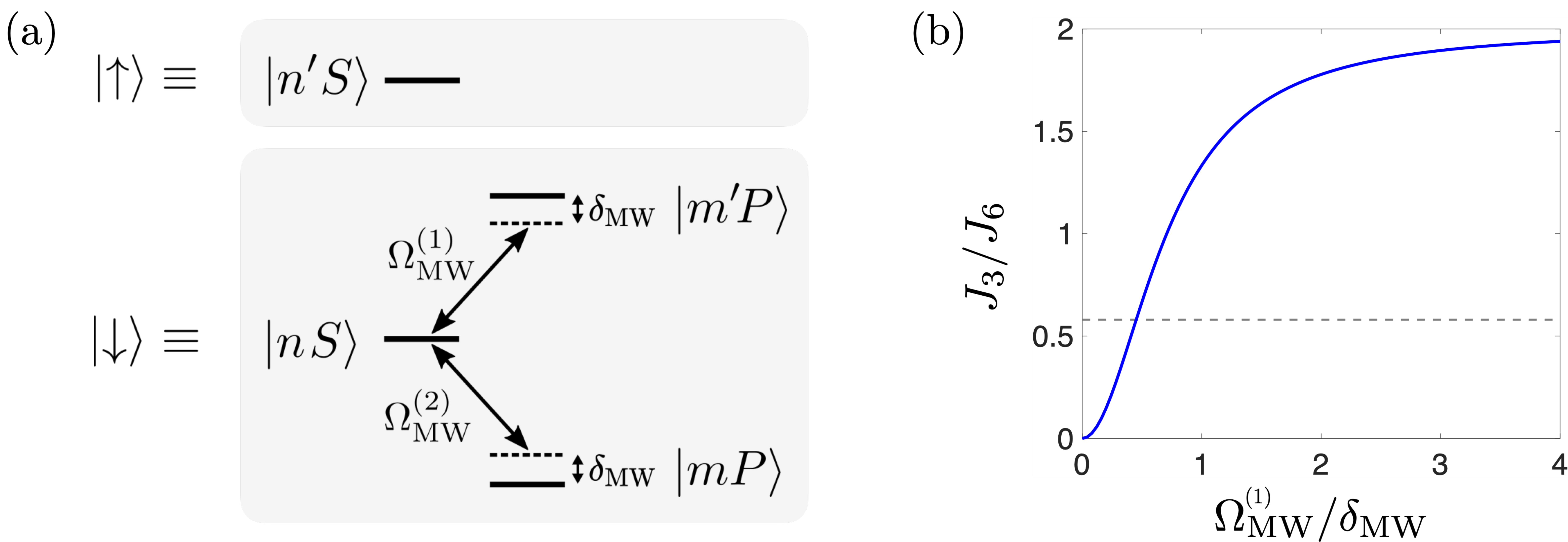}
    \caption{\label{fig:MW} Mechanism for tuning $J_3$. (a) Rydberg levels of a single atom; many encodings are possible, but we choose $\ket{nS}$ and $\ket{mP}$ states for concreteness to illustrate which levels are dipole coupled. For $\Omega_{\textrm{MW}} = 0$, $\ket{\uparrow} = \ket{n'S}$ and $\ket{\downarrow} = \ket{nS}$, which have no exchange interaction. Tuning $\Omega_{\textrm{MW}} > 0$ admixes components of $\ket{m'P}$ and $\ket{mP}$ into $\ket{nS}$, dressing $\ket{\downarrow}$ and inducing an exchange coupling between $\ket{\uparrow}$ and $\ket{\downarrow}$. (b) Exchange coupling between $\ket{\uparrow \downarrow}, \ket{\downarrow \uparrow}$ dependence on microwave control parameters, assuming a bare exchange strength of 4 times the Rydberg blockade strength, equal dipole matrix elements $\bra{m'P} V_{dd} \ket{nS} = \bra{mP} V_{dd} \ket{nS}$, and $\Omega_{\textrm{MW}}^{(1)} = \Omega_{\textrm{MW}}^{(2)}$. Dashed grey line corresponds to DQCP for $\alpha = 0.5$
    }
\end{figure}

\subsection{Cancellation of first order Ising shift due to microwave dressing}


In the previous subsection, we demonstrated the spin-1/2 effective Hamiltonian for encoding scheme using one microwave dressing, and discussed its limitations. In order to overcome these issues, we propose a new scheme where $H_{ZZ}'$ can be cancelled by using a second microwave tone to admix a second state, such as  $mP$, into $nS$, while still maintaining the desired exchange coupling. Since we choose $m<m'<n'$, the transition dipole moment between $\ket{mP}$ and $\ket{n'S}$ is much smaller than the transition dipole moment between $\ket{m'P}$ and $\ket{n'P}$ \cite{Weber_2017}. Therefore, for simplicity we ignore the transition dipole moment between $\ket{mP}$ and $\ket{n'S}$. Like before, we apply microwave drive $\Omega_{2}$  to admix $\ket{mP}$, into $\ket{nS}$ as in \figref{fig:MW}a. 
Whereas $\ket{\uparrow} = \ket{n'S}$, $\ket{\downarrow}$ becomes a dressed state
\begin{equation}
    \ket{\downarrow} = c_0 \ket{nS} + c_1 \ket{n'P} + c_2 \ket{mP},
\end{equation} 
where the coefficients $c_{1,2}$ depends on the strength of microwave drive $\Omega^{(1,2)}_\textrm{MW}$ and detuning $\delta$. One way this can be used to make the ZZ term vanish is following Ref.\,\cite{adjust_dipolar2021}, let us define 
\begin{align}
    \mu_z &\equiv \bra{n'S} d_z \ket{m'P} \nonumber \\
    \mu_+ &\equiv \bra{n'S} d_+ \ket{mP} 
\end{align}
where $d_z$ and $d_+$ are the electric dipole operator in $\hat{z}$ and $+$ polarizations and atomic separations are in $z$-directions (we assume the orbitals are chosen in such a way that dipole moments in other directions vanish). Then, the direct dipolar interaction strength between $\ket{\downarrow}$ at sites $i$ and $j$ is \cite{adjust_dipolar2021}:
\begin{equation}
    \bra{\downarrow_i \downarrow_j} \hat{V}_{dd} \ket{\downarrow_i \downarrow_j} \propto \abs{c_1}^2 \mu_z^2 - \abs{c_2}^2 \mu_+^2
\end{equation}
Therefore, by choosing $\Omega_\textrm{MW}^{(1,2)}$ in such a way that $\abs{c_1} = (\mu_z/\mu_+)\abs{c_2}$, we can always remove the dipolar direct interaction. 

Even more simply, we can choose $\ket{mP}$ to have the same $l, j, m_j$ quantum numbers as $\ket{m'P}$, and admixing both into $\ket{nS}$ with opposite sign detunings as shown in \figref{fig:MW}. $\ket{mP}$ and $\ket{m'P}$ then have identical angular wavefunctions, so the relative strength of their coupling to $\ket{nS}$ is determined by their radial wavefunction matrix elements. As they are admixed with opposite sign, tuning the microwave drives then allows the cancellation of these couplings. This is the scheme we use in the main text, where the effective Hamiltonian is given by 
\begin{equation}
    H_\textrm{tot} = H_{XY} + H_{ZZ},
\end{equation}
which has Ising interactions independent of the degree of admixture (c.f. \eqnref{eq:H_one}).

\section{State preparation protocols} \label{app:preparation}


In a Rydberg quantum simulator, there are a number of different ways to adiabatically prepare the ground state of $\Heff$ at a desired point in the phase diagram of the model Hamiltonian. One approach is to construct a trajectory inspired by  an adiabatic rapid passage (ARP) \cite{beterov2020} protocol, where each ground state $\ket{g}$ is laser coupled to Rydberg states in ${\cal H}_\textrm{Ryd}$ and then boosted in energy above the Rydberg manifold by adjusting the laser's detuning in the rotating frame, as shown in  Fig.3b of the main text. However, simply transferring all the atoms into the individual subspace ${\cal H}_\textrm{Ryd}$ is not sufficient, as we may want to land in a specific total magnetization sector; in the case of DQC, the sector  with the ground state of  of interest is such that $N^\textrm{gs}_\uparrow=L/2$.

To this end, we introduce a general method that adiabatically ramps $\ket{g}^{\otimes L}$ into a final state in ${\cal H}_\textrm{Ryd}^{\otimes L}$ with a desired magnetization $N_\uparrow$.
It turns out, this can be achieved in a surprisingly simple way: an ARP trajectory where we address $N_\uparrow$ sites with $\Omega_\uparrow$, and $L - N_\uparrow$ sites with $\Omega_\downarrow$. Conceptually, during the adiabatic sweep, the laser $\Omega_\uparrow$ ($\Omega_\downarrow$) can only transfer atoms from $\ket{g}$ into $\ket{\uparrow}$ ($\ket{\downarrow}$), and once transferred, the number of atoms in $\ket{\uparrow}$ is conserved by $H_{\textrm{Ryd}}$. At the end of the trajectory then, when no atoms are left in $\ket{g}$, we are guaranteed to achieve the desired $N_\uparrow$.


To prove this, consider the following Hamiltonian, consisting of three levels in each atom: $\ket{\uparrow}$, $\ket{\downarrow}$, and a ground state $\ket{g}$ as in the main Fig.3. Laser $\Omega_{\textrm{L} \uparrow}$ couples $\ket{g}$ to $\ket{\uparrow}$ on a set of sites $A$ of size $N_A$, and laser $\Omega_{\textrm{L} \downarrow}$ couples $\ket{g}$ to $\ket{\downarrow}$ on a set of sites $B$ of size $N_B$, all with detuning $\Delta_\textrm{L}$:
\begin{align} \label{eq:app_adiabatic}
    H = \Heff + \Omega_{\textrm{L} \uparrow} \sum_{i \in A}(\ket{\uparrow}_i \bra{g}_i + h.c.)   + \Omega_{\textrm{L} \downarrow} \sum_{i \in B}(\ket{\downarrow}_i\bra{g}_i + h.c.) + \Delta_\textrm{L} \sum_{i}\ket{g}_i\bra{g}_i.
\end{align}

At the end of the trajectory where $\Delta_L \gg 1$, we can show that the number of spin up and down states $N_\uparrow$ and $N_\downarrow$ are $N_A$ and $N_B$. To prove this,  note that the following quantities are conserved throughout \textit{any} adiabatic trajectory, where $n^{\uparrow}_i$ is the projector onto $\ket{\uparrow}$ on site $i$:
\begin{align}
    \sum_{i \in A \cup B} n^{\uparrow}_i + \sum_{j \in A} n^g_j = N_A \qquad   \sum_{i \in A \cup B} n^{\downarrow}_i  + \sum_{j \in B} n^g_j = N_B  \qquad \Rightarrow \qquad \textrm{Conserved}
\end{align}
\noindent \emph{Proof:} 
\begin{align}
    \qty[\sum_{i \in A \cup B} n^{\uparrow}_i + \sum_{j \in A} n^g_j, H] &= \qty[\sum_{i \in A \cup B} n^{\uparrow}_i + \sum_{j \in A} n^g_j, \Heff] + \qty[\sum_{i \in A \cup B} n^{\uparrow}_i + \sum_{j \in A} n^g_j, H- \Heff] \nonumber \\
    &= \Omega_{\textrm{L} \uparrow} \sum_{i \in A} \bigg[   n^\uparrow_i +  n^g_i,\,\,  \ket{\uparrow}_i\bra{g}_i + \textrm{h.c.}  \bigg] = 0
\end{align}
where the first term in the RHS of the first line disappears as the DQCP Hamiltonian is block diagonal in Rydberg states and conserves $S^z$ within the manifold. The same holds when $A$ and $B$ are exchanged.

Therefore, if we assume that $n^g_i = 0$ at the end of the trajectory, which we physically expect due to large $\Delta_\textrm{L} \gg 1$, we find that $\sum n_i^\uparrow = N_A$ and $\sum n_i^\downarrow = N_B$ no matter what the trajectory is. This provides a general mechanism to prepare a state with desired filling.

Now, there is a separate question of whether the trajectory is gapped along the path (except at the end where the gap necessarily has to close since we are preparing a critical state). Here, we make a perturbative argument for the trajectory being gapped. For simplicity, we ignore terms with projectors in \eqnref{eq:Rydberg}, and treat $H_\textrm{DQCP}$ as a small perturbation. First, ignore $H_\textrm{DQCP}$ in \eqnref{eq:app_adiabatic}. Parametrize $(\Delta, \Omega)$ as $\frac{\Delta}{2} = -\eta \cos \theta$ and $\Omega = \eta \sin \theta$. Then, the energy levels of each odd site is given as
\begin{eqnarray}
    & E_1 = -\eta(1+\cos \theta) \quad &\ket{\varphi_1} = \frac{\sin(\theta/2)}{\sin \theta} \cdot \Big[ -(\cos \theta + 1) \ket{g} +  \sin \theta \ket{\downarrow} \Big] = \ket{\downarrow'_\textrm{odd}} \nonumber \\
    & E_2 = 0\quad & \ket{\varphi_2} = \ket{\uparrow} = \ket{\uparrow'_\textrm{odd}}  \nonumber \\
    & E_3 = \eta(1-\cos \theta) \quad & \ket{\varphi_3} =\frac{\cos(\theta/2)}{\sin \theta} \cdot \Big[ -(\cos \theta - 1) \ket{g} +  \sin \theta \ket{\downarrow} \Big]
\end{eqnarray}
and for even site is given as 
\begin{eqnarray}
    & E_1 = -\eta(1+\cos \theta) \quad &\ket{\varphi_1} = \frac{\sin(\theta/2)}{\sin \theta} \cdot \Big[ -(\cos \theta + 1) \ket{g} +  \sin \theta \ket{\uparrow} \Big] = \ket{\uparrow'_\textrm{even}} \nonumber \\
    & E_2 = 0\quad & \ket{\varphi_2} = \ket{\downarrow} = \ket{\downarrow'_\textrm{even}} \nonumber \\
    & E_3 = \eta(1-\cos \theta) \quad & \ket{\varphi_3} =\frac{\cos(\theta/2)}{\sin \theta} \cdot \Big[ -(\cos \theta - 1) \ket{g} +  \sin \theta \ket{\uparrow} \Big]
\end{eqnarray}
If we project into the lower two levels for each site, we have effective two-level system with the following staggering field:
\begin{equation}
    H'_\textrm{stag} = \sum_n (-1)^n \frac{\eta(1+\cos \theta)}{2} Z'_n
\end{equation}
Furthermore, on this effective spin up and down subspace $\ket{\uparrow'}$ and $\ket{\downarrow'}$, the first-order contribution of $H_\textrm{DQCP}$ is given by the following:
\begin{align}
    H'_\textrm{DQCP} &= \qty[\sin(\frac{\theta}{2})]^2  J_3 \sum_i \Big[ (X'_i X'_{i+1} + Y'_i Y'_{i+1}) + \alpha (X'_i X'_{i+2} + Y'_i Y'_{i+2}) + \dots \Big] \nonumber \\
    & + \qty[\frac{1 + (\sin(\theta/2))^2}{2}]^2  J_6 \sum_i \Big[ Z'_i Z'_{i+1} + \alpha^2 Z'_i Z'_{i+2} + \dots \Big] + \textrm{(chemical potential)}
\end{align}
which, in fact, acts like a zAFM Hamiltonian since the effective strength of the $ZZ$ coupling becomes stronger than the $XY$ couplings for $\theta < \pi$. Since the staggering field induces a strong gap on the DQCP-zAFM phase, we can argue that through the trajectory of $\theta \in [0,\pi)$, the system is gapped.

The ARP-inspired state preparation protocol we describe above is not the only way to prepare the ground state in desired magnetization sector. For example,  if we assume we can prepare any product state in ${\cal H}_\textrm{Ryd}^{\otimes L}$ with high precision, e.g.,~through individually or sub-lattice laser addressing atoms from the ground states, we can  imagine simply beginning  from a state $\ket{\psi_0^\textrm{stag}} = \ket{\uparrow, \downarrow, \uparrow, \downarrow, \dots}$. This can be done by applying a short pulse to the ground state of Rydberg atom array. Consider now implementing a  Hamiltonian with strong alternating $Z$-fields, in addition to the Rydberg Hamiltonian $\Heff$:
\begin{equation}
    H = \Heff + g \sum_n (-1)^n Z_n,
\end{equation}
where $\Heff$ depends on $(J_3/J_6,\alpha)$. We first start from $J_3/J_6 = 0$ and $g > 0$ (assuming $J_6$ is at unity). There, our initial state is a perfect groundstate of the Hamiltonian. Then, we slowly increase the tuning parameter $J_3$ to the critical value for a given $\alpha$ and $J_6$, which can be achieved by modifying the strength of microwave dressing. After which, we  slowly turn off the field $g_2$. Throughout the entire adiabatic path, there is a constant gap except at the DQCP point, which closes as $1/L$.

In both methods,  an explicit bias field is applied that breaks the translation symmetry during the adiabatic evolution, which can lead to  excitations in non-symmetric sectors during  adiabatic evolution. 
In order to overcome this issue, one may envision to implement a symmetric adiabatic evolution. For example, we may start from a symmetric ground state (cat state) of the zAFM phase at $J_3/J_6=0$ for some $\alpha$, and then increase $J_3/J_6$ adiabatically to reach the DQCP ground state. However, this requires the cat state preparation with high fidelity, which is generally nontrivial; note that previously proposed methods of preparing a  cat state  involves an adiabatic trajectory through an Ising quantum phase transition~\cite{cat_state};  as the gap closes there too (in addition to at the DQCP) this entails a significant state preparation time for a high fidelity.
In contrast, our proposed methods involving  an adiabatic evolution in which symmetries are explicitly broken, only encounters a single gap-closing of $\sim 1/L$ near the end of the trajectory, and should thus be more favorable than the symmetric adiabatic evolution protocol. 





\begin{figure}[!t]
    \centering
    \includegraphics[width = 0.6 \textwidth]{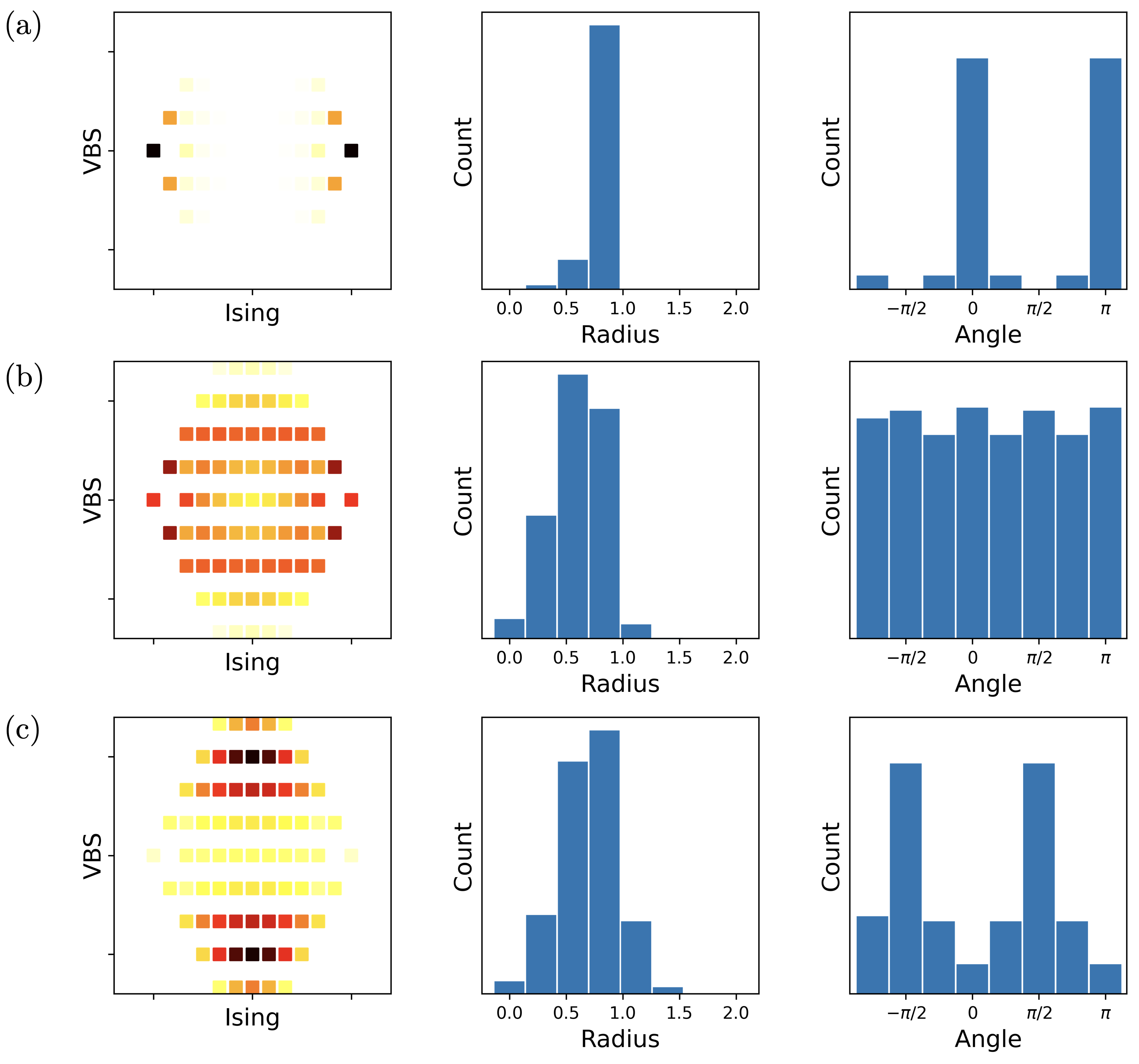}
    \vspace{-10pt}
    \caption{\label{fig:sup_order}  Density plot and radial/angular histograms of measurement outcomes of order parameters, obtained from exact diagonalization at $L=24$ for three different regimes at $\alpha=0.5$: (a) zAFM (b) DQCP (c) VBS. Note that for the radial distribution, the histogram is normalized by the radius so that $N_\textrm{tot} = \int \dd{r} 2\pi r \cdot N(r)$. Therefore, it faithfully represents the radial density of the order parameter. }
\end{figure}

\section{Snapshots of Observables in Finite Systems}
\label{app:observable}

In numerical simulations of finite-size systems, we measured the following observables:
\begin{align}
\bar{O}_\textrm{zAFM}& \equiv  \frac{1}{L} \sum_r (-1)^r Z_r \nonumber \\
\bar{O}'_\textrm{VBS} &\equiv \frac{1}{L} \sum_r  (-1)^r \qty[\frac{Z_{r-1}Z_r - Z_{r} Z_{r+1}}{2} ].
\end{align}
Note that the ground state of the system  satisfies the constraint $\sum_i Z_i = 0$. Therefore, if we take a snapshot of the wavefunction in the computational $z$-basis,  the possible values of $\bar{O}_\textrm{zAFM}$ and $\bar{O}'_\textrm{VBS}$ are spaced by multiples of $\frac{4}{L}$.For a given system size $L=2m$,  we thus have
\begin{align}
    \bar{O}_\textrm{zAFM}, \bar{O}'_\textrm{VBS} &\in \frac{1}{L} \cdot [-2m, -2m+4, ..., 2m]
\end{align}
However, note that while $\expval{\bar{O}_\textrm{zAFM}}$ is $\pm 1$ for the ideal zAFM state ($\ket{\uparrow \downarrow \uparrow ...}$), $\expval{\bar{O}'_\textrm{VBS} }$ is $\pm 1/2$ in the ideal VBS state (product state of singlet dimers).  When we compare these observables in a 2d histogram, we have normalized the VBS order parameter by its value in the ideal VBS state. (Strictly speaking, we can fix the relative normalization of $\Ovbs$ and $\Ozafm$ by, for instance, requiring that their distributions have the same second moment. However, the above normalization procedure  already produced distributions with near circular symmetry). In the histogram plot, each data point is  hence  specified by the data set $(\bar{O}_\textrm{zAFM}, 2\bar{O}'_\textrm{VBS})$. After this rescaling, we  obtain a circular distribution of the order parameters $(\bar{O}_\textrm{zAFM}, 2\bar{O}'_\textrm{VBS})$ at the DQCP, as demonstrated in the  Fig.~3(d) of the main text. In fact, already at system size $L=24$, we can observe that the distribution becomes circular (\figref{fig:sup_order}(b)) at the DQCP, whereas the distribution features two isolated peaks in both VBS and zAFM phases (\figref{fig:sup_order}(a,c)).

In order to identify the rotational invariance of the joint probability distribution, we have shown angular and radial distributions in  Figs.~3(e,f) of the main text. Note that almost uniform angular distribution of $(\bar{O}_\textrm{zAFM}, 2\bar{O}'_\textrm{VBS})$ implies the presence of the emergent $\U(1)$ symmetry, a characteristic feature of DQCP. 
For the radial distribution $\hat{P}(r)$, note that the quantity is calculated by normalizing the number of counts in the bin $[r,r+\delta)$ by $2 \pi (r+\delta/2)$ so that the bias originated from the integration measure is properly removed. Then, the peak location of $\hat{P}(r)$ measures the magnitude of order parameter fluctuation in the system with finite-size $L$, which should decay as $1/L^K$ where $K$ is Luttinger parameter. Although it means that the distribution should shrink to the origin in the thermodynamic limit, the fact that the Luttinger parameter $K$ is small ($\sim$\,$1/3$ for the DQCP plotted in the manuscript Fig.3d) allows us to access a circle with the resolution increasing with the system size. Since the resolution improves as $1/L$ while the radius decays with $1/L^K$ with $K<1$, we can identify the circle better and better as we increase the system size.
Finally, we note that for $L>24$, the data was obtained by simulated measurements in finite DMRG simulations with the periodic boundary condition implemented by the long-range interaction between the first and last sites. In order to improve the statistics, which is limited by the number of measurements ($2 \times 10^4$ for all $L=64,128,192$), we explicitly symmetrize the outcomes with respect to inversion and spin-flip symmetries.


\section{DMRG simulation of long-range Hamiltonian}\label{app:long_range}

In the Rydberg atom array system, the Hamiltonian has long-range dipolar (XY) and  Van der Waals (ZZ) interactions, which exhibit power-law decaying behavior as $1/r^3$ (XY) and $1/r^6$ (ZZ) respectively. However, an exact representation of a power-law decaying long-range interaction in the DMRG simulation is computationally inefficient, while exponentially decaying interactions can be expressed and computed efficiently in the simulation. Therefore, we approximated these power-law decaying interactions with a set of exponential functions~\cite{LR_Zaletel}; since dipolar and van der Waals interactions  decay quickly with large powers, such an approximation can be made very accurate with only a small number of exponential functions.
In \figref{fig:longrange}, we compare the strengths of the actual interactions and exponentially approximated versions. In the DMRG simulations, we use the sum of four exponential functions, where we find that the integrated difference is less than $1\%$ for both dipolar and $0.01\%$ for Van der Waals interactions, and the difference  decreases for smaller $\alpha$.
Such small differences justify the use of DMRG method with a set of exponentially decaying interactions to represent dipolar and van der Waals interactions. 

\begin{figure}[!t]
    \centering
    \includegraphics[width = 0.9 \textwidth]{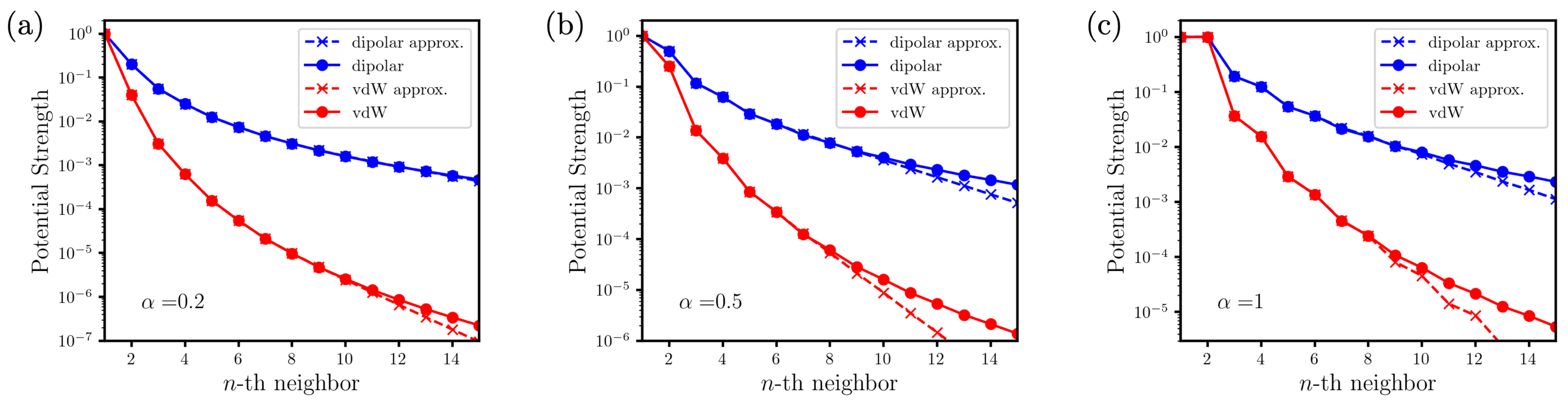}
    \caption{\label{fig:longrange} Comparison between actual dipolar/van der Waals interactions and exponentially fitted approximation using the sum of four exponentials at (a) $\alpha = 0.2$, (b) $\alpha=0.5$, and (c) $\alpha=1.0$. The difference between power-law decaying interactions and their exponential approximations integrated over the entire range is less than 1\% for the range of $\alpha$ we simulated. 
    }
\end{figure}

\end{widetext}


\begin{thebibliography}{79}%
\makeatletter
\providecommand \@ifxundefined [1]{%
 \@ifx{#1\undefined}
}%
\providecommand \@ifnum [1]{%
 \ifnum #1\expandafter \@firstoftwo
 \else \expandafter \@secondoftwo
 \fi
}%
\providecommand \@ifx [1]{%
 \ifx #1\expandafter \@firstoftwo
 \else \expandafter \@secondoftwo
 \fi
}%
\providecommand \natexlab [1]{#1}%
\providecommand \enquote  [1]{``#1''}%
\providecommand \bibnamefont  [1]{#1}%
\providecommand \bibfnamefont [1]{#1}%
\providecommand \citenamefont [1]{#1}%
\providecommand \href@noop [0]{\@secondoftwo}%
\providecommand \href [0]{\begingroup \@sanitize@url \@href}%
\providecommand \@href[1]{\@@startlink{#1}\@@href}%
\providecommand \@@href[1]{\endgroup#1\@@endlink}%
\providecommand \@sanitize@url [0]{\catcode `\\12\catcode `\$12\catcode
  `\&12\catcode `\#12\catcode `\^12\catcode `\_12\catcode `\%12\relax}%
\providecommand \@@startlink[1]{}%
\providecommand \@@endlink[0]{}%
\providecommand \url  [0]{\begingroup\@sanitize@url \@url }%
\providecommand \@url [1]{\endgroup\@href {#1}{\urlprefix }}%
\providecommand \urlprefix  [0]{URL }%
\providecommand \Eprint [0]{\href }%
\providecommand \doibase [0]{http://dx.doi.org/}%
\providecommand \selectlanguage [0]{\@gobble}%
\providecommand \bibinfo  [0]{\@secondoftwo}%
\providecommand \bibfield  [0]{\@secondoftwo}%
\providecommand \translation [1]{[#1]}%
\providecommand \BibitemOpen [0]{}%
\providecommand \bibitemStop [0]{}%
\providecommand \bibitemNoStop [0]{.\EOS\space}%
\providecommand \EOS [0]{\spacefactor3000\relax}%
\providecommand \BibitemShut  [1]{\csname bibitem#1\endcsname}%
\let\auto@bib@innerbib\@empty
\bibitem [{\citenamefont {Senthil}\ \emph
  {et~al.}(2004{\natexlab{a}})\citenamefont {Senthil}, \citenamefont {Balents},
  \citenamefont {Sachdev}, \citenamefont {Vishwanath},\ and\ \citenamefont
  {Fisher}}]{Senthil2004}%
  \BibitemOpen
  \bibfield  {author} {\bibinfo {author} {\bibfnamefont {T.}~\bibnamefont
  {Senthil}}, \bibinfo {author} {\bibfnamefont {Leon}\ \bibnamefont {Balents}},
  \bibinfo {author} {\bibfnamefont {Subir}\ \bibnamefont {Sachdev}}, \bibinfo
  {author} {\bibfnamefont {Ashvin}\ \bibnamefont {Vishwanath}}, \ and\ \bibinfo
  {author} {\bibfnamefont {Matthew P.~A.}\ \bibnamefont {Fisher}},\ }\bibfield
  {title} {\enquote {\bibinfo {title} {Quantum criticality beyond the
  landau-ginzburg-wilson paradigm},}\ }\href {\doibase
  10.1103/PhysRevB.70.144407} {\bibfield  {journal} {\bibinfo  {journal} {Phys.
  Rev. B}\ }\textbf {\bibinfo {volume} {70}},\ \bibinfo {pages} {144407}
  (\bibinfo {year} {2004}{\natexlab{a}})}\BibitemShut {NoStop}%
\bibitem [{\citenamefont {Senthil}\ \emph
  {et~al.}(2004{\natexlab{b}})\citenamefont {Senthil}, \citenamefont
  {Vishwanath}, \citenamefont {Balents}, \citenamefont {Sachdev},\ and\
  \citenamefont {Fisher}}]{Senthil2004_Science}%
  \BibitemOpen
  \bibfield  {author} {\bibinfo {author} {\bibfnamefont {T.}~\bibnamefont
  {Senthil}}, \bibinfo {author} {\bibfnamefont {Ashvin}\ \bibnamefont
  {Vishwanath}}, \bibinfo {author} {\bibfnamefont {Leon}\ \bibnamefont
  {Balents}}, \bibinfo {author} {\bibfnamefont {Subir}\ \bibnamefont
  {Sachdev}}, \ and\ \bibinfo {author} {\bibfnamefont {Matthew P.~A.}\
  \bibnamefont {Fisher}},\ }\bibfield  {title} {\enquote {\bibinfo {title}
  {Deconfined quantum critical points},}\ }\href {\doibase
  10.1126/science.1091806} {\bibfield  {journal} {\bibinfo  {journal}
  {Science}\ }\textbf {\bibinfo {volume} {303}},\ \bibinfo {pages} {1490--1494}
  (\bibinfo {year} {2004}{\natexlab{b}})}\BibitemShut {NoStop}%
\bibitem [{\citenamefont {Nahum}\ \emph {et~al.}(2015)\citenamefont {Nahum},
  \citenamefont {Serna}, \citenamefont {Chalker}, \citenamefont {Ortu\~no},\
  and\ \citenamefont {Somoza}}]{Nahum2015SO5}%
  \BibitemOpen
  \bibfield  {author} {\bibinfo {author} {\bibfnamefont {Adam}\ \bibnamefont
  {Nahum}}, \bibinfo {author} {\bibfnamefont {P.}~\bibnamefont {Serna}},
  \bibinfo {author} {\bibfnamefont {J.~T.}\ \bibnamefont {Chalker}}, \bibinfo
  {author} {\bibfnamefont {M.}~\bibnamefont {Ortu\~no}}, \ and\ \bibinfo
  {author} {\bibfnamefont {A.~M.}\ \bibnamefont {Somoza}},\ }\bibfield  {title}
  {\enquote {\bibinfo {title} {Emergent so(5) symmetry at the n\'eel to
  valence-bond-solid transition},}\ }\href {\doibase
  10.1103/PhysRevLett.115.267203} {\bibfield  {journal} {\bibinfo  {journal}
  {Phys. Rev. Lett.}\ }\textbf {\bibinfo {volume} {115}},\ \bibinfo {pages}
  {267203} (\bibinfo {year} {2015})}\BibitemShut {NoStop}%
\bibitem [{\citenamefont {Metlitski}\ and\ \citenamefont
  {Thorngren}(2018)}]{Thorngren2017}%
  \BibitemOpen
  \bibfield  {author} {\bibinfo {author} {\bibfnamefont {Max~A.}\ \bibnamefont
  {Metlitski}}\ and\ \bibinfo {author} {\bibfnamefont {Ryan}\ \bibnamefont
  {Thorngren}},\ }\bibfield  {title} {\enquote {\bibinfo {title} {Intrinsic and
  emergent anomalies at deconfined critical points},}\ }\href {\doibase
  10.1103/PhysRevB.98.085140} {\bibfield  {journal} {\bibinfo  {journal} {Phys.
  Rev. B}\ }\textbf {\bibinfo {volume} {98}},\ \bibinfo {pages} {085140}
  (\bibinfo {year} {2018})}\BibitemShut {NoStop}%
\bibitem [{\citenamefont {Wang}\ \emph {et~al.}(2017)\citenamefont {Wang},
  \citenamefont {Nahum}, \citenamefont {Metlitski}, \citenamefont {Xu},\ and\
  \citenamefont {Senthil}}]{Wang2017}%
  \BibitemOpen
  \bibfield  {author} {\bibinfo {author} {\bibfnamefont {Chong}\ \bibnamefont
  {Wang}}, \bibinfo {author} {\bibfnamefont {Adam}\ \bibnamefont {Nahum}},
  \bibinfo {author} {\bibfnamefont {Max~A.}\ \bibnamefont {Metlitski}},
  \bibinfo {author} {\bibfnamefont {Cenke}\ \bibnamefont {Xu}}, \ and\ \bibinfo
  {author} {\bibfnamefont {T.}~\bibnamefont {Senthil}},\ }\bibfield  {title}
  {\enquote {\bibinfo {title} {Deconfined quantum critical points: Symmetries
  and dualities},}\ }\href {\doibase 10.1103/PhysRevX.7.031051} {\bibfield
  {journal} {\bibinfo  {journal} {Phys. Rev. X}\ }\textbf {\bibinfo {volume}
  {7}},\ \bibinfo {pages} {031051} (\bibinfo {year} {2017})}\BibitemShut
  {NoStop}%
\bibitem [{\citenamefont {Ma}\ \emph {et~al.}(2018)\citenamefont {Ma},
  \citenamefont {Sun}, \citenamefont {You}, \citenamefont {Xu}, \citenamefont
  {Vishwanath}, \citenamefont {Sandvik},\ and\ \citenamefont
  {Meng}}]{Ma2017DQCP}%
  \BibitemOpen
  \bibfield  {author} {\bibinfo {author} {\bibfnamefont {Nvsen}\ \bibnamefont
  {Ma}}, \bibinfo {author} {\bibfnamefont {Guang-Yu}\ \bibnamefont {Sun}},
  \bibinfo {author} {\bibfnamefont {Yi-Zhuang}\ \bibnamefont {You}}, \bibinfo
  {author} {\bibfnamefont {Cenke}\ \bibnamefont {Xu}}, \bibinfo {author}
  {\bibfnamefont {Ashvin}\ \bibnamefont {Vishwanath}}, \bibinfo {author}
  {\bibfnamefont {Anders~W.}\ \bibnamefont {Sandvik}}, \ and\ \bibinfo {author}
  {\bibfnamefont {Zi~Yang}\ \bibnamefont {Meng}},\ }\bibfield  {title}
  {\enquote {\bibinfo {title} {Dynamical signature of fractionalization at a
  deconfined quantum critical point},}\ }\href {\doibase
  10.1103/PhysRevB.98.174421} {\bibfield  {journal} {\bibinfo  {journal} {Phys.
  Rev. B}\ }\textbf {\bibinfo {volume} {98}},\ \bibinfo {pages} {174421}
  (\bibinfo {year} {2018})}\BibitemShut {NoStop}%
\bibitem [{\citenamefont {Ma}\ \emph {et~al.}(2019)\citenamefont {Ma},
  \citenamefont {You},\ and\ \citenamefont {Meng}}]{MaDQCP2019}%
  \BibitemOpen
  \bibfield  {author} {\bibinfo {author} {\bibfnamefont {Nvsen}\ \bibnamefont
  {Ma}}, \bibinfo {author} {\bibfnamefont {Yi-Zhuang}\ \bibnamefont {You}}, \
  and\ \bibinfo {author} {\bibfnamefont {Zi~Yang}\ \bibnamefont {Meng}},\
  }\bibfield  {title} {\enquote {\bibinfo {title} {Role of noether's theorem at
  the deconfined quantum critical point},}\ }\href {\doibase
  10.1103/PhysRevLett.122.175701} {\bibfield  {journal} {\bibinfo  {journal}
  {Phys. Rev. Lett.}\ }\textbf {\bibinfo {volume} {122}},\ \bibinfo {pages}
  {175701} (\bibinfo {year} {2019})}\BibitemShut {NoStop}%
\bibitem [{\citenamefont {Kuklov}\ \emph {et~al.}(2008)\citenamefont {Kuklov},
  \citenamefont {Matsumoto}, \citenamefont {Prokof'ev}, \citenamefont
  {Svistunov},\ and\ \citenamefont {Troyer}}]{Kuklov2008DQCP}%
  \BibitemOpen
  \bibfield  {author} {\bibinfo {author} {\bibfnamefont {A.~B.}\ \bibnamefont
  {Kuklov}}, \bibinfo {author} {\bibfnamefont {M.}~\bibnamefont {Matsumoto}},
  \bibinfo {author} {\bibfnamefont {N.~V.}\ \bibnamefont {Prokof'ev}}, \bibinfo
  {author} {\bibfnamefont {B.~V.}\ \bibnamefont {Svistunov}}, \ and\ \bibinfo
  {author} {\bibfnamefont {M.}~\bibnamefont {Troyer}},\ }\bibfield  {title}
  {\enquote {\bibinfo {title} {Deconfined criticality: Generic first-order
  transition in the su(2) symmetry case},}\ }\href {\doibase
  10.1103/PhysRevLett.101.050405} {\bibfield  {journal} {\bibinfo  {journal}
  {Phys. Rev. Lett.}\ }\textbf {\bibinfo {volume} {101}},\ \bibinfo {pages}
  {050405} (\bibinfo {year} {2008})}\BibitemShut {NoStop}%
\bibitem [{\citenamefont {Chen}\ \emph {et~al.}(2009)\citenamefont {Chen},
  \citenamefont {Gukelberger}, \citenamefont {Trebst}, \citenamefont {Alet},\
  and\ \citenamefont {Balents}}]{Chen2009DQCP}%
  \BibitemOpen
  \bibfield  {author} {\bibinfo {author} {\bibfnamefont {Gang}\ \bibnamefont
  {Chen}}, \bibinfo {author} {\bibfnamefont {Jan}\ \bibnamefont {Gukelberger}},
  \bibinfo {author} {\bibfnamefont {Simon}\ \bibnamefont {Trebst}}, \bibinfo
  {author} {\bibfnamefont {Fabien}\ \bibnamefont {Alet}}, \ and\ \bibinfo
  {author} {\bibfnamefont {Leon}\ \bibnamefont {Balents}},\ }\bibfield  {title}
  {\enquote {\bibinfo {title} {Coulomb gas transitions in three-dimensional
  classical dimer models},}\ }\href {\doibase 10.1103/PhysRevB.80.045112}
  {\bibfield  {journal} {\bibinfo  {journal} {Phys. Rev. B}\ }\textbf {\bibinfo
  {volume} {80}},\ \bibinfo {pages} {045112} (\bibinfo {year}
  {2009})}\BibitemShut {NoStop}%
\bibitem [{\citenamefont {Lou}\ \emph {et~al.}(2009)\citenamefont {Lou},
  \citenamefont {Sandvik},\ and\ \citenamefont {Kawashima}}]{Lou2009DQCP}%
  \BibitemOpen
  \bibfield  {author} {\bibinfo {author} {\bibfnamefont {Jie}\ \bibnamefont
  {Lou}}, \bibinfo {author} {\bibfnamefont {Anders~W.}\ \bibnamefont
  {Sandvik}}, \ and\ \bibinfo {author} {\bibfnamefont {Naoki}\ \bibnamefont
  {Kawashima}},\ }\bibfield  {title} {\enquote {\bibinfo {title}
  {Antiferromagnetic to valence-bond-solid transitions in two-dimensional
  $\text{SU}(n)$ heisenberg models with multispin interactions},}\ }\href
  {\doibase 10.1103/PhysRevB.80.180414} {\bibfield  {journal} {\bibinfo
  {journal} {Phys. Rev. B}\ }\textbf {\bibinfo {volume} {80}},\ \bibinfo
  {pages} {180414} (\bibinfo {year} {2009})}\BibitemShut {NoStop}%
\bibitem [{\citenamefont {Charrier}\ and\ \citenamefont
  {Alet}(2010)}]{Charrier2010DQCP}%
  \BibitemOpen
  \bibfield  {author} {\bibinfo {author} {\bibfnamefont {D.}~\bibnamefont
  {Charrier}}\ and\ \bibinfo {author} {\bibfnamefont {F.}~\bibnamefont
  {Alet}},\ }\bibfield  {title} {\enquote {\bibinfo {title} {Phase diagram of
  an extended classical dimer model},}\ }\href {\doibase
  10.1103/PhysRevB.82.014429} {\bibfield  {journal} {\bibinfo  {journal} {Phys.
  Rev. B}\ }\textbf {\bibinfo {volume} {82}},\ \bibinfo {pages} {014429}
  (\bibinfo {year} {2010})}\BibitemShut {NoStop}%
\bibitem [{\citenamefont {Nahum}\ \emph {et~al.}(2011)\citenamefont {Nahum},
  \citenamefont {Chalker}, \citenamefont {Serna}, \citenamefont {Ortu\~no},\
  and\ \citenamefont {Somoza}}]{LoopModel2011DQCP}%
  \BibitemOpen
  \bibfield  {author} {\bibinfo {author} {\bibfnamefont {Adam}\ \bibnamefont
  {Nahum}}, \bibinfo {author} {\bibfnamefont {J.~T.}\ \bibnamefont {Chalker}},
  \bibinfo {author} {\bibfnamefont {P.}~\bibnamefont {Serna}}, \bibinfo
  {author} {\bibfnamefont {M.}~\bibnamefont {Ortu\~no}}, \ and\ \bibinfo
  {author} {\bibfnamefont {A.~M.}\ \bibnamefont {Somoza}},\ }\bibfield  {title}
  {\enquote {\bibinfo {title} {3d loop models and the
  ${\mathrm{cp}}^{n\ensuremath{-}1}$ sigma model},}\ }\href {\doibase
  10.1103/PhysRevLett.107.110601} {\bibfield  {journal} {\bibinfo  {journal}
  {Phys. Rev. Lett.}\ }\textbf {\bibinfo {volume} {107}},\ \bibinfo {pages}
  {110601} (\bibinfo {year} {2011})}\BibitemShut {NoStop}%
\bibitem [{\citenamefont {Harada}\ \emph {et~al.}(2013)\citenamefont {Harada},
  \citenamefont {Suzuki}, \citenamefont {Okubo}, \citenamefont {Matsuo},
  \citenamefont {Lou}, \citenamefont {Watanabe}, \citenamefont {Todo},\ and\
  \citenamefont {Kawashima}}]{Harada2013DQCP}%
  \BibitemOpen
  \bibfield  {author} {\bibinfo {author} {\bibfnamefont {Kenji}\ \bibnamefont
  {Harada}}, \bibinfo {author} {\bibfnamefont {Takafumi}\ \bibnamefont
  {Suzuki}}, \bibinfo {author} {\bibfnamefont {Tsuyoshi}\ \bibnamefont
  {Okubo}}, \bibinfo {author} {\bibfnamefont {Haruhiko}\ \bibnamefont
  {Matsuo}}, \bibinfo {author} {\bibfnamefont {Jie}\ \bibnamefont {Lou}},
  \bibinfo {author} {\bibfnamefont {Hiroshi}\ \bibnamefont {Watanabe}},
  \bibinfo {author} {\bibfnamefont {Synge}\ \bibnamefont {Todo}}, \ and\
  \bibinfo {author} {\bibfnamefont {Naoki}\ \bibnamefont {Kawashima}},\
  }\bibfield  {title} {\enquote {\bibinfo {title} {Possibility of deconfined
  criticality in su($n$) heisenberg models at small $n$},}\ }\href {\doibase
  10.1103/PhysRevB.88.220408} {\bibfield  {journal} {\bibinfo  {journal} {Phys.
  Rev. B}\ }\textbf {\bibinfo {volume} {88}},\ \bibinfo {pages} {220408}
  (\bibinfo {year} {2013})}\BibitemShut {NoStop}%
\bibitem [{\citenamefont {Block}\ \emph {et~al.}(2013)\citenamefont {Block},
  \citenamefont {Melko},\ and\ \citenamefont {Kaul}}]{Block2013DQCP}%
  \BibitemOpen
  \bibfield  {author} {\bibinfo {author} {\bibfnamefont {Matthew~S.}\
  \bibnamefont {Block}}, \bibinfo {author} {\bibfnamefont {Roger~G.}\
  \bibnamefont {Melko}}, \ and\ \bibinfo {author} {\bibfnamefont {Ribhu~K.}\
  \bibnamefont {Kaul}},\ }\bibfield  {title} {\enquote {\bibinfo {title} {Fate
  of $\mathbb{C}{\mathbb{p}}^{N\ensuremath{-}1}$ fixed points with $q$
  monopoles},}\ }\href {\doibase 10.1103/PhysRevLett.111.137202} {\bibfield
  {journal} {\bibinfo  {journal} {Phys. Rev. Lett.}\ }\textbf {\bibinfo
  {volume} {111}},\ \bibinfo {pages} {137202} (\bibinfo {year}
  {2013})}\BibitemShut {NoStop}%
\bibitem [{\citenamefont {Bartosch}(2013)}]{DQCP_correction2013}%
  \BibitemOpen
  \bibfield  {author} {\bibinfo {author} {\bibfnamefont {Lorenz}\ \bibnamefont
  {Bartosch}},\ }\bibfield  {title} {\enquote {\bibinfo {title} {Corrections to
  scaling in the critical theory of deconfined criticality},}\ }\href {\doibase
  10.1103/PhysRevB.88.195140} {\bibfield  {journal} {\bibinfo  {journal} {Phys.
  Rev. B}\ }\textbf {\bibinfo {volume} {88}},\ \bibinfo {pages} {195140}
  (\bibinfo {year} {2013})}\BibitemShut {NoStop}%
\bibitem [{\citenamefont {Qin}\ \emph {et~al.}(2017)\citenamefont {Qin},
  \citenamefont {He}, \citenamefont {You}, \citenamefont {Lu}, \citenamefont
  {Sen}, \citenamefont {Sandvik}, \citenamefont {Xu},\ and\ \citenamefont
  {Meng}}]{Meng2017DQCP}%
  \BibitemOpen
  \bibfield  {author} {\bibinfo {author} {\bibfnamefont {Yan~Qi}\ \bibnamefont
  {Qin}}, \bibinfo {author} {\bibfnamefont {Yuan-Yao}\ \bibnamefont {He}},
  \bibinfo {author} {\bibfnamefont {Yi-Zhuang}\ \bibnamefont {You}}, \bibinfo
  {author} {\bibfnamefont {Zhong-Yi}\ \bibnamefont {Lu}}, \bibinfo {author}
  {\bibfnamefont {Arnab}\ \bibnamefont {Sen}}, \bibinfo {author} {\bibfnamefont
  {Anders~W.}\ \bibnamefont {Sandvik}}, \bibinfo {author} {\bibfnamefont
  {Cenke}\ \bibnamefont {Xu}}, \ and\ \bibinfo {author} {\bibfnamefont
  {Zi~Yang}\ \bibnamefont {Meng}},\ }\bibfield  {title} {\enquote {\bibinfo
  {title} {Duality between the deconfined quantum-critical point and the
  bosonic topological transition},}\ }\href {\doibase
  10.1103/PhysRevX.7.031052} {\bibfield  {journal} {\bibinfo  {journal} {Phys.
  Rev. X}\ }\textbf {\bibinfo {volume} {7}},\ \bibinfo {pages} {031052}
  (\bibinfo {year} {2017})}\BibitemShut {NoStop}%
\bibitem [{\citenamefont {Sato}\ \emph {et~al.}(2017)\citenamefont {Sato},
  \citenamefont {Hohenadler},\ and\ \citenamefont {Assaad}}]{Sato2017DQCP}%
  \BibitemOpen
  \bibfield  {author} {\bibinfo {author} {\bibfnamefont {Toshihiro}\
  \bibnamefont {Sato}}, \bibinfo {author} {\bibfnamefont {Martin}\ \bibnamefont
  {Hohenadler}}, \ and\ \bibinfo {author} {\bibfnamefont {Fakher~F.}\
  \bibnamefont {Assaad}},\ }\bibfield  {title} {\enquote {\bibinfo {title}
  {Dirac fermions with competing orders: Non-landau transition with emergent
  symmetry},}\ }\href {\doibase 10.1103/PhysRevLett.119.197203} {\bibfield
  {journal} {\bibinfo  {journal} {Phys. Rev. Lett.}\ }\textbf {\bibinfo
  {volume} {119}},\ \bibinfo {pages} {197203} (\bibinfo {year}
  {2017})}\BibitemShut {NoStop}%
\bibitem [{\citenamefont {Shao}\ \emph {et~al.}(2017)\citenamefont {Shao},
  \citenamefont {Qin}, \citenamefont {Capponi}, \citenamefont {Chesi},
  \citenamefont {Meng},\ and\ \citenamefont {Sandvik}}]{Shao2017DQCP}%
  \BibitemOpen
  \bibfield  {author} {\bibinfo {author} {\bibfnamefont {Hui}\ \bibnamefont
  {Shao}}, \bibinfo {author} {\bibfnamefont {Yan~Qi}\ \bibnamefont {Qin}},
  \bibinfo {author} {\bibfnamefont {Sylvain}\ \bibnamefont {Capponi}}, \bibinfo
  {author} {\bibfnamefont {Stefano}\ \bibnamefont {Chesi}}, \bibinfo {author}
  {\bibfnamefont {Zi~Yang}\ \bibnamefont {Meng}}, \ and\ \bibinfo {author}
  {\bibfnamefont {Anders~W.}\ \bibnamefont {Sandvik}},\ }\bibfield  {title}
  {\enquote {\bibinfo {title} {Nearly deconfined spinon excitations in the
  square-lattice spin-$1/2$ heisenberg antiferromagnet},}\ }\href {\doibase
  10.1103/PhysRevX.7.041072} {\bibfield  {journal} {\bibinfo  {journal} {Phys.
  Rev. X}\ }\textbf {\bibinfo {volume} {7}},\ \bibinfo {pages} {041072}
  (\bibinfo {year} {2017})}\BibitemShut {NoStop}%
\bibitem [{\citenamefont {Ippoliti}\ \emph {et~al.}(2018)\citenamefont
  {Ippoliti}, \citenamefont {Mong}, \citenamefont {Assaad},\ and\ \citenamefont
  {Zaletel}}]{Ippoliti2018DQCP}%
  \BibitemOpen
  \bibfield  {author} {\bibinfo {author} {\bibfnamefont {Matteo}\ \bibnamefont
  {Ippoliti}}, \bibinfo {author} {\bibfnamefont {Roger S.~K.}\ \bibnamefont
  {Mong}}, \bibinfo {author} {\bibfnamefont {Fakher~F.}\ \bibnamefont
  {Assaad}}, \ and\ \bibinfo {author} {\bibfnamefont {Michael~P.}\ \bibnamefont
  {Zaletel}},\ }\bibfield  {title} {\enquote {\bibinfo {title} {Half-filled
  landau levels: A continuum and sign-free regularization for three-dimensional
  quantum critical points},}\ }\href {\doibase 10.1103/PhysRevB.98.235108}
  {\bibfield  {journal} {\bibinfo  {journal} {Phys. Rev. B}\ }\textbf {\bibinfo
  {volume} {98}},\ \bibinfo {pages} {235108} (\bibinfo {year}
  {2018})}\BibitemShut {NoStop}%
\bibitem [{\citenamefont {Lee}\ \emph {et~al.}(2018)\citenamefont {Lee},
  \citenamefont {Wang}, \citenamefont {Zaletel}, \citenamefont {Vishwanath},\
  and\ \citenamefont {He}}]{Lee2018FQH}%
  \BibitemOpen
  \bibfield  {author} {\bibinfo {author} {\bibfnamefont {Jong~Yeon}\
  \bibnamefont {Lee}}, \bibinfo {author} {\bibfnamefont {Chong}\ \bibnamefont
  {Wang}}, \bibinfo {author} {\bibfnamefont {Michael~P.}\ \bibnamefont
  {Zaletel}}, \bibinfo {author} {\bibfnamefont {Ashvin}\ \bibnamefont
  {Vishwanath}}, \ and\ \bibinfo {author} {\bibfnamefont {Yin-Chen}\
  \bibnamefont {He}},\ }\bibfield  {title} {\enquote {\bibinfo {title}
  {Emergent multi-flavor ${\mathrm{qed}}_{3}$ at the plateau transition between
  fractional chern insulators: Applications to graphene heterostructures},}\
  }\href {\doibase 10.1103/PhysRevX.8.031015} {\bibfield  {journal} {\bibinfo
  {journal} {Phys. Rev. X}\ }\textbf {\bibinfo {volume} {8}},\ \bibinfo {pages}
  {031015} (\bibinfo {year} {2018})}\BibitemShut {NoStop}%
\bibitem [{\citenamefont {Zhao}\ \emph {et~al.}(2019)\citenamefont {Zhao},
  \citenamefont {Weinberg},\ and\ \citenamefont {Sandvik}}]{Sandvik2018O4}%
  \BibitemOpen
  \bibfield  {author} {\bibinfo {author} {\bibfnamefont {Bowen}\ \bibnamefont
  {Zhao}}, \bibinfo {author} {\bibfnamefont {Phillip}\ \bibnamefont
  {Weinberg}}, \ and\ \bibinfo {author} {\bibfnamefont {Anders~W.}\
  \bibnamefont {Sandvik}},\ }\bibfield  {title} {\enquote {\bibinfo {title}
  {Symmetry-enhanced discontinuous phase transition in a two-dimensional
  quantum magnet},}\ }\href {\doibase 10.1038/s41567-019-0484-x} {\bibfield
  {journal} {\bibinfo  {journal} {Nature Physics}\ }\textbf {\bibinfo {volume}
  {15}},\ \bibinfo {pages} {678--682} (\bibinfo {year} {2019})}\BibitemShut
  {NoStop}%
\bibitem [{\citenamefont {Serna}\ and\ \citenamefont
  {Nahum}(2019)}]{Nahum2018O4}%
  \BibitemOpen
  \bibfield  {author} {\bibinfo {author} {\bibfnamefont {Pablo}\ \bibnamefont
  {Serna}}\ and\ \bibinfo {author} {\bibfnamefont {Adam}\ \bibnamefont
  {Nahum}},\ }\bibfield  {title} {\enquote {\bibinfo {title} {Emergence and
  spontaneous breaking of approximate $\mathrm{O}(4)$ symmetry at a weakly
  first-order deconfined phase transition},}\ }\href {\doibase
  10.1103/PhysRevB.99.195110} {\bibfield  {journal} {\bibinfo  {journal} {Phys.
  Rev. B}\ }\textbf {\bibinfo {volume} {99}},\ \bibinfo {pages} {195110}
  (\bibinfo {year} {2019})}\BibitemShut {NoStop}%
\bibitem [{\citenamefont {Lee}\ \emph {et~al.}(2019)\citenamefont {Lee},
  \citenamefont {You}, \citenamefont {Sachdev},\ and\ \citenamefont
  {Vishwanath}}]{Lee2019DQCP}%
  \BibitemOpen
  \bibfield  {author} {\bibinfo {author} {\bibfnamefont {Jong~Yeon}\
  \bibnamefont {Lee}}, \bibinfo {author} {\bibfnamefont {Yi-Zhuang}\
  \bibnamefont {You}}, \bibinfo {author} {\bibfnamefont {Subir}\ \bibnamefont
  {Sachdev}}, \ and\ \bibinfo {author} {\bibfnamefont {Ashvin}\ \bibnamefont
  {Vishwanath}},\ }\bibfield  {title} {\enquote {\bibinfo {title} {Signatures
  of a deconfined phase transition on the shastry-sutherland lattice:
  Applications to quantum critical
  ${\mathrm{srcu}}_{2}({\mathrm{bo}}_{3}{)}_{2}$},}\ }\href {\doibase
  10.1103/PhysRevX.9.041037} {\bibfield  {journal} {\bibinfo  {journal} {Phys.
  Rev. X}\ }\textbf {\bibinfo {volume} {9}},\ \bibinfo {pages} {041037}
  (\bibinfo {year} {2019})}\BibitemShut {NoStop}%
\bibitem [{\citenamefont {Huang}\ \emph {et~al.}(2019)\citenamefont {Huang},
  \citenamefont {Lu}, \citenamefont {You}, \citenamefont {Meng},\ and\
  \citenamefont {Xiang}}]{Huang2019}%
  \BibitemOpen
  \bibfield  {author} {\bibinfo {author} {\bibfnamefont {Rui-Zhen}\
  \bibnamefont {Huang}}, \bibinfo {author} {\bibfnamefont {Da-Chuan}\
  \bibnamefont {Lu}}, \bibinfo {author} {\bibfnamefont {Yi-Zhuang}\
  \bibnamefont {You}}, \bibinfo {author} {\bibfnamefont {Zi~Yang}\ \bibnamefont
  {Meng}}, \ and\ \bibinfo {author} {\bibfnamefont {Tao}\ \bibnamefont
  {Xiang}},\ }\bibfield  {title} {\enquote {\bibinfo {title} {Emergent symmetry
  and conserved current at a one-dimensional incarnation of deconfined quantum
  critical point},}\ }\href {\doibase 10.1103/PhysRevB.100.125137} {\bibfield
  {journal} {\bibinfo  {journal} {Phys. Rev. B}\ }\textbf {\bibinfo {volume}
  {100}},\ \bibinfo {pages} {125137} (\bibinfo {year} {2019})}\BibitemShut
  {NoStop}%
\bibitem [{\citenamefont {Jiang}\ and\ \citenamefont
  {Motrunich}(2019)}]{Jiang2019DQCP}%
  \BibitemOpen
  \bibfield  {author} {\bibinfo {author} {\bibfnamefont {Shenghan}\
  \bibnamefont {Jiang}}\ and\ \bibinfo {author} {\bibfnamefont {Olexei}\
  \bibnamefont {Motrunich}},\ }\bibfield  {title} {\enquote {\bibinfo {title}
  {Ising ferromagnet to valence bond solid transition in a one-dimensional spin
  chain: Analogies to deconfined quantum critical points},}\ }\href {\doibase
  10.1103/PhysRevB.99.075103} {\bibfield  {journal} {\bibinfo  {journal} {Phys.
  Rev. B}\ }\textbf {\bibinfo {volume} {99}},\ \bibinfo {pages} {075103}
  (\bibinfo {year} {2019})}\BibitemShut {NoStop}%
\bibitem [{\citenamefont {Mudry}\ \emph {et~al.}(2019)\citenamefont {Mudry},
  \citenamefont {Furusaki}, \citenamefont {Morimoto},\ and\ \citenamefont
  {Hikihara}}]{Mudry2019DQCP}%
  \BibitemOpen
  \bibfield  {author} {\bibinfo {author} {\bibfnamefont {Christopher}\
  \bibnamefont {Mudry}}, \bibinfo {author} {\bibfnamefont {Akira}\ \bibnamefont
  {Furusaki}}, \bibinfo {author} {\bibfnamefont {Takahiro}\ \bibnamefont
  {Morimoto}}, \ and\ \bibinfo {author} {\bibfnamefont {Toshiya}\ \bibnamefont
  {Hikihara}},\ }\bibfield  {title} {\enquote {\bibinfo {title} {Quantum phase
  transitions beyond landau-ginzburg theory in one-dimensional space
  revisited},}\ }\href {\doibase 10.1103/PhysRevB.99.205153} {\bibfield
  {journal} {\bibinfo  {journal} {Phys. Rev. B}\ }\textbf {\bibinfo {volume}
  {99}},\ \bibinfo {pages} {205153} (\bibinfo {year} {2019})}\BibitemShut
  {NoStop}%
\bibitem [{\citenamefont {Huang}\ and\ \citenamefont {Yin}(2020)}]{KZ_DQCP}%
  \BibitemOpen
  \bibfield  {author} {\bibinfo {author} {\bibfnamefont {Rui-Zhen}\
  \bibnamefont {Huang}}\ and\ \bibinfo {author} {\bibfnamefont {Shuai}\
  \bibnamefont {Yin}},\ }\bibfield  {title} {\enquote {\bibinfo {title}
  {Kibble-zurek mechanism for a one-dimensional incarnation of a deconfined
  quantum critical point},}\ }\href {\doibase 10.1103/PhysRevResearch.2.023175}
  {\bibfield  {journal} {\bibinfo  {journal} {Phys. Rev. Research}\ }\textbf
  {\bibinfo {volume} {2}},\ \bibinfo {pages} {023175} (\bibinfo {year}
  {2020})}\BibitemShut {NoStop}%
\bibitem [{\citenamefont {Roberts}\ \emph {et~al.}(2021)\citenamefont
  {Roberts}, \citenamefont {Jiang},\ and\ \citenamefont
  {Motrunich}}]{Robert2021DQCP}%
  \BibitemOpen
  \bibfield  {author} {\bibinfo {author} {\bibfnamefont {Brenden}\ \bibnamefont
  {Roberts}}, \bibinfo {author} {\bibfnamefont {Shenghan}\ \bibnamefont
  {Jiang}}, \ and\ \bibinfo {author} {\bibfnamefont {Olexei~I.}\ \bibnamefont
  {Motrunich}},\ }\bibfield  {title} {\enquote {\bibinfo {title}
  {One-dimensional model for deconfined criticality with
  ${\mathbb{z}}_{3}\ifmmode\times\else\texttimes\fi{}{\mathbb{z}}_{3}$
  symmetry},}\ }\href {\doibase 10.1103/PhysRevB.103.155143} {\bibfield
  {journal} {\bibinfo  {journal} {Phys. Rev. B}\ }\textbf {\bibinfo {volume}
  {103}},\ \bibinfo {pages} {155143} (\bibinfo {year} {2021})}\BibitemShut
  {NoStop}%
\bibitem [{\citenamefont {Zou}\ and\ \citenamefont {He}(2020)}]{Zou2020}%
  \BibitemOpen
  \bibfield  {author} {\bibinfo {author} {\bibfnamefont {Liujun}\ \bibnamefont
  {Zou}}\ and\ \bibinfo {author} {\bibfnamefont {Yin-Chen}\ \bibnamefont
  {He}},\ }\bibfield  {title} {\enquote {\bibinfo {title} {Field-induced
  ${\mathrm{qcd}}_{3}$-chern-simons quantum criticalities in kitaev
  materials},}\ }\href {\doibase 10.1103/PhysRevResearch.2.013072} {\bibfield
  {journal} {\bibinfo  {journal} {Phys. Rev. Research}\ }\textbf {\bibinfo
  {volume} {2}},\ \bibinfo {pages} {013072} (\bibinfo {year}
  {2020})}\BibitemShut {NoStop}%
\bibitem [{\citenamefont {Slagle}\ \emph {et~al.}(2022)\citenamefont {Slagle},
  \citenamefont {Liu}, \citenamefont {Aasen}, \citenamefont {Pichler},
  \citenamefont {Mong}, \citenamefont {Chen}, \citenamefont {Endres},\ and\
  \citenamefont {Alicea}}]{aliceaIsing}%
  \BibitemOpen
  \bibfield  {author} {\bibinfo {author} {\bibfnamefont {Kevin}\ \bibnamefont
  {Slagle}}, \bibinfo {author} {\bibfnamefont {Yue}\ \bibnamefont {Liu}},
  \bibinfo {author} {\bibfnamefont {David}\ \bibnamefont {Aasen}}, \bibinfo
  {author} {\bibfnamefont {Hannes}\ \bibnamefont {Pichler}}, \bibinfo {author}
  {\bibfnamefont {Roger S.~K.}\ \bibnamefont {Mong}}, \bibinfo {author}
  {\bibfnamefont {Xie}\ \bibnamefont {Chen}}, \bibinfo {author} {\bibfnamefont
  {Manuel}\ \bibnamefont {Endres}}, \ and\ \bibinfo {author} {\bibfnamefont
  {Jason}\ \bibnamefont {Alicea}},\ }\href {\doibase 10.48550/ARXIV.2204.00013}
  {\enquote {\bibinfo {title} {Quantum spin liquids bootstrapped from ising
  criticality in rydberg arrays},}\ } (\bibinfo {year} {2022})\BibitemShut
  {NoStop}%
\bibitem [{\citenamefont {Bernien}\ \emph {et~al.}(2017)\citenamefont
  {Bernien}, \citenamefont {Schwartz}, \citenamefont {Keesling}, \citenamefont
  {Levine}, \citenamefont {Omran}, \citenamefont {Pichler}, \citenamefont
  {Choi}, \citenamefont {Zibrov}, \citenamefont {Endres}, \citenamefont
  {Greiner}, \citenamefont {Vuleti{\'c}},\ and\ \citenamefont
  {Lukin}}]{scar2017}%
  \BibitemOpen
  \bibfield  {author} {\bibinfo {author} {\bibfnamefont {Hannes}\ \bibnamefont
  {Bernien}}, \bibinfo {author} {\bibfnamefont {Sylvain}\ \bibnamefont
  {Schwartz}}, \bibinfo {author} {\bibfnamefont {Alexander}\ \bibnamefont
  {Keesling}}, \bibinfo {author} {\bibfnamefont {Harry}\ \bibnamefont
  {Levine}}, \bibinfo {author} {\bibfnamefont {Ahmed}\ \bibnamefont {Omran}},
  \bibinfo {author} {\bibfnamefont {Hannes}\ \bibnamefont {Pichler}}, \bibinfo
  {author} {\bibfnamefont {Soonwon}\ \bibnamefont {Choi}}, \bibinfo {author}
  {\bibfnamefont {Alexander~S.}\ \bibnamefont {Zibrov}}, \bibinfo {author}
  {\bibfnamefont {Manuel}\ \bibnamefont {Endres}}, \bibinfo {author}
  {\bibfnamefont {Markus}\ \bibnamefont {Greiner}}, \bibinfo {author}
  {\bibfnamefont {Vladan}\ \bibnamefont {Vuleti{\'c}}}, \ and\ \bibinfo
  {author} {\bibfnamefont {Mikhail~D.}\ \bibnamefont {Lukin}},\ }\bibfield
  {title} {\enquote {\bibinfo {title} {Probing many-body dynamics on a 51-atom
  quantum simulator},}\ }\href {\doibase 10.1038/nature24622} {\bibfield
  {journal} {\bibinfo  {journal} {Nature}\ }\textbf {\bibinfo {volume} {551}},\
  \bibinfo {pages} {579--584} (\bibinfo {year} {2017})}\BibitemShut {NoStop}%
\bibitem [{\citenamefont {de~Léséleuc}\ \emph {et~al.}(2019)\citenamefont
  {de~Léséleuc}, \citenamefont {Lienhard}, \citenamefont {Scholl},
  \citenamefont {Barredo}, \citenamefont {Weber}, \citenamefont {Lang},
  \citenamefont {Büchler}, \citenamefont {Lahaye},\ and\ \citenamefont
  {Browaeys}}]{SPT2019}%
  \BibitemOpen
  \bibfield  {author} {\bibinfo {author} {\bibfnamefont {Sylvain}\ \bibnamefont
  {de~Léséleuc}}, \bibinfo {author} {\bibfnamefont {Vincent}\ \bibnamefont
  {Lienhard}}, \bibinfo {author} {\bibfnamefont {Pascal}\ \bibnamefont
  {Scholl}}, \bibinfo {author} {\bibfnamefont {Daniel}\ \bibnamefont
  {Barredo}}, \bibinfo {author} {\bibfnamefont {Sebastian}\ \bibnamefont
  {Weber}}, \bibinfo {author} {\bibfnamefont {Nicolai}\ \bibnamefont {Lang}},
  \bibinfo {author} {\bibfnamefont {Hans~Peter}\ \bibnamefont {Büchler}},
  \bibinfo {author} {\bibfnamefont {Thierry}\ \bibnamefont {Lahaye}}, \ and\
  \bibinfo {author} {\bibfnamefont {Antoine}\ \bibnamefont {Browaeys}},\
  }\bibfield  {title} {\enquote {\bibinfo {title} {Observation of a
  symmetry-protected topological phase of interacting bosons with rydberg
  atoms},}\ }\href {\doibase 10.1126/science.aav9105} {\bibfield  {journal}
  {\bibinfo  {journal} {Science}\ }\textbf {\bibinfo {volume} {365}},\ \bibinfo
  {pages} {775--780} (\bibinfo {year} {2019})}\BibitemShut {NoStop}%
\bibitem [{\citenamefont {Keesling}\ \emph {et~al.}(2019)\citenamefont
  {Keesling}, \citenamefont {Omran}, \citenamefont {Levine}, \citenamefont
  {Bernien}, \citenamefont {Pichler}, \citenamefont {Choi}, \citenamefont
  {Samajdar}, \citenamefont {Schwartz}, \citenamefont {Silvi}, \citenamefont
  {Sachdev}, \citenamefont {Zoller}, \citenamefont {Endres}, \citenamefont
  {Greiner}, \citenamefont {Vuleti{\'c}},\ and\ \citenamefont
  {Lukin}}]{KZ2019}%
  \BibitemOpen
  \bibfield  {author} {\bibinfo {author} {\bibfnamefont {Alexander}\
  \bibnamefont {Keesling}}, \bibinfo {author} {\bibfnamefont {Ahmed}\
  \bibnamefont {Omran}}, \bibinfo {author} {\bibfnamefont {Harry}\ \bibnamefont
  {Levine}}, \bibinfo {author} {\bibfnamefont {Hannes}\ \bibnamefont
  {Bernien}}, \bibinfo {author} {\bibfnamefont {Hannes}\ \bibnamefont
  {Pichler}}, \bibinfo {author} {\bibfnamefont {Soonwon}\ \bibnamefont {Choi}},
  \bibinfo {author} {\bibfnamefont {Rhine}\ \bibnamefont {Samajdar}}, \bibinfo
  {author} {\bibfnamefont {Sylvain}\ \bibnamefont {Schwartz}}, \bibinfo
  {author} {\bibfnamefont {Pietro}\ \bibnamefont {Silvi}}, \bibinfo {author}
  {\bibfnamefont {Subir}\ \bibnamefont {Sachdev}}, \bibinfo {author}
  {\bibfnamefont {Peter}\ \bibnamefont {Zoller}}, \bibinfo {author}
  {\bibfnamefont {Manuel}\ \bibnamefont {Endres}}, \bibinfo {author}
  {\bibfnamefont {Markus}\ \bibnamefont {Greiner}}, \bibinfo {author}
  {\bibfnamefont {Vladan}\ \bibnamefont {Vuleti{\'c}}}, \ and\ \bibinfo
  {author} {\bibfnamefont {Mikhail~D.}\ \bibnamefont {Lukin}},\ }\bibfield
  {title} {\enquote {\bibinfo {title} {Quantum kibble--zurek mechanism and
  critical dynamics on a programmable rydberg simulator},}\ }\href {\doibase
  10.1038/s41586-019-1070-1} {\bibfield  {journal} {\bibinfo  {journal}
  {Nature}\ }\textbf {\bibinfo {volume} {568}},\ \bibinfo {pages} {207--211}
  (\bibinfo {year} {2019})}\BibitemShut {NoStop}%
\bibitem [{\citenamefont {Semeghini}\ \emph {et~al.}(2021)\citenamefont
  {Semeghini}, \citenamefont {Levine}, \citenamefont {Keesling}, \citenamefont
  {Ebadi}, \citenamefont {Wang}, \citenamefont {Bluvstein}, \citenamefont
  {Verresen}, \citenamefont {Pichler}, \citenamefont {Kalinowski},
  \citenamefont {Samajdar}, \citenamefont {Omran}, \citenamefont {Sachdev},
  \citenamefont {Vishwanath}, \citenamefont {Greiner}, \citenamefont
  {Vuletić},\ and\ \citenamefont {Lukin}}]{SL2021}%
  \BibitemOpen
  \bibfield  {author} {\bibinfo {author} {\bibfnamefont {G.}~\bibnamefont
  {Semeghini}}, \bibinfo {author} {\bibfnamefont {H.}~\bibnamefont {Levine}},
  \bibinfo {author} {\bibfnamefont {A.}~\bibnamefont {Keesling}}, \bibinfo
  {author} {\bibfnamefont {S.}~\bibnamefont {Ebadi}}, \bibinfo {author}
  {\bibfnamefont {T.~T.}\ \bibnamefont {Wang}}, \bibinfo {author}
  {\bibfnamefont {D.}~\bibnamefont {Bluvstein}}, \bibinfo {author}
  {\bibfnamefont {R.}~\bibnamefont {Verresen}}, \bibinfo {author}
  {\bibfnamefont {H.}~\bibnamefont {Pichler}}, \bibinfo {author} {\bibfnamefont
  {M.}~\bibnamefont {Kalinowski}}, \bibinfo {author} {\bibfnamefont
  {R.}~\bibnamefont {Samajdar}}, \bibinfo {author} {\bibfnamefont
  {A.}~\bibnamefont {Omran}}, \bibinfo {author} {\bibfnamefont
  {S.}~\bibnamefont {Sachdev}}, \bibinfo {author} {\bibfnamefont
  {A.}~\bibnamefont {Vishwanath}}, \bibinfo {author} {\bibfnamefont
  {M.}~\bibnamefont {Greiner}}, \bibinfo {author} {\bibfnamefont
  {V.}~\bibnamefont {Vuletić}}, \ and\ \bibinfo {author} {\bibfnamefont
  {M.~D.}\ \bibnamefont {Lukin}},\ }\bibfield  {title} {\enquote {\bibinfo
  {title} {Probing topological spin liquids on a programmable quantum
  simulator},}\ }\href {\doibase 10.1126/science.abi8794} {\bibfield  {journal}
  {\bibinfo  {journal} {Science}\ }\textbf {\bibinfo {volume} {374}},\ \bibinfo
  {pages} {1242--1247} (\bibinfo {year} {2021})}\BibitemShut {NoStop}%
\bibitem [{\citenamefont {Graham}\ \emph {et~al.}(2022)\citenamefont {Graham},
  \citenamefont {Song}, \citenamefont {Scott}, \citenamefont {Poole},
  \citenamefont {Phuttitarn}, \citenamefont {Jooya}, \citenamefont {Eichler},
  \citenamefont {Jiang}, \citenamefont {Marra}, \citenamefont {Grinkemeyer},
  \citenamefont {Kwon}, \citenamefont {Ebert}, \citenamefont {Cherek},
  \citenamefont {Lichtman}, \citenamefont {Gillette}, \citenamefont {Gilbert},
  \citenamefont {Bowman}, \citenamefont {Ballance}, \citenamefont {Campbell},
  \citenamefont {Dahl}, \citenamefont {Crawford}, \citenamefont {Blunt},
  \citenamefont {Rogers}, \citenamefont {Noel},\ and\ \citenamefont
  {Saffman}}]{Saffman}%
  \BibitemOpen
  \bibfield  {author} {\bibinfo {author} {\bibfnamefont {T.~M.}\ \bibnamefont
  {Graham}}, \bibinfo {author} {\bibfnamefont {Y.}~\bibnamefont {Song}},
  \bibinfo {author} {\bibfnamefont {J.}~\bibnamefont {Scott}}, \bibinfo
  {author} {\bibfnamefont {C.}~\bibnamefont {Poole}}, \bibinfo {author}
  {\bibfnamefont {L.}~\bibnamefont {Phuttitarn}}, \bibinfo {author}
  {\bibfnamefont {K.}~\bibnamefont {Jooya}}, \bibinfo {author} {\bibfnamefont
  {P.}~\bibnamefont {Eichler}}, \bibinfo {author} {\bibfnamefont
  {X.}~\bibnamefont {Jiang}}, \bibinfo {author} {\bibfnamefont
  {A.}~\bibnamefont {Marra}}, \bibinfo {author} {\bibfnamefont
  {B.}~\bibnamefont {Grinkemeyer}}, \bibinfo {author} {\bibfnamefont
  {M.}~\bibnamefont {Kwon}}, \bibinfo {author} {\bibfnamefont {M.}~\bibnamefont
  {Ebert}}, \bibinfo {author} {\bibfnamefont {J.}~\bibnamefont {Cherek}},
  \bibinfo {author} {\bibfnamefont {M.~T.}\ \bibnamefont {Lichtman}}, \bibinfo
  {author} {\bibfnamefont {M.}~\bibnamefont {Gillette}}, \bibinfo {author}
  {\bibfnamefont {J.}~\bibnamefont {Gilbert}}, \bibinfo {author} {\bibfnamefont
  {D.}~\bibnamefont {Bowman}}, \bibinfo {author} {\bibfnamefont
  {T.}~\bibnamefont {Ballance}}, \bibinfo {author} {\bibfnamefont
  {C.}~\bibnamefont {Campbell}}, \bibinfo {author} {\bibfnamefont {E.~D.}\
  \bibnamefont {Dahl}}, \bibinfo {author} {\bibfnamefont {O.}~\bibnamefont
  {Crawford}}, \bibinfo {author} {\bibfnamefont {N.~S.}\ \bibnamefont {Blunt}},
  \bibinfo {author} {\bibfnamefont {B.}~\bibnamefont {Rogers}}, \bibinfo
  {author} {\bibfnamefont {T.}~\bibnamefont {Noel}}, \ and\ \bibinfo {author}
  {\bibfnamefont {M.}~\bibnamefont {Saffman}},\ }\bibfield  {title} {\enquote
  {\bibinfo {title} {Multi-qubit entanglement and algorithms on a neutral-atom
  quantum computer},}\ }\href {\doibase 10.1038/s41586-022-04603-6} {\bibfield
  {journal} {\bibinfo  {journal} {Nature}\ }\textbf {\bibinfo {volume} {604}},\
  \bibinfo {pages} {457--462} (\bibinfo {year} {2022})}\BibitemShut {NoStop}%
\bibitem [{SM()}]{SM}%
  \BibitemOpen
  \href@noop {} {}\bibinfo {note} {See Supplemental Material}\BibitemShut
  {NoStop}%
\bibitem [{\citenamefont {Lieb}\ \emph {et~al.}(1961)\citenamefont {Lieb},
  \citenamefont {Schultz},\ and\ \citenamefont {Mattis}}]{LSM_original}%
  \BibitemOpen
  \bibfield  {author} {\bibinfo {author} {\bibfnamefont {Elliott}\ \bibnamefont
  {Lieb}}, \bibinfo {author} {\bibfnamefont {Theodore}\ \bibnamefont
  {Schultz}}, \ and\ \bibinfo {author} {\bibfnamefont {Daniel}\ \bibnamefont
  {Mattis}},\ }\bibfield  {title} {\enquote {\bibinfo {title} {Two soluble
  models of an antiferromagnetic chain},}\ }\href {\doibase
  https://doi.org/10.1016/0003-4916(61)90115-4} {\bibfield  {journal} {\bibinfo
   {journal} {Annals of Physics}\ }\textbf {\bibinfo {volume} {16}},\ \bibinfo
  {pages} {407 -- 466} (\bibinfo {year} {1961})}\BibitemShut {NoStop}%
\bibitem [{\citenamefont {Lieb}\ \emph {et~al.}(2004)\citenamefont {Lieb},
  \citenamefont {Schultz},\ and\ \citenamefont {Mattis}}]{LSM}%
  \BibitemOpen
  \bibfield  {author} {\bibinfo {author} {\bibfnamefont {E.}~\bibnamefont
  {Lieb}}, \bibinfo {author} {\bibfnamefont {T.}~\bibnamefont {Schultz}}, \
  and\ \bibinfo {author} {\bibfnamefont {D.}~\bibnamefont {Mattis}},\
  }\href@noop {} {\emph {\bibinfo {title} {Condensed Matter Physics and Exactly
  Soluble Models}}}\ (\bibinfo  {publisher} {Springer, New York},\ \bibinfo
  {year} {2004})\BibitemShut {NoStop}%
\bibitem [{\citenamefont {Hastings}(2005)}]{Hastings2005}%
  \BibitemOpen
  \bibfield  {author} {\bibinfo {author} {\bibfnamefont {M.~B.}\ \bibnamefont
  {Hastings}},\ }\bibfield  {title} {\enquote {\bibinfo {title} {Sufficient
  conditions for topological order in insulators},}\ }\href
  {http://stacks.iop.org/0295-5075/70/i=6/a=824} {\bibfield  {journal}
  {\bibinfo  {journal} {EPL (Europhysics Letters)}\ }\textbf {\bibinfo {volume}
  {70}},\ \bibinfo {pages} {824} (\bibinfo {year} {2005})}\BibitemShut
  {NoStop}%
\bibitem [{\citenamefont {Oshikawa}(2000)}]{Oshikawa2000}%
  \BibitemOpen
  \bibfield  {author} {\bibinfo {author} {\bibfnamefont {Masaki}\ \bibnamefont
  {Oshikawa}},\ }\bibfield  {title} {\enquote {\bibinfo {title}
  {Commensurability, excitation gap, and topology in quantum many-particle
  systems on a periodic lattice},}\ }\href {\doibase
  10.1103/PhysRevLett.84.1535} {\bibfield  {journal} {\bibinfo  {journal}
  {Phys. Rev. Lett.}\ }\textbf {\bibinfo {volume} {84}},\ \bibinfo {pages}
  {1535--1538} (\bibinfo {year} {2000})}\BibitemShut {NoStop}%
\bibitem [{\citenamefont {Parreira}\ \emph {et~al.}(1997)\citenamefont
  {Parreira}, \citenamefont {Bolina},\ and\ \citenamefont
  {Perez}}]{Parreira1997}%
  \BibitemOpen
  \bibfield  {author} {\bibinfo {author} {\bibfnamefont {J~Rodrigo}\
  \bibnamefont {Parreira}}, \bibinfo {author} {\bibfnamefont {O}~\bibnamefont
  {Bolina}}, \ and\ \bibinfo {author} {\bibfnamefont {J~Fernando}\ \bibnamefont
  {Perez}},\ }\bibfield  {title} {\enquote {\bibinfo {title} {N{\'{e}}el order
  in the ground state of heisenberg antiferromagnetic chains with long-range
  interactions},}\ }\href {\doibase 10.1088/0305-4470/30/4/012} {\bibfield
  {journal} {\bibinfo  {journal} {Journal of Physics A: Mathematical and
  General}\ }\textbf {\bibinfo {volume} {30}},\ \bibinfo {pages} {1095--1100}
  (\bibinfo {year} {1997})}\BibitemShut {NoStop}%
\bibitem [{\citenamefont {Laflorencie}\ \emph {et~al.}(2005)\citenamefont
  {Laflorencie}, \citenamefont {Affleck},\ and\ \citenamefont
  {Berciu}}]{Laflorencie2005}%
  \BibitemOpen
  \bibfield  {author} {\bibinfo {author} {\bibfnamefont {Nicolas}\ \bibnamefont
  {Laflorencie}}, \bibinfo {author} {\bibfnamefont {Ian}\ \bibnamefont
  {Affleck}}, \ and\ \bibinfo {author} {\bibfnamefont {Mona}\ \bibnamefont
  {Berciu}},\ }\bibfield  {title} {\enquote {\bibinfo {title} {Critical
  phenomena and quantum phase transition in long range heisenberg
  antiferromagnetic chains},}\ }\href {\doibase
  10.1088/1742-5468/2005/12/p12001} {\bibfield  {journal} {\bibinfo  {journal}
  {Journal of Statistical Mechanics: Theory and Experiment}\ }\textbf {\bibinfo
  {volume} {2005}},\ \bibinfo {pages} {P12001--P12001} (\bibinfo {year}
  {2005})}\BibitemShut {NoStop}%
\bibitem [{\citenamefont {Maghrebi}\ \emph {et~al.}(2017)\citenamefont
  {Maghrebi}, \citenamefont {Gong},\ and\ \citenamefont
  {Gorshkov}}]{absenceCSB2017}%
  \BibitemOpen
  \bibfield  {author} {\bibinfo {author} {\bibfnamefont {Mohammad~F.}\
  \bibnamefont {Maghrebi}}, \bibinfo {author} {\bibfnamefont {Zhe-Xuan}\
  \bibnamefont {Gong}}, \ and\ \bibinfo {author} {\bibfnamefont {Alexey~V.}\
  \bibnamefont {Gorshkov}},\ }\bibfield  {title} {\enquote {\bibinfo {title}
  {Continuous symmetry breaking in 1d long-range interacting quantum
  systems},}\ }\href {\doibase 10.1103/PhysRevLett.119.023001} {\bibfield
  {journal} {\bibinfo  {journal} {Phys. Rev. Lett.}\ }\textbf {\bibinfo
  {volume} {119}},\ \bibinfo {pages} {023001} (\bibinfo {year}
  {2017})}\BibitemShut {NoStop}%
\bibitem [{\citenamefont {Zarubin}\ \emph {et~al.}(2020)\citenamefont
  {Zarubin}, \citenamefont {Kassan-Ogly},\ and\ \citenamefont
  {Proshkin}}]{Zarubin2020}%
  \BibitemOpen
  \bibfield  {author} {\bibinfo {author} {\bibfnamefont {A.V.}\ \bibnamefont
  {Zarubin}}, \bibinfo {author} {\bibfnamefont {F.A.}\ \bibnamefont
  {Kassan-Ogly}}, \ and\ \bibinfo {author} {\bibfnamefont {A.I.}\ \bibnamefont
  {Proshkin}},\ }\bibfield  {title} {\enquote {\bibinfo {title} {Frustrations
  in the ising chain with the third-neighbor interactions},}\ }\href {\doibase
  10.1016/j.jmmm.2020.167144} {\bibfield  {journal} {\bibinfo  {journal}
  {Journal of Magnetism and Magnetic Materials}\ }\textbf {\bibinfo {volume}
  {514}},\ \bibinfo {pages} {167144} (\bibinfo {year} {2020})}\BibitemShut
  {NoStop}%
\bibitem [{\citenamefont {Majumdar}\ and\ \citenamefont
  {Ghosh}(1969)}]{MG1970}%
  \BibitemOpen
  \bibfield  {author} {\bibinfo {author} {\bibfnamefont {Chanchal~K.}\
  \bibnamefont {Majumdar}}\ and\ \bibinfo {author} {\bibfnamefont {Dipan~K.}\
  \bibnamefont {Ghosh}},\ }\bibfield  {title} {\enquote {\bibinfo {title} {On
  next‐nearest‐neighbor interaction in linear chain. i},}\ }\href {\doibase
  10.1063/1.1664978} {\bibfield  {journal} {\bibinfo  {journal} {Journal of
  Mathematical Physics}\ }\textbf {\bibinfo {volume} {10}},\ \bibinfo {pages}
  {1388--1398} (\bibinfo {year} {1969})}\BibitemShut {NoStop}%
\bibitem [{\citenamefont {White}(1992)}]{DMRG1}%
  \BibitemOpen
  \bibfield  {author} {\bibinfo {author} {\bibfnamefont {Steven~R.}\
  \bibnamefont {White}},\ }\bibfield  {title} {\enquote {\bibinfo {title}
  {Density matrix formulation for quantum renormalization groups},}\ }\href
  {\doibase 10.1103/PhysRevLett.69.2863} {\bibfield  {journal} {\bibinfo
  {journal} {Phys. Rev. Lett.}\ }\textbf {\bibinfo {volume} {69}},\ \bibinfo
  {pages} {2863--2866} (\bibinfo {year} {1992})}\BibitemShut {NoStop}%
\bibitem [{\citenamefont {White}(1993)}]{DMRG2}%
  \BibitemOpen
  \bibfield  {author} {\bibinfo {author} {\bibfnamefont {Steven~R.}\
  \bibnamefont {White}},\ }\bibfield  {title} {\enquote {\bibinfo {title}
  {Density-matrix algorithms for quantum renormalization groups},}\ }\href
  {\doibase 10.1103/PhysRevB.48.10345} {\bibfield  {journal} {\bibinfo
  {journal} {Phys. Rev. B}\ }\textbf {\bibinfo {volume} {48}},\ \bibinfo
  {pages} {10345--10356} (\bibinfo {year} {1993})}\BibitemShut {NoStop}%
\bibitem [{\citenamefont {Schollw\"ock}(2005)}]{DMRG3}%
  \BibitemOpen
  \bibfield  {author} {\bibinfo {author} {\bibfnamefont {U.}~\bibnamefont
  {Schollw\"ock}},\ }\bibfield  {title} {\enquote {\bibinfo {title} {The
  density-matrix renormalization group},}\ }\href {\doibase
  10.1103/RevModPhys.77.259} {\bibfield  {journal} {\bibinfo  {journal} {Rev.
  Mod. Phys.}\ }\textbf {\bibinfo {volume} {77}},\ \bibinfo {pages} {259--315}
  (\bibinfo {year} {2005})}\BibitemShut {NoStop}%
\bibitem [{iDM()}]{iDMRG_misc}%
  \BibitemOpen
  \href@noop {} {}\bibinfo {note} {In the DMRG simulation for infinite systems,
  the variationally optimized matrix product state explicitly breaks the
  symmetry in a spontaneously symmetry broken phase. Therefore, the order
  parameter expectation values for the resulting groundstate can take finite
  values}\BibitemShut {NoStop}%
\bibitem [{\citenamefont {Calabrese}\ and\ \citenamefont
  {Cardy}(2009)}]{Calabrese_2009}%
  \BibitemOpen
  \bibfield  {author} {\bibinfo {author} {\bibfnamefont {Pasquale}\
  \bibnamefont {Calabrese}}\ and\ \bibinfo {author} {\bibfnamefont {John}\
  \bibnamefont {Cardy}},\ }\bibfield  {title} {\enquote {\bibinfo {title}
  {Entanglement entropy and conformal field theory},}\ }\href {\doibase
  10.1088/1751-8113/42/50/504005} {\bibfield  {journal} {\bibinfo  {journal}
  {Journal of Physics A: Mathematical and Theoretical}\ }\textbf {\bibinfo
  {volume} {42}},\ \bibinfo {pages} {504005} (\bibinfo {year}
  {2009})}\BibitemShut {NoStop}%
\bibitem [{\citenamefont {{Berezinski{\v{i}}}}(1971)}]{BKT_1}%
  \BibitemOpen
  \bibfield  {author} {\bibinfo {author} {\bibfnamefont {V.~L.}\ \bibnamefont
  {{Berezinski{\v{i}}}}},\ }\bibfield  {title} {\enquote {\bibinfo {title}
  {{Destruction of Long-range Order in One-dimensional and Two-dimensional
  Systems having a Continuous Symmetry Group I. Classical Systems}},}\
  }\href@noop {} {\bibfield  {journal} {\bibinfo  {journal} {Soviet Journal of
  Experimental and Theoretical Physics}\ }\textbf {\bibinfo {volume} {32}},\
  \bibinfo {pages} {493} (\bibinfo {year} {1971})}\BibitemShut {NoStop}%
\bibitem [{\citenamefont {Kosterlitz}\ and\ \citenamefont
  {Thouless}(1973)}]{BKT_2}%
  \BibitemOpen
  \bibfield  {author} {\bibinfo {author} {\bibfnamefont {J~M}\ \bibnamefont
  {Kosterlitz}}\ and\ \bibinfo {author} {\bibfnamefont {D~J}\ \bibnamefont
  {Thouless}},\ }\bibfield  {title} {\enquote {\bibinfo {title} {Ordering,
  metastability and phase transitions in two-dimensional systems},}\ }\href
  {\doibase 10.1088/0022-3719/6/7/010} {\bibfield  {journal} {\bibinfo
  {journal} {Journal of Physics C: Solid State Physics}\ }\textbf {\bibinfo
  {volume} {6}},\ \bibinfo {pages} {1181--1203} (\bibinfo {year}
  {1973})}\BibitemShut {NoStop}%
\bibitem [{BKT()}]{BKT_misc}%
  \BibitemOpen
  \href@noop {} {}\bibinfo {note} {Similarly, the QPT between XY and VBS is
  consistent with a BKT transition.}\BibitemShut {Stop}%
\bibitem [{\citenamefont {Nomura}\ and\ \citenamefont
  {Okamoto}(1994)}]{Nomura_1994}%
  \BibitemOpen
  \bibfield  {author} {\bibinfo {author} {\bibfnamefont {K}~\bibnamefont
  {Nomura}}\ and\ \bibinfo {author} {\bibfnamefont {K}~\bibnamefont
  {Okamoto}},\ }\bibfield  {title} {\enquote {\bibinfo {title} {Critical
  properties of s= 1/2 antiferromagnetic {XXZ} chain with
  next-nearest-neighbour interactions},}\ }\href {\doibase
  10.1088/0305-4470/27/17/012} {\bibfield  {journal} {\bibinfo  {journal}
  {Journal of Physics A: Mathematical and General}\ }\textbf {\bibinfo {volume}
  {27}},\ \bibinfo {pages} {5773--5788} (\bibinfo {year} {1994})}\BibitemShut
  {NoStop}%
\bibitem [{\citenamefont {Ueda}\ and\ \citenamefont
  {Oshikawa}(2021)}]{OshikawaKT2021}%
  \BibitemOpen
  \bibfield  {author} {\bibinfo {author} {\bibfnamefont {Atsushi}\ \bibnamefont
  {Ueda}}\ and\ \bibinfo {author} {\bibfnamefont {Masaki}\ \bibnamefont
  {Oshikawa}},\ }\bibfield  {title} {\enquote {\bibinfo {title} {Resolving the
  berezinskii-kosterlitz-thouless transition in the two-dimensional xy model
  with tensor-network-based level spectroscopy},}\ }\href {\doibase
  10.1103/PhysRevB.104.165132} {\bibfield  {journal} {\bibinfo  {journal}
  {Phys. Rev. B}\ }\textbf {\bibinfo {volume} {104}},\ \bibinfo {pages}
  {165132} (\bibinfo {year} {2021})}\BibitemShut {NoStop}%
\bibitem [{\citenamefont {Haldane}(1982)}]{Haldane1982DQCP}%
  \BibitemOpen
  \bibfield  {author} {\bibinfo {author} {\bibfnamefont {F.~D.~M.}\
  \bibnamefont {Haldane}},\ }\bibfield  {title} {\enquote {\bibinfo {title}
  {Spontaneous dimerization in the $s=\frac{1}{2}$ heisenberg antiferromagnetic
  chain with competing interactions},}\ }\href {\doibase
  10.1103/PhysRevB.25.4925} {\bibfield  {journal} {\bibinfo  {journal} {Phys.
  Rev. B}\ }\textbf {\bibinfo {volume} {25}},\ \bibinfo {pages} {4925--4928}
  (\bibinfo {year} {1982})}\BibitemShut {NoStop}%
\bibitem [{\citenamefont {Giamarchi}(2004)}]{Giamarchi2004}%
  \BibitemOpen
  \bibfield  {author} {\bibinfo {author} {\bibfnamefont {Thierry}\ \bibnamefont
  {Giamarchi}},\ }\href {\doibase 10.1093/acprof:oso/9780198525004.001.0001}
  {\emph {\bibinfo {title} {{Quantum physics in one dimension}}}},\
  International series of monographs on physics\ (\bibinfo  {publisher}
  {Clarendon Press},\ \bibinfo {address} {Oxford},\ \bibinfo {year}
  {2004})\BibitemShut {NoStop}%
\bibitem [{LR_()}]{LR_misc}%
  \BibitemOpen
  \href@noop {} {}\bibinfo {note} {We ignored long-range terms $1/r^{m}$ with
  $m\geq 3$; if one writes down such a long-range interaction in terms of
  operators with positive scaling dimensions, the interaction can be shown to
  be always irrelevant.}\BibitemShut {Stop}%
\bibitem [{\citenamefont {Omran}\ \emph {et~al.}(2019)\citenamefont {Omran},
  \citenamefont {Levine}, \citenamefont {Keesling}, \citenamefont {Semeghini},
  \citenamefont {Wang}, \citenamefont {Ebadi}, \citenamefont {Bernien},
  \citenamefont {Zibrov}, \citenamefont {Pichler}, \citenamefont {Choi},
  \citenamefont {Cui}, \citenamefont {Rossignolo}, \citenamefont {Rembold},
  \citenamefont {Montangero}, \citenamefont {Calarco}, \citenamefont {Endres},
  \citenamefont {Greiner}, \citenamefont {Vuletić},\ and\ \citenamefont
  {Lukin}}]{cat_state}%
  \BibitemOpen
  \bibfield  {author} {\bibinfo {author} {\bibfnamefont {A.}~\bibnamefont
  {Omran}}, \bibinfo {author} {\bibfnamefont {H.}~\bibnamefont {Levine}},
  \bibinfo {author} {\bibfnamefont {A.}~\bibnamefont {Keesling}}, \bibinfo
  {author} {\bibfnamefont {G.}~\bibnamefont {Semeghini}}, \bibinfo {author}
  {\bibfnamefont {T.~T.}\ \bibnamefont {Wang}}, \bibinfo {author}
  {\bibfnamefont {S.}~\bibnamefont {Ebadi}}, \bibinfo {author} {\bibfnamefont
  {H.}~\bibnamefont {Bernien}}, \bibinfo {author} {\bibfnamefont {A.~S.}\
  \bibnamefont {Zibrov}}, \bibinfo {author} {\bibfnamefont {H.}~\bibnamefont
  {Pichler}}, \bibinfo {author} {\bibfnamefont {S.}~\bibnamefont {Choi}},
  \bibinfo {author} {\bibfnamefont {J.}~\bibnamefont {Cui}}, \bibinfo {author}
  {\bibfnamefont {M.}~\bibnamefont {Rossignolo}}, \bibinfo {author}
  {\bibfnamefont {P.}~\bibnamefont {Rembold}}, \bibinfo {author} {\bibfnamefont
  {S.}~\bibnamefont {Montangero}}, \bibinfo {author} {\bibfnamefont
  {T.}~\bibnamefont {Calarco}}, \bibinfo {author} {\bibfnamefont
  {M.}~\bibnamefont {Endres}}, \bibinfo {author} {\bibfnamefont
  {M.}~\bibnamefont {Greiner}}, \bibinfo {author} {\bibfnamefont
  {V.}~\bibnamefont {Vuletić}}, \ and\ \bibinfo {author} {\bibfnamefont
  {M.~D.}\ \bibnamefont {Lukin}},\ }\bibfield  {title} {\enquote {\bibinfo
  {title} {Generation and manipulation of schrodinger cat states in rydberg
  atom arrays},}\ }\href {\doibase 10.1126/science.aax9743} {\bibfield
  {journal} {\bibinfo  {journal} {Science}\ }\textbf {\bibinfo {volume}
  {365}},\ \bibinfo {pages} {570--574} (\bibinfo {year} {2019})}\BibitemShut
  {NoStop}%
\bibitem [{\citenamefont {Young}\ \emph {et~al.}(2021)\citenamefont {Young},
  \citenamefont {Bienias}, \citenamefont {Belyansky}, \citenamefont {Kaufman},\
  and\ \citenamefont {Gorshkov}}]{adjust_dipolar2021}%
  \BibitemOpen
  \bibfield  {author} {\bibinfo {author} {\bibfnamefont {Jeremy~T.}\
  \bibnamefont {Young}}, \bibinfo {author} {\bibfnamefont {Przemyslaw}\
  \bibnamefont {Bienias}}, \bibinfo {author} {\bibfnamefont {Ron}\ \bibnamefont
  {Belyansky}}, \bibinfo {author} {\bibfnamefont {Adam~M.}\ \bibnamefont
  {Kaufman}}, \ and\ \bibinfo {author} {\bibfnamefont {Alexey~V.}\ \bibnamefont
  {Gorshkov}},\ }\bibfield  {title} {\enquote {\bibinfo {title} {Asymmetric
  blockade and multiqubit gates via dipole-dipole interactions},}\ }\href
  {\doibase 10.1103/PhysRevLett.127.120501} {\bibfield  {journal} {\bibinfo
  {journal} {Phys. Rev. Lett.}\ }\textbf {\bibinfo {volume} {127}},\ \bibinfo
  {pages} {120501} (\bibinfo {year} {2021})}\BibitemShut {NoStop}%
\bibitem [{\citenamefont {Beterov}\ \emph {et~al.}(2020)\citenamefont
  {Beterov}, \citenamefont {Tretyakov}, \citenamefont {Entin}, \citenamefont
  {Yakshina}, \citenamefont {Ryabtsev}, \citenamefont {Saffman},\ and\
  \citenamefont {Bergamini}}]{beterov2020}%
  \BibitemOpen
  \bibfield  {author} {\bibinfo {author} {\bibfnamefont {I~I}\ \bibnamefont
  {Beterov}}, \bibinfo {author} {\bibfnamefont {D~B}\ \bibnamefont
  {Tretyakov}}, \bibinfo {author} {\bibfnamefont {V~M}\ \bibnamefont {Entin}},
  \bibinfo {author} {\bibfnamefont {E~A}\ \bibnamefont {Yakshina}}, \bibinfo
  {author} {\bibfnamefont {I~I}\ \bibnamefont {Ryabtsev}}, \bibinfo {author}
  {\bibfnamefont {M}~\bibnamefont {Saffman}}, \ and\ \bibinfo {author}
  {\bibfnamefont {S}~\bibnamefont {Bergamini}},\ }\bibfield  {title} {\enquote
  {\bibinfo {title} {Application of adiabatic passage in rydberg atomic
  ensembles for quantum information processing},}\ }\href {\doibase
  10.1088/1361-6455/ab8719} {\bibfield  {journal} {\bibinfo  {journal} {Journal
  of Physics B: Atomic, Molecular and Optical Physics}\ }\textbf {\bibinfo
  {volume} {53}},\ \bibinfo {pages} {182001} (\bibinfo {year}
  {2020})}\BibitemShut {NoStop}%
\bibitem [{\citenamefont {Kibble}(1976)}]{Kibble_1976}%
  \BibitemOpen
  \bibfield  {author} {\bibinfo {author} {\bibfnamefont {T~W~B}\ \bibnamefont
  {Kibble}},\ }\bibfield  {title} {\enquote {\bibinfo {title} {Topology of
  cosmic domains and strings},}\ }\href {\doibase 10.1088/0305-4470/9/8/029}
  {\bibfield  {journal} {\bibinfo  {journal} {Journal of Physics A:
  Mathematical and General}\ }\textbf {\bibinfo {volume} {9}},\ \bibinfo
  {pages} {1387--1398} (\bibinfo {year} {1976})}\BibitemShut {NoStop}%
\bibitem [{\citenamefont {Zurek}(1985)}]{Zurek1985}%
  \BibitemOpen
  \bibfield  {author} {\bibinfo {author} {\bibfnamefont {W.~H.}\ \bibnamefont
  {Zurek}},\ }\bibfield  {title} {\enquote {\bibinfo {title} {Cosmological
  experiments in superfluid helium?}}\ }\href {\doibase 10.1038/317505a0}
  {\bibfield  {journal} {\bibinfo  {journal} {Nature}\ }\textbf {\bibinfo
  {volume} {317}},\ \bibinfo {pages} {505--508} (\bibinfo {year}
  {1985})}\BibitemShut {NoStop}%
\bibitem [{Sin()}]{Singlet_misc}%
  \BibitemOpen
  \href@noop {} {}\bibinfo {note} {From the field-theoretic
  derivation~\cite{Orignac1998}, $O'_\textrm{VBS}$ can be shown to behave as
  $\sin 2\phi$ as well. Also, due to the emergent singlet nature of the VBS
  phase, correlations in $X$\,$,$\,$Y$\,$,$\,$Z$ are approximately similar
  which we verified numerically throughout the phase.}\BibitemShut {Stop}%
\bibitem [{\citenamefont {Stewart}(2001)}]{NFL_review1}%
  \BibitemOpen
  \bibfield  {author} {\bibinfo {author} {\bibfnamefont {G.~R.}\ \bibnamefont
  {Stewart}},\ }\bibfield  {title} {\enquote {\bibinfo {title}
  {Non-fermi-liquid behavior in $d$- and $f$-electron metals},}\ }\href
  {\doibase 10.1103/RevModPhys.73.797} {\bibfield  {journal} {\bibinfo
  {journal} {Rev. Mod. Phys.}\ }\textbf {\bibinfo {volume} {73}},\ \bibinfo
  {pages} {797--855} (\bibinfo {year} {2001})}\BibitemShut {NoStop}%
\bibitem [{\citenamefont {Lee}(2018)}]{NFL_review2}%
  \BibitemOpen
  \bibfield  {author} {\bibinfo {author} {\bibfnamefont {Sung-Sik}\
  \bibnamefont {Lee}},\ }\bibfield  {title} {\enquote {\bibinfo {title} {Recent
  developments in non-fermi liquid theory},}\ }\href {\doibase
  10.1146/annurev-conmatphys-031016-025531} {\bibfield  {journal} {\bibinfo
  {journal} {Annual Review of Condensed Matter Physics}\ }\textbf {\bibinfo
  {volume} {9}},\ \bibinfo {pages} {227--244} (\bibinfo {year}
  {2018})}\BibitemShut {NoStop}%
\bibitem [{\citenamefont {Mukhopadhyay}\ \emph
  {et~al.}(2001{\natexlab{a}})\citenamefont {Mukhopadhyay}, \citenamefont
  {Kane},\ and\ \citenamefont {Lubensky}}]{crosssliding_NFL}%
  \BibitemOpen
  \bibfield  {author} {\bibinfo {author} {\bibfnamefont {Ranjan}\ \bibnamefont
  {Mukhopadhyay}}, \bibinfo {author} {\bibfnamefont {C.~L.}\ \bibnamefont
  {Kane}}, \ and\ \bibinfo {author} {\bibfnamefont {T.~C.}\ \bibnamefont
  {Lubensky}},\ }\bibfield  {title} {\enquote {\bibinfo {title} {Crossed
  sliding luttinger liquid phase},}\ }\href {\doibase
  10.1103/PhysRevB.63.081103} {\bibfield  {journal} {\bibinfo  {journal} {Phys.
  Rev. B}\ }\textbf {\bibinfo {volume} {63}},\ \bibinfo {pages} {081103}
  (\bibinfo {year} {2001}{\natexlab{a}})}\BibitemShut {NoStop}%
\bibitem [{\citenamefont {Mukhopadhyay}\ \emph
  {et~al.}(2001{\natexlab{b}})\citenamefont {Mukhopadhyay}, \citenamefont
  {Kane},\ and\ \citenamefont {Lubensky}}]{SlidingLL2001}%
  \BibitemOpen
  \bibfield  {author} {\bibinfo {author} {\bibfnamefont {Ranjan}\ \bibnamefont
  {Mukhopadhyay}}, \bibinfo {author} {\bibfnamefont {C.~L.}\ \bibnamefont
  {Kane}}, \ and\ \bibinfo {author} {\bibfnamefont {T.~C.}\ \bibnamefont
  {Lubensky}},\ }\bibfield  {title} {\enquote {\bibinfo {title} {Sliding
  luttinger liquid phases},}\ }\href {\doibase 10.1103/PhysRevB.64.045120}
  {\bibfield  {journal} {\bibinfo  {journal} {Phys. Rev. B}\ }\textbf {\bibinfo
  {volume} {64}},\ \bibinfo {pages} {045120} (\bibinfo {year}
  {2001}{\natexlab{b}})}\BibitemShut {NoStop}%
\bibitem [{\citenamefont {Leviatan}\ and\ \citenamefont
  {Mross}(2020)}]{Leviatan2020}%
  \BibitemOpen
  \bibfield  {author} {\bibinfo {author} {\bibfnamefont {Eyal}\ \bibnamefont
  {Leviatan}}\ and\ \bibinfo {author} {\bibfnamefont {David~F.}\ \bibnamefont
  {Mross}},\ }\bibfield  {title} {\enquote {\bibinfo {title} {Unification of
  parton and coupled-wire approaches to quantum magnetism in two dimensions},}\
  }\href {\doibase 10.1103/PhysRevResearch.2.043437} {\bibfield  {journal}
  {\bibinfo  {journal} {Phys. Rev. Research}\ }\textbf {\bibinfo {volume}
  {2}},\ \bibinfo {pages} {043437} (\bibinfo {year} {2020})}\BibitemShut
  {NoStop}%
\bibitem [{\citenamefont {Orignac}\ and\ \citenamefont
  {Giamarchi}(1998)}]{Orignac1998}%
  \BibitemOpen
  \bibfield  {author} {\bibinfo {author} {\bibfnamefont {E.}~\bibnamefont
  {Orignac}}\ and\ \bibinfo {author} {\bibfnamefont {T.}~\bibnamefont
  {Giamarchi}},\ }\bibfield  {title} {\enquote {\bibinfo {title} {Weakly
  disordered spin ladders},}\ }\href {\doibase 10.1103/PhysRevB.57.5812}
  {\bibfield  {journal} {\bibinfo  {journal} {Phys. Rev. B}\ }\textbf {\bibinfo
  {volume} {57}},\ \bibinfo {pages} {5812--5829} (\bibinfo {year}
  {1998})}\BibitemShut {NoStop}%
\bibitem [{\citenamefont {Whitlock}\ \emph {et~al.}(2017)\citenamefont
  {Whitlock}, \citenamefont {Glaetzle},\ and\ \citenamefont
  {Hannaford}}]{Whitlock_2017}%
  \BibitemOpen
  \bibfield  {author} {\bibinfo {author} {\bibfnamefont {Shannon}\ \bibnamefont
  {Whitlock}}, \bibinfo {author} {\bibfnamefont {Alexander~W}\ \bibnamefont
  {Glaetzle}}, \ and\ \bibinfo {author} {\bibfnamefont {Peter}\ \bibnamefont
  {Hannaford}},\ }\bibfield  {title} {\enquote {\bibinfo {title} {Simulating
  quantum spin models using rydberg-excited atomic ensembles in magnetic
  microtrap arrays},}\ }\href {\doibase 10.1088/1361-6455/aa6149} {\bibfield
  {journal} {\bibinfo  {journal} {Journal of Physics B: Atomic, Molecular and
  Optical Physics}\ }\textbf {\bibinfo {volume} {50}},\ \bibinfo {pages}
  {074001} (\bibinfo {year} {2017})}\BibitemShut {NoStop}%
\bibitem [{Note1()}]{Note1}%
  \BibitemOpen
  \bibinfo {note} {More precisely, one investigates the finite-size scaling
  behaviors of an \protect \emph {squared} order parameter $O$ detecting
  symmetry-breaking, since $\expval {O}=0$ in the ground state of a finite size
  system, which is symmetric.}\BibitemShut {Stop}%
\bibitem [{\citenamefont {Di~Francesco}\ \emph {et~al.}(1997)\citenamefont
  {Di~Francesco}, \citenamefont {Mathieu},\ and\ \citenamefont
  {Senechal}}]{DiFrancesco:1997nk}%
  \BibitemOpen
  \bibfield  {author} {\bibinfo {author} {\bibfnamefont {P.}~\bibnamefont
  {Di~Francesco}}, \bibinfo {author} {\bibfnamefont {P.}~\bibnamefont
  {Mathieu}}, \ and\ \bibinfo {author} {\bibfnamefont {D.}~\bibnamefont
  {Senechal}},\ }\href {\doibase 10.1007/978-1-4612-2256-9} {\emph {\bibinfo
  {title} {{Conformal Field Theory}}}},\ Graduate Texts in Contemporary
  Physics\ (\bibinfo  {publisher} {Springer-Verlag},\ \bibinfo {address} {New
  York},\ \bibinfo {year} {1997})\BibitemShut {NoStop}%
\bibitem [{Note2()}]{Note2}%
  \BibitemOpen
  \bibinfo {note} {This follows from a radial quantization in the CFT through a
  logarithmic mapping which transforms the plane to the cylinder.}\BibitemShut
  {Stop}%
\bibitem [{Note3()}]{Note3}%
  \BibitemOpen
  \bibinfo {note} {Note that the above field-theoretic analysis cannot capture
  the QzAFM phase described in the main text; while the QzAFM phase breaks the
  translation symmetry in such a way that it is invariant under multiple of
  $T_x^4$, the spin variables written in terms of $(\phi ,\theta )$ in
  Eq.\protect \tmspace +\thinmuskip {.1667em}\protect \textup {\hbox
  {\mathsurround \z@ \protect \normalfont (\ignorespaces \ref
  {eq:operator_mapping}\unskip \@@italiccorr )}} are always symmetric under
  $T_x^2$.}\BibitemShut {Stop}%
\bibitem [{\citenamefont {Browaeys}\ \emph {et~al.}(2016)\citenamefont
  {Browaeys}, \citenamefont {Barredo},\ and\ \citenamefont
  {Lahaye}}]{Browaeys_2016}%
  \BibitemOpen
  \bibfield  {author} {\bibinfo {author} {\bibfnamefont {Antoine}\ \bibnamefont
  {Browaeys}}, \bibinfo {author} {\bibfnamefont {Daniel}\ \bibnamefont
  {Barredo}}, \ and\ \bibinfo {author} {\bibfnamefont {Thierry}\ \bibnamefont
  {Lahaye}},\ }\bibfield  {title} {\enquote {\bibinfo {title} {Experimental
  investigations of dipole{\textendash}dipole interactions between a few
  rydberg atoms},}\ }\href {\doibase 10.1088/0953-4075/49/15/152001} {\bibfield
   {journal} {\bibinfo  {journal} {Journal of Physics B: Atomic, Molecular and
  Optical Physics}\ }\textbf {\bibinfo {volume} {49}},\ \bibinfo {pages}
  {152001} (\bibinfo {year} {2016})}\BibitemShut {NoStop}%
\bibitem [{Note4()}]{Note4}%
  \BibitemOpen
  \bibinfo {note} {Appreciable second order exchange couplings \cite
  {Whitlock_2017} can arise via dipole-dipole coupling of, for example, both
  $\ket {nSn'S}$ and $\ket {n'SnS}$ to states like $\ket {nPnP}$, if $\Delta E
  = E_{nSn'S} + E_{n'SnS} - 2 E_{nPnP} \approx 0 $. As this could conspire to
  give nonzero exchange couplings even when the microwave drives are zero,
  making it difficult to access the regime of small $J_3/J_6$, for simplicity
  we assume this is negligible. We can ensure access to the small $J_3/J_6$
  regime in a few ways, including either by choosing states such that $\Delta
  E$ is sufficiently large (for example in alkalis by choosing $\ket {\uparrow
  }$ to be a $D$ state with very different quantum defect), or such that the
  dipole coupling on the microwave driven transition is small}\BibitemShut
  {NoStop}%
\bibitem [{\citenamefont {Weber}\ \emph {et~al.}(2017)\citenamefont {Weber},
  \citenamefont {Tresp}, \citenamefont {Menke}, \citenamefont {Urvoy},
  \citenamefont {Firstenberg}, \citenamefont {Büchler},\ and\ \citenamefont
  {Hofferberth}}]{Weber_2017}%
  \BibitemOpen
  \bibfield  {author} {\bibinfo {author} {\bibfnamefont {Sebastian}\
  \bibnamefont {Weber}}, \bibinfo {author} {\bibfnamefont {Christoph}\
  \bibnamefont {Tresp}}, \bibinfo {author} {\bibfnamefont {Henri}\ \bibnamefont
  {Menke}}, \bibinfo {author} {\bibfnamefont {Alban}\ \bibnamefont {Urvoy}},
  \bibinfo {author} {\bibfnamefont {Ofer}\ \bibnamefont {Firstenberg}},
  \bibinfo {author} {\bibfnamefont {Hans~Peter}\ \bibnamefont {Büchler}}, \
  and\ \bibinfo {author} {\bibfnamefont {Sebastian}\ \bibnamefont
  {Hofferberth}},\ }\bibfield  {title} {\enquote {\bibinfo {title} {Calculation
  of rydberg interaction potentials},}\ }\href {\doibase
  10.1088/1361-6455/aa743a} {\bibfield  {journal} {\bibinfo  {journal} {Journal
  of Physics B: Atomic, Molecular and Optical Physics}\ }\textbf {\bibinfo
  {volume} {50}},\ \bibinfo {pages} {133001} (\bibinfo {year}
  {2017})}\BibitemShut {NoStop}%
\bibitem [{\citenamefont {Zaletel}\ \emph {et~al.}(2015)\citenamefont
  {Zaletel}, \citenamefont {Mong}, \citenamefont {Karrasch}, \citenamefont
  {Moore},\ and\ \citenamefont {Pollmann}}]{LR_Zaletel}%
  \BibitemOpen
  \bibfield  {author} {\bibinfo {author} {\bibfnamefont {Michael~P.}\
  \bibnamefont {Zaletel}}, \bibinfo {author} {\bibfnamefont {Roger S.~K.}\
  \bibnamefont {Mong}}, \bibinfo {author} {\bibfnamefont {Christoph}\
  \bibnamefont {Karrasch}}, \bibinfo {author} {\bibfnamefont {Joel~E.}\
  \bibnamefont {Moore}}, \ and\ \bibinfo {author} {\bibfnamefont {Frank}\
  \bibnamefont {Pollmann}},\ }\bibfield  {title} {\enquote {\bibinfo {title}
  {Time-evolving a matrix product state with long-ranged interactions},}\
  }\href {\doibase 10.1103/PhysRevB.91.165112} {\bibfield  {journal} {\bibinfo
  {journal} {Phys. Rev. B}\ }\textbf {\bibinfo {volume} {91}},\ \bibinfo
  {pages} {165112} (\bibinfo {year} {2015})}\BibitemShut {NoStop}%
\end{thebibliography}
\end{document}